\documentclass[a4]{article}
\usepackage[a4paper]{geometry}
\usepackage{cite}
\usepackage[pdftex]{graphicx}
\usepackage[cmex10]{amsmath}

\usepackage{times}
\usepackage{array}
\usepackage[tight,footnotesize]{subfigure}
\usepackage[font=footnotesize]{subfig}
\usepackage{graphicx}
\usepackage{mathtools}
\usepackage{amsmath}
\usepackage{enumitem}
\def\union{{\cup}}
\def\intersect{{\cap}}

\usepackage{alltt}

\def\boxx{{\vcenter{\vbox{\hrule height.3pt
          \hbox{\vrule width.3pt height6pt
          \kern6pt\vrule width.3pt}\hrule height.3pt}}\;}}

\def\impos{{\;\vcenter{\hbox{\rule{5mm}{0.2mm}}} \vcenter{\hbox{\rule{1.5mm}{1.5mm}}} \;}}

\def\lrarrow{\leftrightarrow \kern-8pt \rightarrow}

\def\Rows{\rm Rows}
\def\Columns{\rm Columns}

\def\thing{{\rm Thing}}

\def\2{\frac{1}{2}}

\def\adj{{\rm adj}}    % A transpose
\def\beq{\begin{eqnarray}}
\def\eeq{\end{eqnarray}}
\def\2{\frac{1}{2}}

\newtheorem{assumption}{Assumption}

\newtheorem{lemma}{Lemma}

\newtheorem{proposition}{Proposition}
\newtheorem{definition}{Definition}

\def\lrarrow{\leftrightarrow \kern-8pt \rightarrow}

\def\frightarrow{\rightarrow \kern-11pt /~~}
\def\reducesto{\simeq \kern -3pt >}
\def\intersection{\cap}

\usepackage{fancyhdr}					% To enable comprehensive headers and footers
\begin{document}
\newcommand{\strust}[1]{\stackrel{\tau:#1}{\longrightarrow}}
\newcommand{\trust}[1]{\stackrel{#1}{{\rm\bf ~Trusts~}}}
\newcommand{\promise}[1]{\xrightarrow{#1}}
\newcommand{\revpromise}[1]{\xleftarrow{#1} }
\newcommand{\assoc}[1]{{\xrightharpoondown{#1}} }
\newcommand{\imposition}[1]{\stackrel{#1}{\impos}}
\newcommand{\scopepromise}[2]{\xrightarrow[#2]{#1}}
\newcommand{\handshake}[1]{\xleftrightarrow{#1} \kern-8pt \xrightarrow{} }
\newcommand{\cpromise}[1]{\stackrel{#1}{\frightarrow}}
\newcommand{\policy}{\stackrel{P}{\equiv}}
\newcommand{\field}[1]{\mathbf{#1}}
\newcommand{\bundle}[1]{\stackrel{#1}{\Longrightarrow}}

\title{Spacetimes with Semantics (I)\\\small Notes on Theory and Formalism}

\author{Mark Burgess}

\maketitle

\begin{abstract}
  Relationships between objects constitute our notion of space.  When
  these relationships change we interpret this as the passage of time.
  Observer interpretations are essential to the way we understand
  these relationships. Hence observer semantics are an integral part of what
  we mean by spacetime.

  Semantics make up the essential difference in how one describes and
  uses the concept of space in physics, chemistry, biology and
  technology.  In these notes, I have tried to assemble what seems to
  be a set of natural, and pragmatic, considerations about discrete,
  finite spacetimes, to unify descriptions of these areas.

  It reviews familiar notions of spacetime, and brings them together
  into a less familiar framework of promise theory (autonomous
  agents), in order to illuminate the goal of encoding the semantics
  of observers into a description of spacetime itself.
  Autonomous agents provide an exacting atomic and local model for
  finite spacetime, which quickly reveals the issues of incomplete
  information and non-locality. From this we should be able to
  reconstruct all other notions of spacetime.

  The aim of this exercise is to apply related tools and ideas to an
  initial unification of real and artificial spaces, e.g. databases
  and information webs with natural spacetime.  By reconstructing
  these spaces from autonomous agents, we may better understand naming
  and coordinatization of semantic spaces, from crowds and swarms to
  datacentres and libraries, as well as the fundamental arena of natural science.
\end{abstract}

\tableofcontents

%%%%%%%%%%%%%%%%%%%%%%%%%%%%%%%%%%%%%%%%%%%%%%%%%%%%%%%%%%%%%%%%%%%%%%%%%%%%%%%%%%%%%%%

\section{Introduction} 

Descriptions of how things are arranged, and how they change,
give us our notions of space and time.  Between any two identifiable
objects, we may identify various kinds of relationship.  Of these many
relationships, spatial and temporal relationships have a special
meaning to us. We identify the idea of {\em location} with the elements of a space and
{\em direction} with the relationships between them.

Traditionally, theories of the world treat spacetime as a smooth
background latticework on which the things within it (matter and energy) move.  In
other words, spacetime is treated as something separate from matter
and energy, a kind of measuring apparatus into which we embed
processes, replete with coordinate labels.  However, in the physics
and chemistry of materials, the notion of place within a material is
not distinguished from the materials themselves, as it would be a
burden to do so.  Moreover, our representation of space in other cases
comes from measuring it along side a set of states we arrange for the
purpose. Thus the perception of distance requires an elapsed time, so
they two cannot be separated.  Indeed, measurement implies the ability
of an observer to compare different notions of distance by signalling
over an elapsed time, so what we are missing is a more general
observer view of spacetime.

One problem with current models of space and time is that they are
inconsistent with respect to locality. Space and time are laid out
with global properties and structure {\em a priori}, only then does
one make corrections the account for locality instead of the other way
around.  This is because observers are already considered to be
looking at the macroscopic scale in a smooth continuum approximation.
When spacetime is considered as arising from fundamental local
interactions at the smallest scales, the picture becomes different,
and familiar notions have to emerge by defining a continuum limit.
Indeed, at this level, there seems to be a fundamental non-locality
to information that is unavoidable.

Einstein was the first to illustrate how observer locality has
unexpected consequences, once one took seriously the relationship
between space and time. In mainstream information systems there are
even more pronounced effects from variations in the `speed' at which
communication can connect different bodies.  This viewpoint has held
dominion for three centuries, but there is something missing,
especially from a practical perspective.  It is is viewpoint that
leads to problems even in physics (cf. Zeno's paradox, the infinities
of quantum field theory, for example).  Moreover, it is not a useful
parameterization of dynamical systems for the information sciences.

Setting aside whether it might have fundamental significance, or
merely traditionalist convenience, we need an intentional theory of
spacetime that includes functional aspects of the world we live in,
i.e. a way to bring the observer and semantics into the picture.  The
fact that this is missing from conventional descriptions means that we
cannot understand all aspects of relativity that stem from situational
differences of observers.

Until recently, science has eschewed the notion that agency (that
quality which adds subjectivity and intent) should be a part of a
description of the world. This has led to manifest contradictions and
paradoxes in relativity and quantum mechanics.  Starting with
Einstein, we have been forced to take seriously the role of the
observer and allow for differences between their perceptions.

The idea of connectivity is not unique to the physical sciences, but
plays a major role in information systems too. There, spatial
relationships and topologies are often functional and imply semantics.
Semantics and intentions are subjective and require agency, so
objectifying the notion of spacetime will never offer a satisfactory
description for information science. If one starts with an agent view
of the world, the idea of an absolute spacetime seems untenable or at
best artificial. Thus we need a version of relativity that takes into
account how autonomous agents learn information about the world.  This
has some simple but profound implications for the resulting view of
space and time, which can no longer be seen as independent concepts.

Promise Theory already exists to address questions like these; but,
one of the pressing questions we face (which happens to be of great
practical importance in information systems) is how do we scale from
small clusters of promises to large fabrics, spaces, or even manifolds.

A descriptions of even an objective spacetime is potentially a highly
technical topic that has filled lifetimes with mathematics, so we must
have modest ambitions here. I shall try to sketch only an outline of
how such a picture of spatial relationships works in this paper.
No reference will be made to string or brane theories of spacetime,
which suffer from the same defects as Euclidean space.
Details can be studied in more depth elsewhere and by others.

%%%%%%%%%%%%%%%%%%%%%%%%%%%%%%%%%%%%%%%%%%%%%%%%%%%%%%%%%%%%%%%%%%%%%%%%%%%%%%%%%%%%%%%

\section{Models of spacetime}

Let us begin with the idea that space and time are observed
quantities. We take the existence of `elements' (points, regions,
agents) of location as given, though we shall not always
elaborate on their nature.  Changes in the state of these elements can
include changes in their relative relationships. Relative position is
one such relationship. As states change, we mark out different
versions of these relationships, which is how we measure the
progression of time.

\begin{definition}[Spatial element]
  An element of a spacetime is a member in a set, which expresses the
  property of location.  It can be named, and satisfies all the
  properties of set theory.
\end{definition}

\begin{figure}[ht]
\begin{center}
\includegraphics[width=8.5cm]{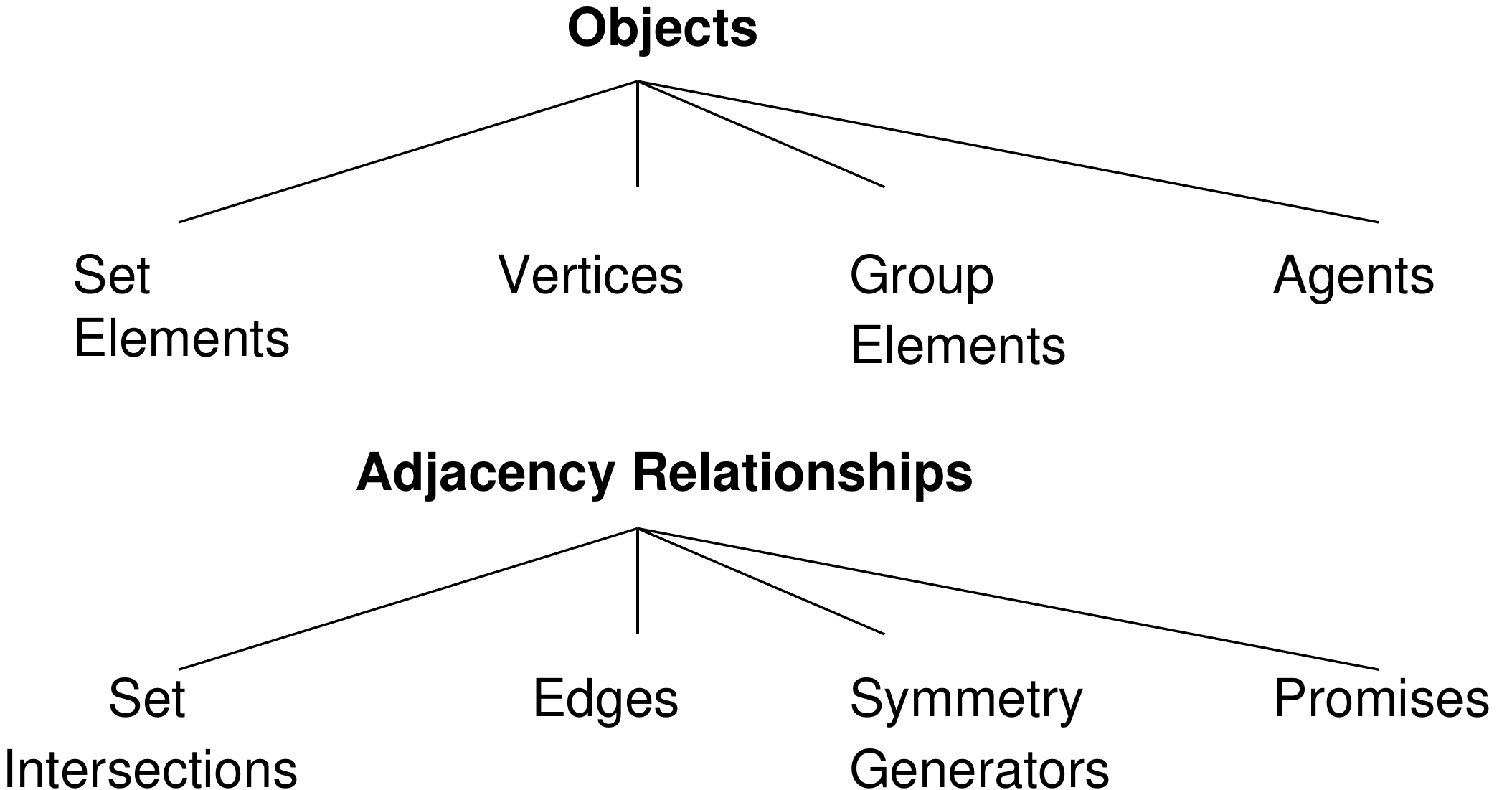}
\caption{Spatial relationships between elements\label{things}}
\end{center}
\end{figure}

\begin{assumption}[Adjacency represents space]
Space is the expression of a connectivity of its base constituent elements. 
\end{assumption}
We often speak of points for elements, but I'll reserve that for manifolds or
Euclidean space and infinitesimal dots that form a continuum. There is no need for that
to be the nature of space (see for instance a larger point element in fig. \ref{elephant}). 

\begin{assumption}[Clocks measure time]
  Time measures and is measured by change. Any changing system may be
  regarded as a clock that marks the passage of time. Without distinguishable change
there is no time.
\end{assumption}
Physicists are used to stydying large systems over long times compared
to Caesium transitions, so one many take for granted the ability to
have a seemingly independent clock. This cannot be taken for granted
in a finite system of states.  From the assumption of clocks, we can
immediately say:
\begin{lemma}
If the variety of states available in a system is finite, then only
a finite number of times can be measured. Hence time is finite in a finite space.
\end{lemma}

If we ignore what we think we know about spacetime from experience,
these basic assumptions already contain must insight. Space is about
structural patterns. How much structure do we need to be able to
represent patterns? What is the medium of such a pattern in empty
space?  If space is discrete and finite, then it possesses a finite
lifetime, which may or may not proceed as an ordered sequence of
states. The next step must be to elucidate the meaning of space using
well known formalisms.

There are different approaches to describing space.  In the Euclidean
view, one underlines the notion of expanse.  Then there is the
container view in which everything is placed within some kind of
bounded region and one element can be inside another (nested like
Russian dolls).  These views are complementary but quite different. 
They may be related through vector formulae such as Gauss's law or
the divergence theorem, etc.

The role of semantics now enters swiftly into the discussion.
How we choose to parameterize spatial relationships is an observer
choice that requires intent (e.g. do we choose Cartesian or polar
coordinates for the naming convention?), thus we need to unify intent
with description to understand spatial relationships.

\subsection{Elements with names}

How we label the elements and collections of elements in a space is
an important issue. If elements cannot be identified, then they do not
really exist to an observer's universe.  This does not necessarily
imply that every element must be individually distinguishable: there
may be symmetries between elements which are interchangeable.
Collective names (categories) may be used for such redundant parts.
These symmetries in turn affect an observer's ability to tell time.

Names may further have {\em scope}, or a domain of validity. In
computer science one speaks of private and public names. In vector
spaces, with metric scales, we commonly use Cartesian vector
coordinates, based on tuples, to identify elements (points in that
instance). Thus a name might also be a vector. In general any property
that can pinpoint an element may be called its name, whether an
identifying mark, a proper string, or a coordinate vector.

Any element might have more than one name, and there is no particular reason
to assume that names will remain constant. We need to understand the
consequences of these considerations.

Sets of elements may also collectively have names, representing common
properties. The term scope, {\em namespace}, {\em container}, {\em
class}, and even {\em group} are used for this. Some of these names
have specific technical meanings.  Each represents a set of
neighbouring elements, or a `patch' of space in which an element's name
has validity.  The functional role of an element can easily
be used as a name, even if that changes, making a connection with semantics.

There is a long tradition of naming elements in hierarchical
namespaces, called taxonomies. The coordinates of a name within a
hierarchy are given by paths through the graph from some reference
root element, through each branch to the specific element.  Tree-like
hierarchy is not a necessity, but in general one needs a collection of
spanning sets to give every element ownership to a named scope, even
if those scopes overlap.  The naming of elements into independent
classes or regions leads to the concept of dimensionality. This is
very familiar to us in the case of a locally Euclidean vector space,
but understanding the nature of dimensions is quite hard for
topologies such as graphs.  As we shall see, especially in connection
with graphs, dimension is something an observer can choose to perceive
in a discrete spacetime.  This issue is about concepts like {\em
  orthogonality} of vectors or {\em independence} of sets, scopes or
namespaces, as well as adjacency. In this regard, we shall meet the
additional generalized concepts of linear independence, matroids and
spanning trees, to mention a few.

\begin{itemize}
\item How do we talk about where things are, and how those relationships change?

\item A difference of two locations (two names) is considered to be both a vector and
a derivative on a topological space. This is how we define {\em rate of change}
or gradient. 

\item If all of the names for an element are removed, then it effectively disappears from
the universe of things as an independent.

\item The limits of a space are effectively the size of the list of names that
it contains, whether they be coordinates or other labels. Thus we have the notion
of boundedness as the dimension of a set $|S|$. 
\end{itemize}

\subsection{Models of space}

\begin{figure}[ht]
\begin{center}
\includegraphics[width=6.5cm]{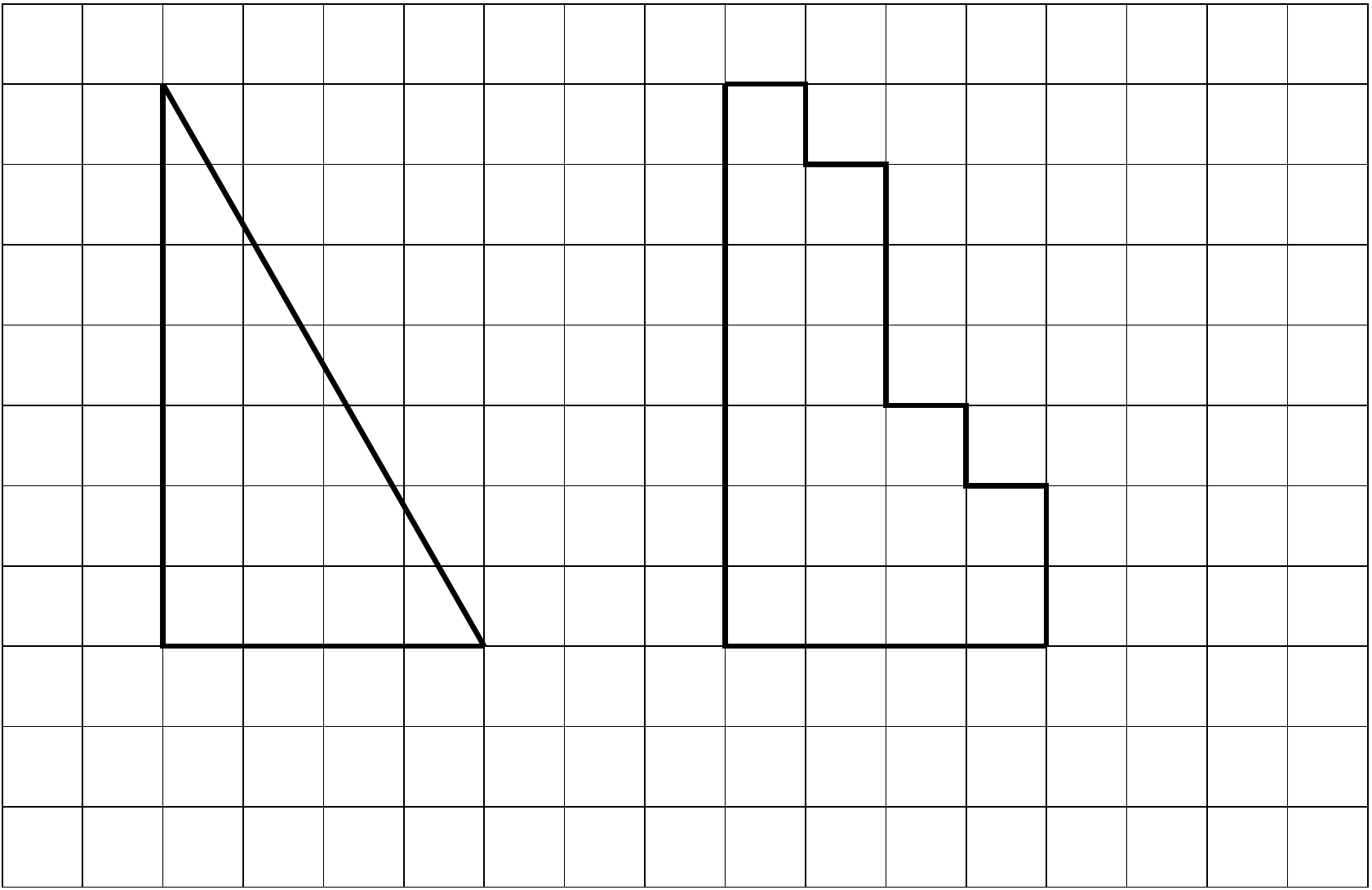}
\caption{We are used to assuming that a triangle exists as straight
  lines in smooth Euclidean space, but in a finite, discrete space
  there are no diagonal paths.\label{pythag}}
\end{center}
\end{figure}

Our notion of extent begins with sets of elements and nearest
neighbours within those sets. The concept of a neighbourhood is the
basis of topology. We understand extent as {\em transitivity}.  i.e.
if $\{ A, B , C\}$ are elements of a partially ordered set (poset),
where we consider the $A$ is a nearest neighbour of and $B$ is a
nearest neighbour of $C$, then in order to go from $A$ to $C$, we must
go through $B$. In general, to get from one element to another, we
might need to make several {\em hops} from element to element. This
might seem trivial (certainly commonplace), but there is no reason why
this has to be the case. One could also imaging directly teleporting
to any element from any other with equal cost i.e. distance. This
would imply a `complete graph' structure. Conversely, discreteness
of elements invalidates assumptions about distance like Pythagorean
relationships (see fig. \ref{pythag}). Thus
the assumption of transitivity might turn out to be wrong, but that will
be the assumption for the remainder of this paper.

From the mathematics of sets, one may add a succession of refinements with
increasing amounts of structure. Some key concepts include:
\begin{itemize}
\item Groupoid or Magma:  a set with a closed operation.
\item Semi-group: a magma whose operation is associative (a directed graph).
\item Monoid:  a semi-group with an identity element (a directed graph).
\item Group:  a monoid with an inverse (an undirected graph).
\item Lattice: a tiling formed by the generators of a discrete group.
\end{itemize}
Our traditional notions of spacetime usually begin with groups and the
regular lattices they generate. From there, we invoke a continuum
hypothesis, approximating a large countable set by a non-countable
distribution of values to obtain a smooth manifold (see below).
Here we shall end up with a more primitive (elementary) view.

\subsection{Models of time}

Time, for any observer, is measured by clocks. A clock is a mechanism
which changes some measurable phenomenon at a rate which has to be
assumed locally constant. In principle there is only one clock, which
is the state of the entire universe; however, not all agents can
observe the entire clock, thus observers partition off their own set
of distinguishable states which they use to measure their perception
of change.

Measurable phenomena from which we might build clocks include those
that require a change of spatial position (e.g. a rotary clock), but
could also include spin, charge and other `quantum numbers'. Any
finite (countable) state machine will do as a clock.

More importantly, every distinguishable spatial motion behaves as a
clock, thus one cannot define space and time and speed independently.
They are dependent concepts. The best one can do is to fix one of the
values and work from there. In Einsteinian relativity, one fixes the
speed of light.

We assume that the passage of time is monotonic, but this is merely an
assumption and we shall see that consistency requires us to allow time
to run forwards and backwards in finite systems.

In a finite system of states, time is inseparably bound to those
states, including states of position. In order to make a clock that
can measure time, we need to distinguish observable states.  A change
in the states {\em is} a change in time, as there are no other
measurable qualities. If a state change is repeated, by cycle or reversal,
then time also repeats, as measured by this clock.

\begin{proposition}
In a finite system, which acts as its own clock, a change of state
cannot happen independently of a change of time, since that change of
state (the movement of the clock) {\em is} what we mean by {\em proper time}.
The number of possible times is equal to the number of distinct states.
The measurable velocities are $0 < dx/dt < \infty$.
\end{proposition}
For the latter, if space measures proper time and a unit change of space
is therefore a unit change of time, then there is only a single speed
at which all motion takes place: $\Delta x = \Delta t =\Delta x /\Delta t  = 1$.
Any different speed can only come about by choosing to measure
$\Delta x, \Delta t$ differently. See section \ref{promisemotion}.

The notion of proper time here is essentially what computer science
refers to as a version of an information source.  Each `spatial
hypersurface' containing a unique microstate is a `version', and a
single position in time. If the microstate is reversed to an earlier
configuration, then it is indistinguishable from that time, and
effectively the system is reverted to that time. This can only seem
like a strange idea to an observer who has an independent
clock\footnote{The reason we don't think of this in physics is that we
  do not measure the proper time of the universe. Space is too big.
  However, in a closed system, external states are irrelevant except
  to an outside observer with access to them. Our familiar notion of
  time must be understood as a continuum hypothesis of a finite state
  system.}.

If we compose spaces with more states, then all of the state
configurations contribute to possible times in the system. As the
state space grows, time can extend combinatorially as the number of
distinguishable configurations.  Thus time is linked to the level of
entanglement of the different elements forming the space. The
connectivity of space must be linked to the extent of possible times.
However, necessarily in a space with a countable number of elements,
time is also countable and finite. This follows from the closure of
the space under whatever operation relates neighbouring elements.
Topology is therefore the origin of time.

A note is in order here about the traditions of physics, where the
complete state of a system is thought to be canonically defined by two
quantities: position and velocity. In a state model, where time is
measured by space itself, there cannot be an independent notion of
velocity, since there is no independent clock; thus a phase-space
description has to be an emergent model from an underlying state model
in the view of this work. Velocity is replaced by a knowledge of
transitions between states. However, we cannot consider this to be a
measure of time, as this will not change what observers see, only the
sequence with which they potentially see outcome. There is, surely,
additional information in transitions that is not understood in this
explanation, but it cannot change the perception of time experienced
by observers, who are part of the states themselves.

\subsection{Scaling of descriptions - renormalization}

There is every reason to expect that the fundamental nature of a
spacetime changes at different scales, so eventually we shall need to
define what we mean by scale. The different mathematical structures
represented in this paper have quite different properties in this
respect.

\begin{figure}[ht]
\begin{center}
\includegraphics[width=5.5cm]{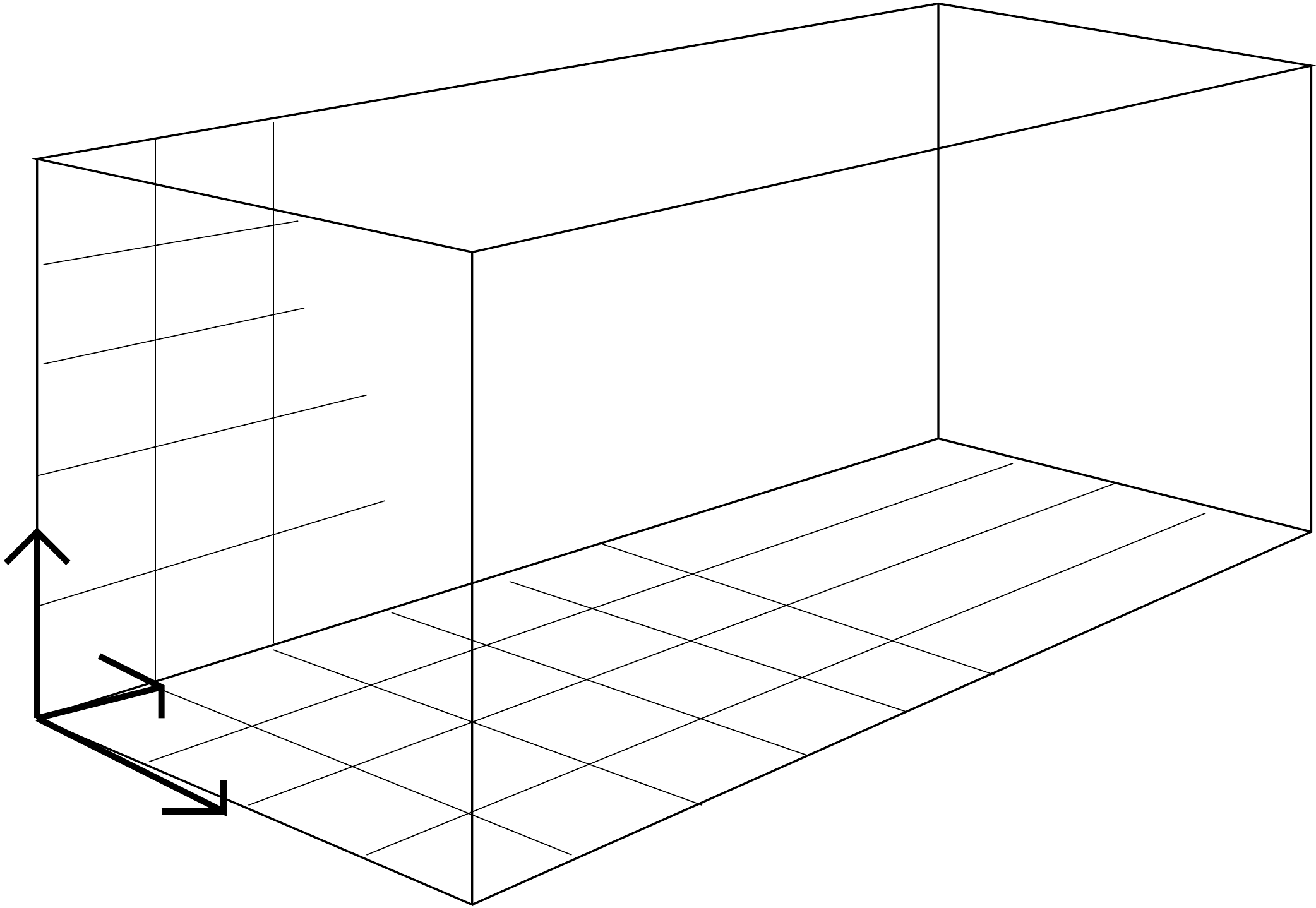}
\caption{Regular lattice scaling - the dimensionality is constant at all scales due to the
group structure.\label{scale1}}
\end{center}
\end{figure}

Defining scale is more complicated than it sounds in a general space.
In Euclidean spaces that map to $R^n$, scaling is a trivial group of
multiplicative factors because the structure of space is self-similar
under different multiplicative maps, even when they are anisotropic or
inhomogeneous.  Euclidean spaces and lattices have scale invariant
behaviour down the minimum cutoff distance (see fig \ref{scale1}).
However, graphs which are not regular lattices have no such regular
scaling properties. The number of nearest neighbours at one scale and
one location might be very different than at a different scale or
location.  

\begin{figure}[ht]
\begin{center}
\includegraphics[width=9.5cm]{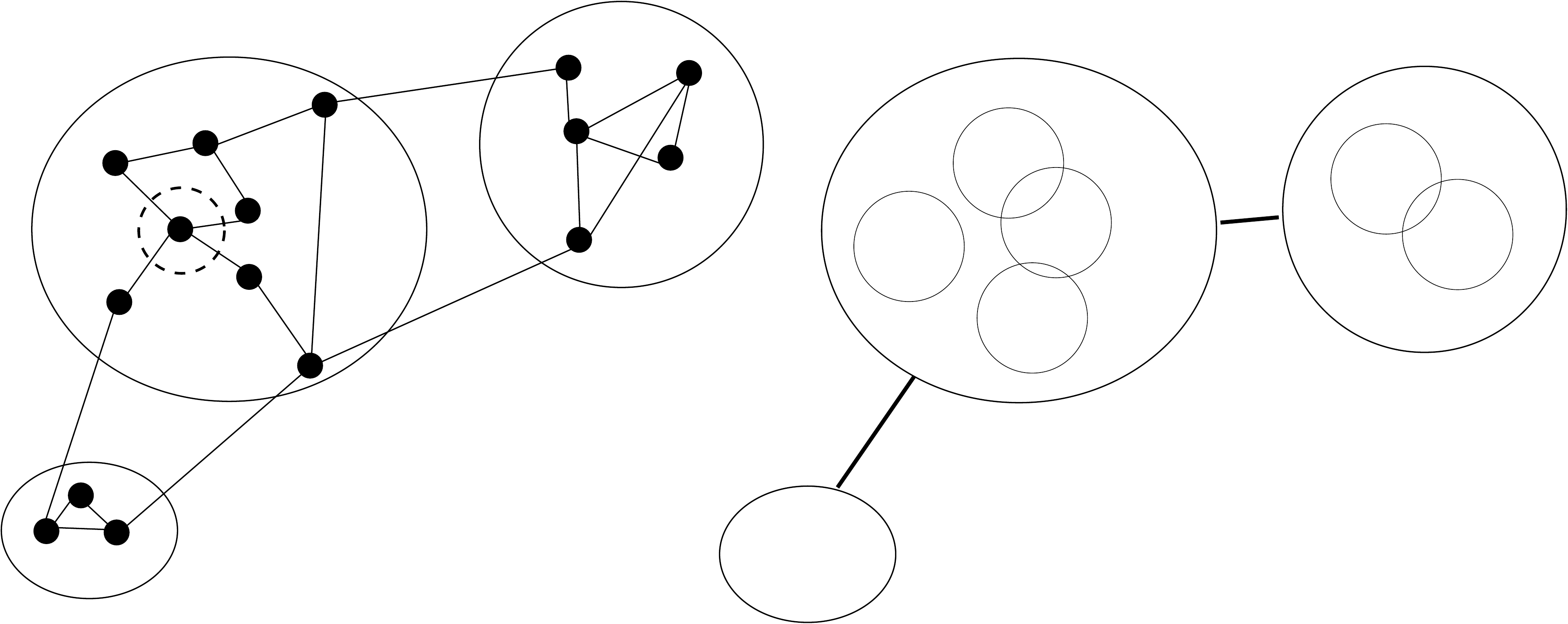}
\caption{Scaling in a graph. At a large scale the arrangement of
  points seems 1 dimensional. A local observer at a smaller grain size
  experiences more degrees of freedom at each point, suggesting a
  higher dimensionality. The dimension of spacetime is only one aspect
that can change under a scale transformation.\label{scale2}}
\end{center}
\end{figure}

Spacetimes are expected to be inhomogeneous and non-scale-invariant.
Several approaches to spacetime already account for this. One thinks
of the Regge calculus\cite{regge}, curved Riemann
manifolds\cite{weinberg1972gravitation}, fuzzy sets and topologies\cite{fuzzy}, and also studies of
large scale networks (the Internet, the World Wide Web, etc)\cite{graphpaper}.  The
meaning of scaling in a graph has been defined through Strongly
Connected Components (SCC) in \cite{graphpaper} (see fig.
\ref{scale2}).

For a general set topology, we may take a collection of {\em independent
subsets} $\sigma_i$ (sets where each subset has unique elements) and form a single
set by their union:
\beq
S = \union_i \sigma_i
\eeq
This coarse-grains (or renormalizes) a high resolution picture (small
scale) into a lower resolution element (large scale). This view of scale is
sufficient for the present paper. The matter of scaling is highly
technical.

Dimensionality is another area one struggles to define uniquely. In a
non-isotropic, inhomogeneous space, like a graph, dimensionality can
vary from place to place and from scale to scale. In other words the
ability to translate position from element to element, depends on the
scale at which we define translation and how far we intend to go.
This is a very different picture of space than the one we learn in school.

\subsection{Models of adjacency, communication and linkage}

How can we model spacetime without the prejudice of earlier ideas
intruding too much? The most primitive concept is that of {\em
  locations} that can be occupied by certain information (i.e.
properties), and their {\em adjacencies} or {\em neighbourhoods}.
Within a neighbourhood, elements may be considered adjacent in a
number of ways: either by a string-like communication channel (like a
graph), or in the manner of intersecting open regions (see fig.
\ref{patches}), as in algebraic topology. In either case, the
connectivity implied by adjacency leads to an implicit graph
structure. Whether that graph structure is real or merely
a map (like a coordinatization) might not be possible to say.

As an idealization, Shannon's model of source, sink and channel is particularly useful
here, as it unifies the idea of set intersection with that of an
graph-like adjacency channel\cite{shannon1}.  
\begin{figure}[ht]
\begin{center}
\includegraphics[width=5.5cm]{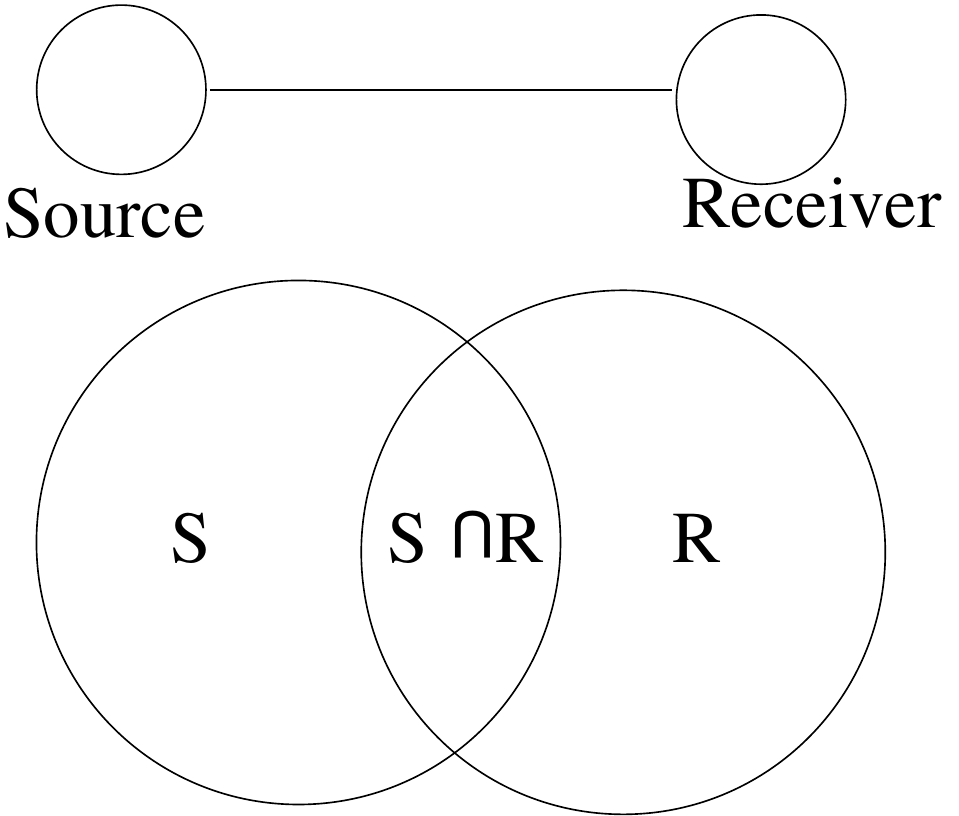}
\caption{Shannon's communication channel unifies the overlap of sent
  and received message domains with a simple graphical view of what
  adjacency means in terms of information. Adjacency is a channel by which
information can be shared.\label{shannon}}
\end{center}
\end{figure}

Any source element may be considered a transmitter or a receiver. If
there is a communication channel between them, we may consider them to
be adjacent. A communication channel could be thought of as a
signalling channel, as is the case in communication networks, or the
virtual particle interactions of Feynman diagrams. The nature of a
signal from one to another may be assumed discrete and symbolic, as in
Shannon's model of the discrete channel\cite{shannon1}.

%%%%%%%%%%%%%%%%%%%%%%%%%%%%%%%%%%%%%%%%%%%%%%%%%%%%%%%%%%%%%%%%%%%%%%%%%%%%%%%%%%%%%%%

\section{Mathematical structures of spacetime}

Let us now review how these concepts emerge in a number of well known mathematical
formalisms for space. This helps to compare and contrast the different ideas, preparing
the way for an autonomous agent viewpoint which will be a basis for all other notions.

\subsection{Sets and Topology}

\begin{figure}[ht]
\begin{center}
\includegraphics[width=5.5cm]{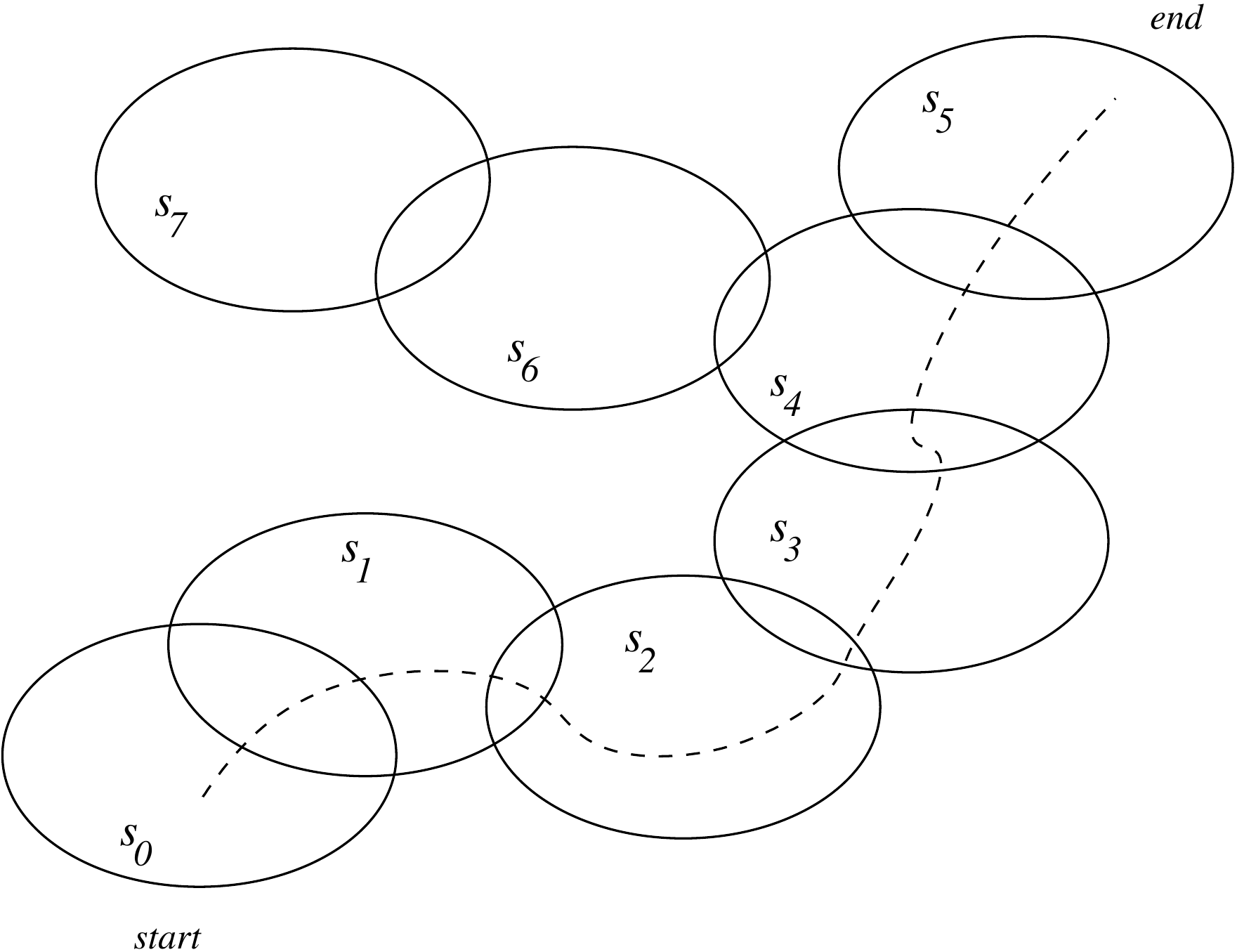}
\caption{Overlapping patches form a notion of connected spatial regions\label{patches}}
\end{center}
\end{figure}

The most basic mathematical notion of a space comes from elementary
topology.  Topology deals with sets of elements, often called
`points'. These elements are grouped into sets which may overlap of
intersect (see fig. \ref{patches}), and thus cover space.
Sets of points allows us to talk about three important properties of
spatial extent: continuity, connectedness and limiting convergence or
path ordering. A topology is then a set of points with neighbourhoods
that satisfy certain axioms.

At such an elementary level, it is not possible to assign unambiguous
meanings to basic elements and concepts (this is one reason why we
need a semantic theory of space).  The `size' of an element (e.g. a
`point') is not defined in this description; we imagine points as
semantic primitives, and say that questions of the size or extent of
points are meaningless. They may or may not have internal structure;
but, if they do, it is irrelevant to the notion of the space they
form. Points are merely `agents' that exude the property of {\em
  location}.

\begin{figure}[ht]
\begin{center}
\includegraphics[width=5.5cm]{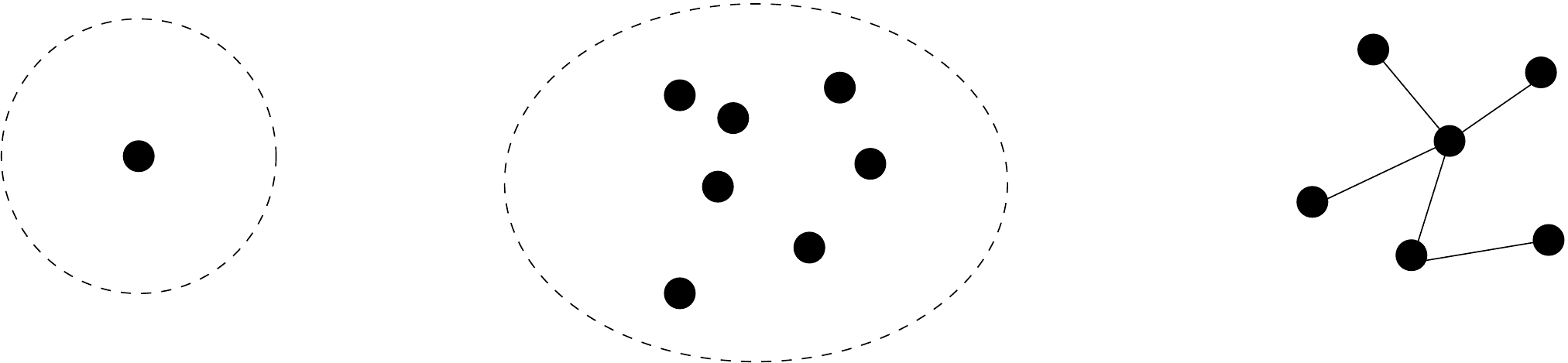}
\caption{Neighbourhoods and open sets\label{neighbours}}
\end{center}
\end{figure}

A neighbourhood of any element $p$ in a set $S$, is a subset $N$ of
$S$ that incorporates an {\em open set} $U$ including $p$: $p \in N
\subseteq S$.  We may equivalently say that $p$ is in the interior of
the neighbourhood $N$.  An {\em open set} is one that does not include
any of its boundary points, for some notion of a boundary.  It is
usually the generalization of the open interval in the real numbers
$R^1$, but this is concept is meant to be used flexibly. For a
manifold, a neighbourhood is essentially an open ball (see fig.
\ref{neighbours})\footnote{The concept of an open set or open ball in
  algebraic topology is also subject to some ambiguity, which
  ultimately arises from our expectation about points in space having
  no size. Such a ball represents a unit element of space.}.

Three concepts are particularly important:
\begin{itemize}
\item Continuity of functions, which means that small changes of
the space lead to small changes of the function.
\item Connectedness, which implies that a space is not a union of disjoint open
sets.
\item Compactness generalizes the notion of closure of the set (including boundary).
If formed from bounded sets, the the points are naturally close together in some sense.
If unbounded, all points are defined to lie within some fixed `distance' of one another.
Compactness is also a flexible concept, depending on the notion of boundedness. Clearly
there is no clear notion of distance either at this stage, so it is a heuristic.
For example $[0,\infty)$ is not compact since infinity is unbounded.
\end{itemize}

\subsubsection{Vectors between elements and direction}

\begin{figure}[ht]
\begin{center}
\includegraphics[width=6.5cm]{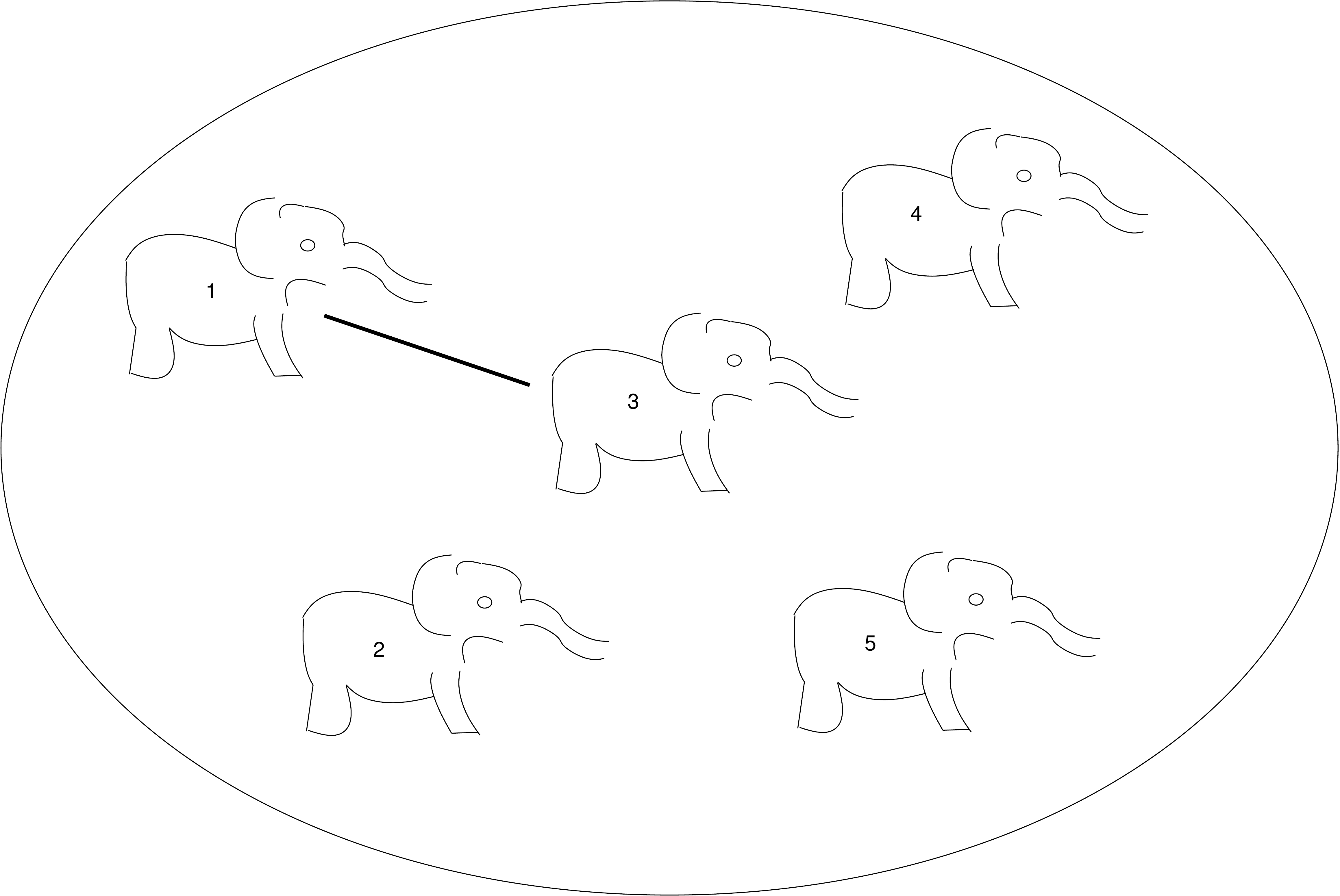}
\caption{A vector is any ordered pair of similar elements in a space\label{elephant}}
\end{center}
\end{figure}

Directionality is normally associated with vectors, which in turn are
normally associated with Euclidean-Cartesian vector spaces, but we are
free to define the idea of a vector as a direction by associating any
pair of elements in any space.  For instance, an elephant
vector is shown in fig.  \ref{elephant} from a set of different
elephants.

{\em Unit vectors} are always defined from adjacent named elements.
The elements of space we consider adjacent must be
defined at a given scale.  If there is internal structure within an
element, this can be ignored.  In a set $S$ with adjacent elements
$s_i \in S$, a vector is some function of the two elements $\vec s(ij)
= f(s_i,s_j )$.

If two elements are not adjacent, we may still write a vector as two points,
consisting of potentially several transits (often called hops), but we
should be careful of the issue depicted in fig. \ref{pythag} of assuming that
vectors represent straight line routes from one element to another.
We might call such a vector a {\em path} or a {\em route}, i.e. a string of
transitions belonging to the graph of adjacencies. We are used to being able
to assume the structure of intermediate points, by the homogeneity, isotropy
and apparent continuity of a Euclidean spacetime lattice. No such assumption
can be made about general spacetimes.

\subsubsection{Coorinates, bases, matroids and independent dimensions}

In Euclidean space we have a clear notion of dimensionality from
everyday convention.  However, if we are constructing space from sets
of elements, dimensionality is by no means given. If we are given a
set of bricks, we may arrange them into a line, a plane or a three
dimensional block at will. Thus, dimensionality is a matter of {\em
  intent}, i.e. semantics, as we build from the bottom up (see appendix \ref{appendix}).

Our understanding of dimensionality of space is based on independence
of vectors, or what is called a {\em basis} or spanning set. Note that
this is coordinate semantics, hence dimensionality will emerge to be a choice
rather than a definite reality that ultimately is limited only by the 
maximum number of elements in the base set.

The analogue of a basis for sets is called a {\em matroid}.  A matroid
captures and generalizes the notion of linear independence familiar in
vector spaces (see below). There are several independent ways to
define matroids. Given a finite set $E$, one may choose a family of
subsets $I$ of $E$, called the independent sets.

Just as independent vectors don't need to be orthogonal, the
independent sets may overlap as long as each subset in $I$ contains
points unique to itself. This mimics the idea that two vectors can be
independent as long as they are not collinear.  One can exchange points
between the independent sets in a matroid, just as one may alter the angle
between two vectors.

\begin{figure}[ht]
\begin{center}
\includegraphics[width=5.5cm]{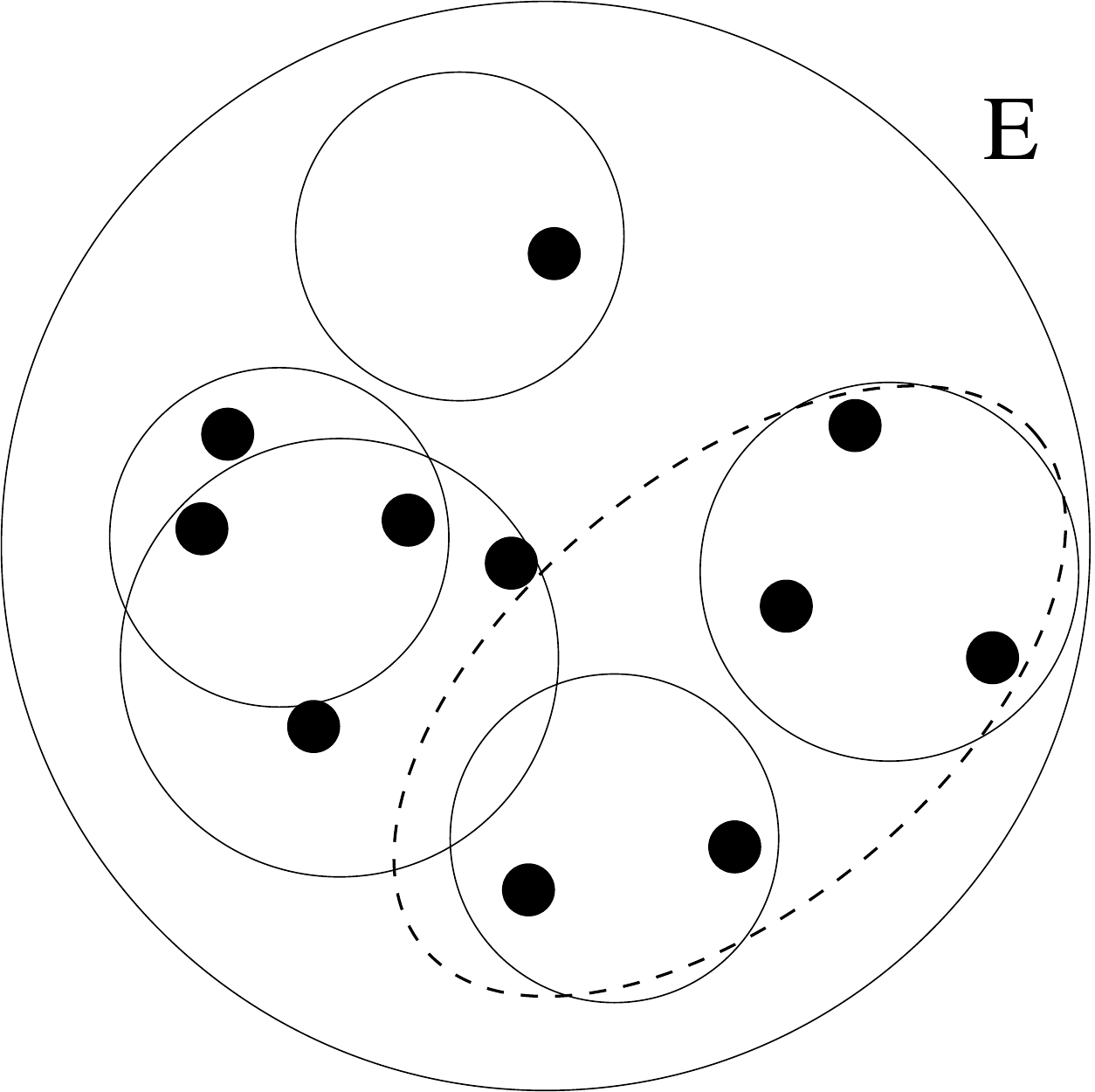}
\caption{A matroid of rank 5 on $E$. The dotted subset is not independent of
the others, as it is spanned by the intersection of two independent sets.\label{matroid}}
\end{center}
\end{figure}

A matroid is thus a pair $(E,I)$ of a set of elements and a spanning
set of subset patches.  For a finite set $E$, there is a limit to have
many sets can be formed containing unique elements and without
overlapping completely.  The {\em rank} of a matroid is the number of
independent sets. As set is maximal if it becomes dependent by adding
any element of $E$. Clearly the rank cannot exceed the size of base
set $E$. Just as we may organize bricks into several different
dimensional arrangements, so the dimensionality of a set is defined by
a specific matroid {\em basis}. This non-uniqueness of dimension is
unfamiliar from our ideas about vector spaces.

An element of the rank $r$ dimensional set is not a single element of $E$, but
a tuple of elements from each set in $I$ that become associated through this
spanning basis. Thus, the dimensionality of a set is a matter of choice,
provided there are sufficient base elements to construct tuples, with each
coordinate component represented by an element from a named independent set.

\subsection{Categories and algebras}

Category Theory is a very general way of classifying structures and
arrangements. A category comprises a collection of things (objects)
and arrows (morphisms) $f$ which behave as functions.
Arrows map from a domain to a co-domain via a function
\beq
f: A \rightarrow B\\
g: B \rightarrow C\\
h: C \rightarrow D
\eeq
Under composition, the arrows must be associative
\beq
h \circ (g \circ f) = (h \circ g) \circ f
\eeq
There is operator ordering from right to left. 
There are right and left identity arrows, for any arrow $f$:
\beq
id_B \circ f = f ~~~{\rm and}~~~~~~  f \circ id_A = f
\eeq
Categories have names, e.g. the category of sets is called {\bf Set},
when arrows represent total functions.
A {\em monoid} $M$ is a set with a structure in its domain,
that satisfies 
\beq
(x\cdot y) \cdot z = x\cdot (y \cdot z), ~~~\forall x,y,z \in M
\eeq
and has identity $e \in M$. It is a homomorphism (it is also a semigroup that has an identity.
A partially ordered set or {\em poset} $(P, \leq_P)$ has a relation $\leq$ which preserves
ordering, i.e. can represent monotonic functions.

\begin{enumerate}
\item The category $\bf 0$ has no objects and no arrows. 
\item The category $\bf 1$ has one object and one arrow (the identity).
\item The category $\bf 2$ has two objects,  two identity arrows and one arrow from one object to the other (see fig. \ref{cat2})

\begin{figure}[ht]
\begin{center}
\includegraphics[width=6.5cm]{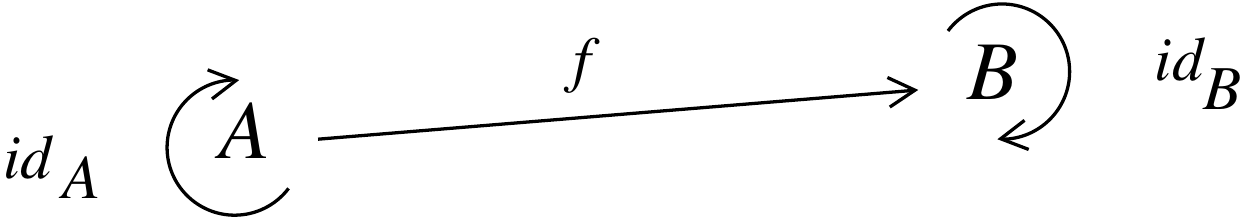}
\caption{Category with two objects $\bf 2$\label{cat2}}
\end{center}
\end{figure}

\item The category $\bf 3$ has three objects,  three identity arrows and two arrows between the objects (see fig. \ref{cat3})

\begin{figure}[ht]
\begin{center}
\includegraphics[width=6.5cm]{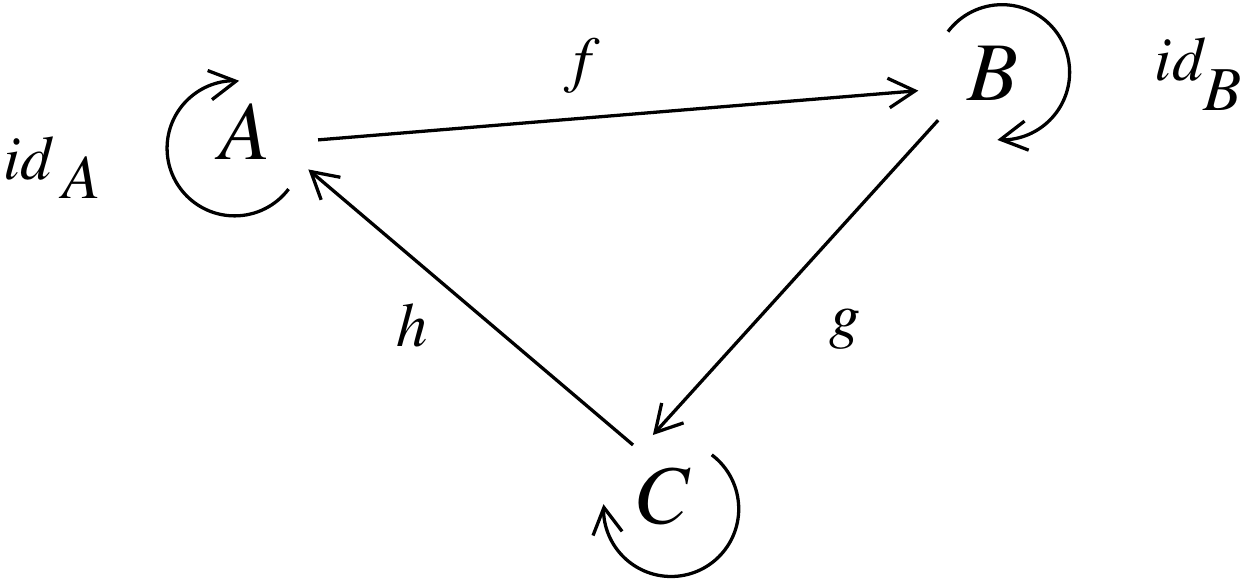}
\caption{Category with three objects\label{cat3}}
\end{center}
\end{figure}

\end{enumerate}

\subsubsection{Operators and total functions}\label{totalf}

In connecting discrete objects together we shall frequently have use
for the notion of an operator (such as in matrix multiplication and
map composition) as a total function.

An inner product of mappings may be defined if we have operators (functions) $O_1$ and $O_2$,
where $O_1$ maps from a domain $D$ to co-domain $I$, and $O_2$ maps from domain $I$ to co-domain
$C$:
\beq
O_1: D \rightarrow I\\
O_2: I \rightarrow C
\eeq
Then these two operator functions may be joined into a single total function
so that their product maps directly from the domain $D$ to co-domain $C$, eliminating all
reference to the intermediate range $I$:
\beq
O_2 \circ O_1 : I\rightarrow K,\label{innercategory}
\eeq

We find examples of this kind of product in tuple spaces of
matrices, as well as in graphs and the more complex bi-graphs.  The
inner product is possible when transformations are compatible with one
another, by being smoothly joinable across an intermediate domain.

An inner product can be performed on any two operations, but there is
a special significance to automorphic functions that take one domain
onto an image of itself, such as with `square matrices'. These are
particularly important for representing global symmetries of spacetime.

\subsubsection{Diagrams as categories}

Diagrams are graphs with functionally labelled edges, and are
categories if each object is labelled, and the directed edges are
consistently labelled with edges $f,g,h, \ldots$, etc, with domains
and co-domains. Diagrams that
form categories should not be confused with diagrams of categories.

\begin{figure}[ht]
\begin{center}
\includegraphics[width=3.5cm]{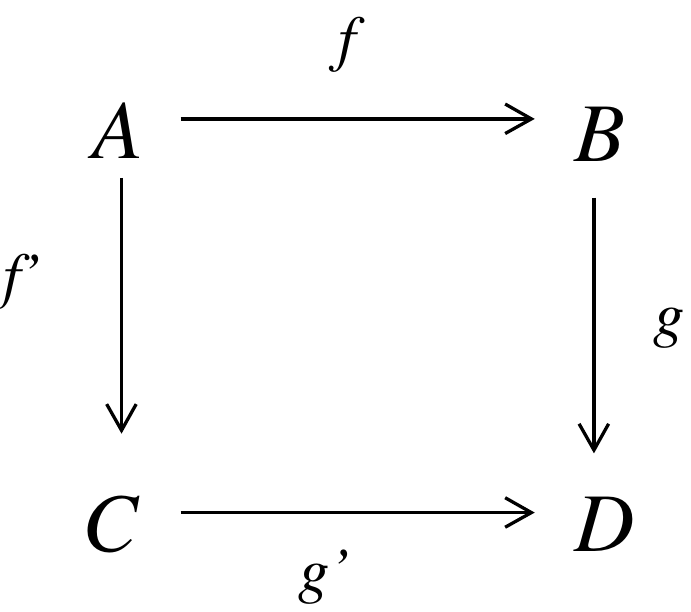}
\caption{Diagrams and commutativity. This commutes if $f\circ g = f'\circ g'$ \label{diagram}}
\end{center}
\end{figure}

A diagram is said to commute if all the paths from one object to
another are equivalent and equal In figure \ref{diagram}, the diagram
commutes if the compositions of morphisms from domain $A$
to co-domain $D$ are equal, i.e. $f\circ g = f'\circ g'$, despite the fact that the two
functional routes pass through different intermediate co-domains $B$ and $C$.

An object is initial (a source) if it has exactly one arrow to a neighbour, and it is final
(a sink) if it receives one arrow from a neighbour.

\subsection{Vector spaces}

The notion of a vector space is what we normally associate with
Euclidean space of the natural world, with tuples $p_n$ of coordinates
$(x,y,z,...)$. A vector is essentially defined by two points $\vec v =
(p_1, p_2)$. This is a vector rather than merely a tuple space because
the key structural properties lie in the transitions between the
coordinates (the vectors) rather than the points themselves.

A vector space has the algebraic structure of a group, and satisfies group axioms for
two kinds of operation: $+$ and product. If $V$ is a vector space, containing vectors, then
it satisfies the following axiom under combination:
\begin{enumerate}
\item Closure: $\vec u + \vec v \in V, ~~~~~\forall \vec u, \vec v \in V$.

\item Associativity: $\vec u + (\vec v + \vec w) = (\vec u + \vec v) + \vec w$.

\item Identity exists: $\vec u + \vec 0 = \vec u$.

\item Inverse exists: $\vec u + (-\vec u) = \vec 0$.

\item Commutation: $\vec u + \vec v = \vec v + \vec u$.
\end{enumerate}
Under scaling by a field $\lambda$, the vectors satisfy:
\begin{enumerate}
\item Closure: $\lambda \vec u \in V, ~~~~~\forall \lambda, \forall \vec u \in V$

\item Associativity: $\lambda (\vec u + \vec v) = \lambda\vec u + \lambda\vec v$,
and $(\lambda + \mu)\vec u = \lambda\vec u + \mu\vec u$. 

\item Identity exists: $1\,\vec u = \vec u$.

\end{enumerate}
There is no formal restriction on the commutativity of scalar multipliers, but this
is generally assumed as given.
The associativity under $+$ implies that we can enumerate the inverse:
\beq
(1 + (-1))\vec u = \vec 0\nonumber\\
(-1)\vec u = -\vec u.
\eeq

\subsubsection{Group lattice}

Vector spaces are special in that they have an assumed group structure,
which we take very much for granted. This is sometimes called a lattice structure.
A lattice is an ordered, regular {\em tiling} of vectors translations that leads to
a simple tuple structure. For many it is hard to set aside this
intuitive notion of dimensionality, as it models the world as we experience
it as humans at the macroscopic scale.

Every point is homogeneous with respect to the options for translating
along a vector to a new point.  Homogeneity of a space implies uniform
properties as one follows a single direction.  Isotropy implies the
appearance of uniform properties in all directions around a point of
observation.

\subsubsection{Naming of points on a vector space: coordinate systems and bases}

Points in an $n$-dimensional vector space are named numerically in tuples.
A coordinate tuple is often written as a column vector\footnote{A tuple is
sometime loosely called a vector, where it is assumed to be measured relative
to the origin or tuple composed of all zeros.}, here shown decomposed
as a linear combination of orthonormal basis vectors:
\beq
\vec v = \left( 
\begin{array}{c}
1\\
2\\
0\\
7
\end{array}
\right)
=
1\left( 
\begin{array}{c}
1\\
0\\
0\\
0
\end{array}
\right)
+2
\left( 
\begin{array}{c}
0\\
1\\
0\\
0
\end{array}
\right)
+ 0\left( 
\begin{array}{c}
0\\
0\\
1\\
0
\end{array}
\right)
+7\left( 
\begin{array}{c}
0\\
0\\
0\\
1
\end{array}
\right)
\eeq
This has the structure
\beq
\vec v = (\lambda_1, \lambda_2,\ldots,\lambda_n ) = \sum_i^n \lambda_i \vec e_i 
\eeq
where $\vec e_i$ are called basis vectors, which satisfy the orthonormality property,
and $\vec \lambda$ is the tuple of vector components measured relative to that basis.
\beq
\vec e_i \circ \vec e_j = \delta_{ij}\label{orthonorm}
\eeq
where $\delta_{ij}$ is the Kronecker delta ($\delta = 1$ iff $i=j$, else zero).

\subsubsection{Rank, linear independence and dimensionality}

A Euclidean space (or generally a manifold) of dimension $n$ is formed
by the outer product of independent sets isomorphic to the real line
$R^1$, thus $R^n = R^1 \otimes R^1 \otimes \ldots \otimes R^1$.  From
the generalized viewpoint of matroids, or coordinate bases, these
real-line sets are the independent sets of the matroid, and a point
in the $n$-dimensional space is really an association of $n$-elements from the $n$
spanning sets.

Normally, one expresses this in the following way.  A vector space is
said to be $n$-dimensional if there exist $n$ linearly independent
vectors $\vec e_i$ (basis vectors or generators), which are
pairs of tuples that originate from an arbitrary point (called the origin),
and point directly to each independent set. Linear independence means
that there is no non-zero value of
$\lambda_i$ such that 

\beq
\sum_{i=1}^n \lambda_i \vec e_i = 0,\\
\Rightarrow \lambda_i = 0, ~~~~~~~\forall _i\label{rank} 
\eeq

We are so familiar with the idea of Euclidean dimension that the
dimensions of other structures are often defined implicitly by
immersion into a Euclidean space of some dimension. The smallest
number of Euclidean dimensions required to embed something then
becomes a definition of dimensionality. However, this might be interesting
in some geometrical sense, but it is not a true measure of the
number of degrees of freedom available to an element at some location (see appendix \ref{embed}).

In group theory, the algebras which generate the equational structure
of the group may also be drawn as vectors known as the roots or
weights of the algebra in a given dimension.  The rank of the algebra
is the number of independent generators, however there might be more 
members in a tuple which represent the algebra, with some that are
not independent.

\subsubsection{Tensors, order and rank}

Tensors are indexed arrays of values that transform correctly between different coordinate
bases. Vectors are tensors of order 1. Matrices have order 2.
Objects of higher dimension can also be formed:
\beq
T_i ~~~~~{\rm (tuple)}\\
T_{ij} ~~~~~{\rm (matrix)}\\
T_{ijk\ldots mn}  ~~~~~{\rm (higher~order)}
\eeq
Rank is distinct from order. Rank refers (as with coordinate spans)
to the number of independent sets composing the object.
A matrix (which has order 2) has rank one if it can be written as an outer product
of two non-zero rank-1 vectors:
\beq
T_{ij} = v_iw_j
\eeq
The rank $r$ is the smallest number of such outer products that can be summed
to produce the tensor.
\beq
T_{ij} = \sum_{n=1}^r v_i^{(n)}w_j^{(n)}
\eeq

\subsubsection{Distance and the inner product}

Vectors have inner products that reduce a matrix
of dimension $l \times n$ and $n \times m$ and produces an element
of dimension $l \times m$.
The inner product of two vectors $\vec a, \vec b$ with components $a_i,
a_b$ relative to some basis $\vec e$ with components $e_i$ is:
\beq
a\circ b = \sum_i a_ib_i = L_{ab}
\eeq
This has a geometric interpretation as a relative length. If $\vec a = \vec b$, then this
reduces to Pythagoras' result for the square length of the vector. For orthogonal vectors it vanishes,
providing the orthonormality condition in equation \ref{orthonorm}. In all other
cases, this inner product gives a scalar result that is the scaled projection of
one vector along the other, somewhat like a geometric average of their extent.
This is an example of an inner product as described in equation \ref{innercategory}.

Fig. \ref{pythag} shows how we could potentially define distance by
different measures in a vector space. If the space is continuous, we
may define effective distance by Pythagoras' theorem. If there is not
direct connection along the hypotenuse of the triangle, then the
discrete distance is found by summing the bond lengths of every atom
along the edge however. In a discrete space, there is no way to take
the direct Pythagorean route, but we might still interpret this as an `average'
measure of the effective distance.

\subsubsection{Matrices and transformations}

One of the first places for the appearance of semantics in spatial models arises
with the introduction of structure through transformations, usually
symmetries. Variable observability, with transformations to map between
them, implies a relativity for local observers.

For a matrix with components $M_{ij}$, $i = 1\ldots \Rows(M)$, and
$j = 1\ldots \Columns(M)$. In a square matrix $\Rows(M) = \Columns(M)$.
Square matrices are a necessary condition for closure
under multiplication.

Matrix multiplication is an inner product (see section \ref{totalf}),
defined by $M\circ N = \sum_j M_{ij}N_{jk}$, which is possible if the number of
rows in the first is equal to the number of columns of the second matrix.

\beq \vec X = \left(
\begin{array}{c}
X_1 \\
X_2 \\
\vdots \\
X_n
\end{array} 
\right).
\eeq
In refs. \cite{jan60, burgess04analytical, burgesstheory}, a bounded n-dimensional Euclidean model was used to model the memory
of a computer system.

As large as current systems may be, they remain finite and can be modelled by finite vectors. 
We define relative operators for individual parameters in a state as 
\beq
\left(
\begin{array}{cc}
a & b \\
c & d \\
\end{array} 
\right) \circ
\left(
\begin{array}{c}
A \\
B \\
\end{array} 
\right) 
=
\left(
\begin{array}{c}
aA+bB \\
cA+dB \\
\end{array} 
\right) 
\eeq

\beq
\left(
\begin{array}{cc}
a & b \\
c & d \\
\end{array} 
\right) \circ
\left(
\begin{array}{cc}
A & B \\
C & D \\
\end{array} 
\right) 
=
\left(
\begin{array}{cc}
aA+bC & aB+bD \\
cA+dC & cB +dD \\
\end{array} 
\right) 
\eeq
Matrix multiplication is associative, and may or may not be commutative (path dependent).

\subsubsection{Derivatives and vectors}

A vector is a difference of tuple elements. If $T$ is a generator of
a translation, we may write:
\beq
\vec v = p_2 - p_1 = (T-1)p_1
\eeq
A derivative is a measure of how quickly a function changes with respect to 
a constrain parameter. It measures differences. In Euclidean space, we have
the classic Newton-Leibniz definition of a derivative

\beq 
\frac{df}{dx}\Bigg|_x = \lim_{dx\rightarrow 0} \frac{f(x +dx)-f(x)}{dx}
\eeq 
This can be written mode simply in terms of atomic succession operator $+$ like this:
\beq
df_x = f_{x+} - f_x
\eeq
The length scale implied by the difference between $x$ and $x+$ is now
irrelevant as it is assumed to be the nearest neighbour spacing.
On a lattice, the smallest distance between neighbours has to be considered
constant, as there is no objective way to measure it from inside the lattice.

Thus unless we can define distance in a simple way, we cannot
measure rate of change, or derivative. Ultimately, these definitions
all become mutually intertwined, more self-consistent web of ideas
than a hierarchy of axioms\footnote{This too kind of self-consistent
  web will become important and explicable in the context of semantic
  networks below.}.

\subsubsection{Boundaries, subspaces and immersion of structures}

We may embed subspaces $h_i$ of lower dimension into a Euclidean space
$R^n$ by immersion. These are often called hypersurfaces and may be
defined in the form of additional constraints of the spanning set of
independent vectors: 
\beq 
\chi(\vec v) = 0 
\eeq 
By adding such additional
constraints with such an equation (analogous to (\ref{rank})), we may
reduce the number of independent vectors leading to a smaller solution
set for $\vec v$ with reduced dimensionality.  For example, a two
dimensional sphere $S^2$ (the surface of a three dimensional ball) may
be embedded within $R^3$ by direct immersion.
Moreover, by finding equivalence classes of a group generator, a vector space
can be decomposed into subspaces or non-overlapping hypersurfaces.
\begin{figure}[ht]
\begin{center}
\includegraphics[width=5.5cm]{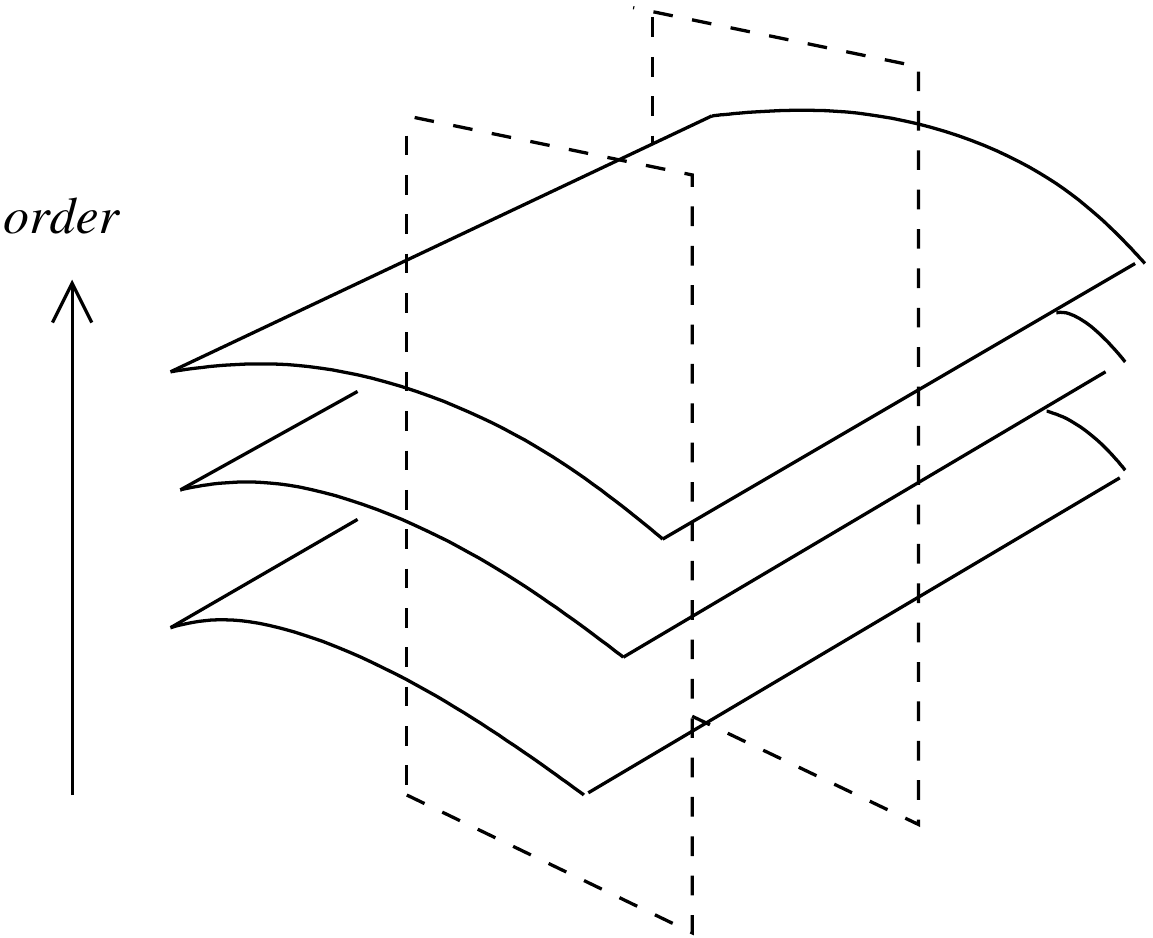}
\caption{Hyperplane (subspace) decomposition of a space\label{hyperplanes}}
\end{center}
\end{figure}
This property of embedding is often used to define the dimensions of
elements who dimensionality is harder to understand (like graphs).

\subsection{Manifolds and Minkowski spacetime}\label{manifold}

If we try to add an independent description of time to a Euclidean
space, then we must introduce a coordinate $t$. Thus one immediately
has the velocity of one observer with respect to an element as $\vec v
= \frac{d\vec x}{dt}$. Again, the velocity of everything except
massless signals depends on the way one chooses to measure space.
The constancy of the speed of light seems to indicate that this is a
primitive form of transmission, as one expects in a discrete spacetime.

In Newtonian mechanics the Galilean group generates the
transformations that relate different observer speeds and directions.
Taking into account Einstein's considerations, based on our knowledge
that the speed of light and other massless signals in a vacuum
appears constant, one finds that the Poincar\'e group generates the
correct transformations in a 4 dimensional representation (see section
\ref{manifold}).

A manifold is the generalization of a locally Euclidean space, whose global
structure might possess curvature. The surface of a sphere is an example.
Pythagoras rule, and sum of angles in a triangle do not object Euclidean
rules.

In modern formulations of Einstein's special relativity, one treats
the constancy of signal propagation (the speed of light) as a
constant, which we've established is entirely natural in a discrete
spacetime.  Expressing this as an invariant inner product on Euclidean
space $R^3$
\beq
\sum_i dx_i dx_i = c^2 dt^2
\eeq
allows a four dimensional space to be defined with inner product
\beq
ds^2 \equiv (-cdt, \vec dx) \circ (cdt, \vec dx)^T &=& 0\\
ds^2 = -c^2dt^2 + \sum_i dx_i dx_i &=& 0
\eeq
This is the spacetime generalization of Pythagoras' theorem, but unlike
Pythagoras which is a spherical constraint this is a rescaling of spatial
elements with respect to elements of time.
It is sometime written symmetrically using an imaginary time coordinate $idt$,
or one may introduce a metric tensor, analogous to the Kronecker delta:
\beq
\eta_{\mu\nu} &=& {\rm diag}(-1,1,1,1),\\
dx^\mu \eta_\mu^{~\nu}dx_\nu &\equiv& (cdt, \vec dx) \circ (cdt, \vec dx)^T = 0
\eeq
where $\mu,\nu = 0,1,2,3$ and the zeroth dimension is now time scaled
by the speed of light.  This metric tensor is useful as it generalizes
to curved spacetimes used in general relativity and other problems of
curvilinear coordinates. Alas, that is a large topic that goes beyond the scope of
this paper.

The symmetry group structure that makes this 4-dimensional extension
of Euclidean space behave like local observer transformations in
Minkowski spacetime is generated by the Poincar\'e group
transformations.

Notice that, by extending time into the very fabric of a spatial coordinate
description, we have made a fundamental choice to embed a clock into
the coordinate system itself.  The Poincar\'e transformations describe
apparent changes in the clock rates for different observers because,
the assumption of a fixed universal measuring scale is incompatible
with the constancy of the speed of light. Formally, the negative sign
for time in $ds^2$ makes this an equation of constraint, which means
that spacetime is a constrained surface. 
This is a reasonable starting point on which to build a
model of spacetime, but if one looks more carefully from a local-observer
viewpoint, it is a simplification of the issues. It homogenizes all
observers to have the same idea of time and space with minor
adjustments. In other words, it imposes uniform semantics on to
spacetime.

\subsection{Graphs or networks}

The first discrete notion of a space that generalizes lattices, and
has minimal initial semantics, may be found in graph theory.  A graph
$G = (V,E)$ is a collection of nodes or vertices $V$ and edges $E$
between nodes. The space is the tuple of vertices, and their structure
and dimensionality are determined from the topology of the edges.  A
graph is said to be acyclic if it contains no loops.  A forest is a
collection of possibly disconnected tree-like or acyclic graphs.

In a general graph, there is no implied regularity of structure as there is
in the tiling patterns of a group lattice.  Graphs may be directed
(with arrows marked in one direction) or undirected (without arrows)
in which case both directions are implied (see the example in fig. \ref{graph}).

The familiar concepts of dimensionality and direction begin to unravel
when we turn to graphs. On the other hand, graphs help us to see the
issues of space more clearly than any other representation, because we
are forced to separate our notion of dimension and direction (as
defined by a basis) from our ideas of adjacency, change and connectivity.

\subsubsection{Adjacency and matrix representation}

\begin{figure}[ht]
\begin{center}
\includegraphics[width=4.5cm]{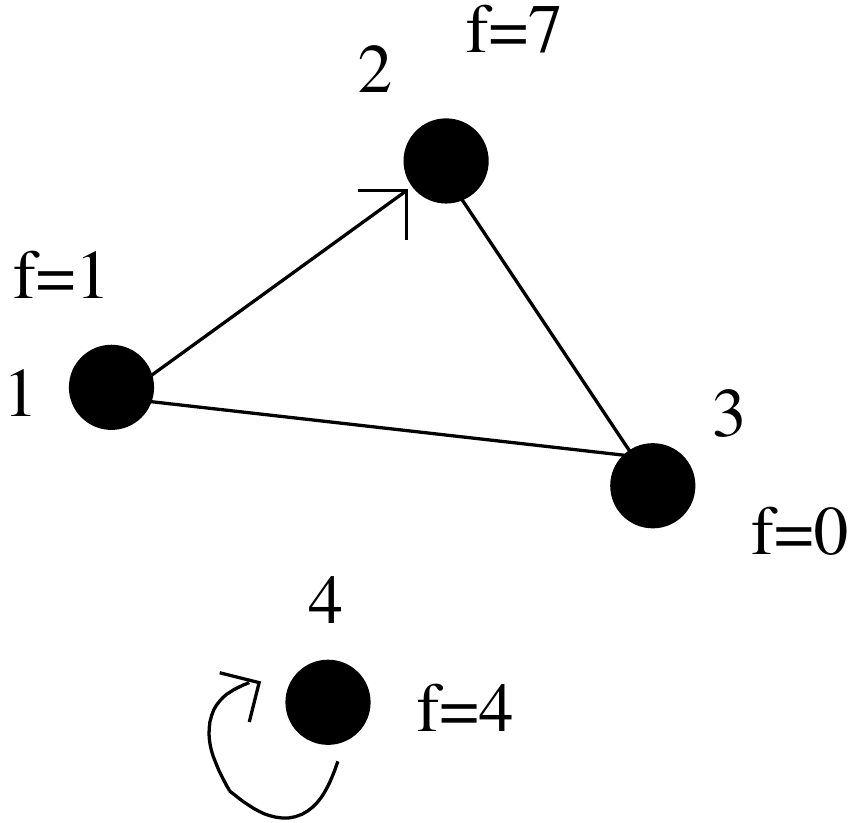}
\caption{A four vertex graph $G_4$, with a function f(G).\label{graph}}
\end{center}
\end{figure}

A graph, may be represented as an ordered tuple of nodes or vertices.
The connectivity (adjacency) of nodes in the graph may be either
directed or undirected.  In either case, the entire graph may be
represented by its adjacency matrix $A$, whose rows and columns
completely describe the graph.  For example fig. \ref{graph}) has
adjacency matrix 
\beq A_{ij} = \left(
\begin{array}{cccc}
0 & 1 & 1 & 0\\
0 & 0 & 1 & 0\\
1 & 1 & 0 & 0\\
0 & 0 & 0 & 1
\end{array} 
\right)
\eeq
where $i,j$ run over the rows and columns, i.e. the ordered vertices.

The matrix element $A_{rc}$ of the adjacency matrix $A$ may be thought
of as the generator of a translation, or a {\em degree of freedom}
$A_{r\to c}$. That is, $A_{rc} = 1$ if $r$ points to $c$.  The symbol
$\lambda$ is used as an unspecified eigenvalue of a matrix. $\rho$ is
reserved for the principal eigenvalue, or spectral radius of a
strongly connected graph.  A graph which has repeated eigenvalues in
its spectrum is said to be {\em degenerate}.

The in-degrees of the nodes, i.e. the numbers of links that point to
each node, is denoted by the vector $\vec k^{\rm in}$; the out-degree vector,
or number of outgoing links per node, is denoted $\vec k^{\rm out}$.
Recalling that $A_{rc}$ implies $r$ = row and $c$ = column, then
in a directed graph, the row-sum of the adjacency matrix is the
out-degree of the node whose row is summed:

\beq
\sum_{c}\; A_{rc} = k_r^{\rm out}.\label{n1}
\eeq
Similarly, the column sum is the in-degree of the node:
\beq
\sum_r\;A_{rc} = k_c^{\rm in}.\label{n2}
\eeq
A square, leading-diagonal sub-matrix of $A$ is written $a_{rc}$, with
$r,c$ now running over a limited set of values. A non-square,
off-diagonal sub-matrix of $A$ is written $\ell_{rc}$.

\subsubsection{Strongly connected components}

The concept of an `SCC' is useful in describing regions of a directed graph\cite{graphpaper}.
\begin{definition}[SCC - Strongly Connected Component]
Let $G$ be a directed graph. A strongly connected component of G
is a maximal subgraph, $g$, in which there exists a directed path from
every node in $g$ to every other node in $g$, by some route.
\end{definition}
It is implicit in this
definition that a path follows the direction of the arrows
(links). The set of SCCs is uniquely determined by the graph, and
every node belongs to one and only one SCC. The number of SCCs for a
graph with $N$ nodes may be as large as $N$ for a directed acyclic
graph or DAG, or as small as one. In the latter case we simply say
that the graph is strongly connected (SC).

The weaker property of being {\em connected} simply requires that
every node in a connected subgraph has some path to every other, {\em
disregarding} the direction of the links (ie, treating them as
undirected). We will discuss only connected graphs in this paper.

More important is the notion of a complete subgraph, or Complete Connected Component (CCC).
\begin{definition}[CCC - Completely Connected Component]
Let $G$ be a directed graph. A strongly connected component of G
is a maximal subgraph, $g$, in which there exists a non-directed path from
every node in $g$ to every other node in $g$, by direct adjacency.
\end{definition}

The Perron-Frobenius theorem addresses the property of {\em reducibility}
in a graph\cite{varga1,minc1}. Each irreducible region can be associated with an SCC,
for the following reason.
A real $N\times N$ matrix $M$, with non-negative entries (such as an
adjacency matrix), is said to be irreducible if every element,
labelled as a row-column pair $(r,c)$, is greater than zero in some
finite power of the matrix, i.e., for every pair $(r,c)$, there is a $p$
such that $(M^p)_{rc} > 0$. The adjacency matrix $A$ of a
strongly connected graph is irreducible, since $(A^p)_{rc}$ is just
the number of paths from $r$ to $c$ of length $p$. Conversely, if a
graph is not strongly connected, then it is reducible.

\subsubsection{Functions $f(x)$ on a graph}

A graph may be represented in a functional form by grouping the
vertices into an $n$-tuple.  The naming of vertices is accomplished by
treating them as a poset, i.e. by simply numbering them uniquely.
This is a global operation. Indeed, graphs do not support the
concept of locality even in disconnected components.

The values in the tuple the represent
the values of a function on the vertices, analogous to $f(x)$:
\beq
f(G_4) = \left(
\begin{array}{c}
1\\
7\\
0\\
4
\end{array}
\right)
\eeq

The edges of the graph may or may not have a realization, i.e. a
specific interpretation.  Any association will do to relate to
elements of the graph.  Graphs are formed by many structures. In
statistical analyses, correlations matrices lead to graphical
structure for example. Analysis of their properties has considerable
importance to revealing trends and clustering properties in the space
of correlation distance. In such a case, the row and column elements
are not 1 or 0, but a value of the correlation function, acting as a
relative weight or distance in the graph.

The edges of the graph may also be viewed as generators of a transformation
from a tuple of names for the vertices onto its image. Since the tuple is
a representation of the space, the operation of the adjacency matrix acts
to generate a rotation of the space in the direction of the arrows.
\beq
A: V \rightarrow V
\eeq
The fixed points of this map lead to an eigenvalue distribution
over the graph, whose principal eigenfunction describes the
relative centrality of the nodes (see \cite{burgesslongversion,graphpaper}).

Self-loops (like identity elements of a category) are important for
global stability of automorphic functions on graphs, as they
avoid singularities by allowing `pumping' of functional value
at a source node, and `orbiting' at sink nodes. Without such nodes
directed graphs that point to sinks act as singularities that suck up
all distribution value.

\subsubsection{Distance on a graph (hops)}

Apart from the weighting of links on the edges of the graph, one can measure
the distance between any two nodes as the length of the path one follows
from one node to another, in `hops', by summing the values of $A_{rc}$ along the path.

\subsubsection{Linear independence, matroids and spanning trees}

A notion of linear independence does not exist for a graph, but we may
use the matroid concept to define dimensionality of a region in terms
of sets of edges.

\begin{figure}[ht]
\begin{center}
\includegraphics[width=7.5cm]{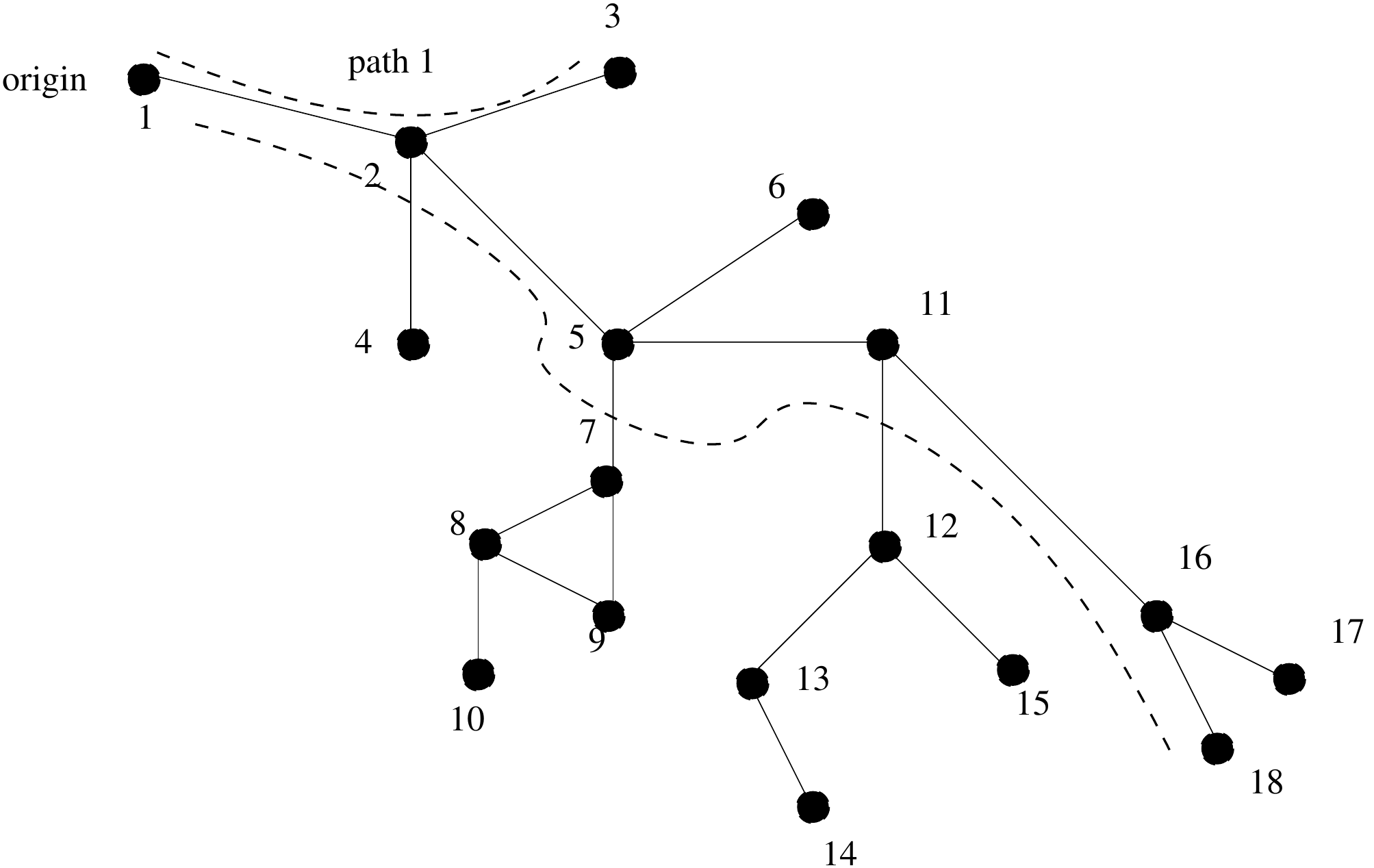}
\caption{A graph and the tuples measured along its spanning tree.\label{graphdimension}}
\end{center}
\end{figure}

Suppose we take the base set $E$ to be set of edges in a graph
$G=(V,E)$.  Then we may form a matroid basis of given rank.  The
individual edges may be considered independent sets provided they form
a forest, i.e. provided there are no cycles in the graph.
This is simply the definition of a {\em spanning tree} in network parlance.

If there are any loops, the extra edge simply short-circuits two nodes
so that there are two routes to the same location, implying that they
are dependent. Each independent edge has to take us somewhere new
in order to be an independent edge (see appendix \ref{appex1}).

The dimensionality of a graph is sometimes defined by the immersion of the
graph into a Euclidean space. Another definition involves the number
of independent colours needed to label adjacent nodes without the same colour
being directly adjacent. However, if we follow the pattern of basing
dimensionality on matroid rank, and the association of tuples spanned by the
independent sets then the interpretation of dimension is clear.
The tuples of an $n$-dimensional graph are spanned by the edges from some origin node or tree root
along the unique paths of a spanning tree, of rank or length $n$. 

The independent sets are thus rank $r$ unique paths $p_k$ from some
origin to the leaves in a spanning tree for the graph. These
constitute the independent {\em directions} in the graph.  Then a
tuple has the form $(p_1, p_2, \ldots ,p_r)$, where the path $\vec p_k
= \{ v_p\}$ is the set of SCC effective vertices that lie along it.
This means we must label all vectors between non-adjacent neighbours
by the independent path by which we intend to traverse the graph order
to label their independent role, e.g. $\vec e^{p_1}_{1n}$.  This is
the equivalent of specifying a basis vector with a fixed tuple label
in a lattice.  For leaf nodes, this is unique.

This is only one coordinatization of the graph. Just as one may have
Cartesian or polar coordinates in Euclidean space, so there are many
options for parameterizing a graph as a multi-dimensional object.  See
appendix \ref{appendix} for an discussion of graph dimensionality with
examples. The key finding is that, unlike the case of a regular lattice,
dimensionality and adjacency by symmetry generator are independent
concepts -- they are not interchangeable as they are in a group lattice.

Graphs that exhibit no special symmetries are fundamentally bounded
structures.  Since the different paths form strings composed of
different numbers of elements, spacetime is bounded {\em
  anisotropically}, i.e.  the size of a graph space is different along
its different directions.  Moreover, these properties depend on the
specific choice of root node or origin, so space is also {\em
  inhomogeneous}. In a directed graph, there is moreover the curious
property of singularities in space into which one may move but never
return, or from which something can emerge but not be absorbed.
This is reminiscent of the idea of monopole and charges in physics, or
singleton objects in information technology.

The length of the paths is important in deciding the range over which
the tuple values run (since one SCC might container more vertices than
another), as well as the number of vertices in any strongly connected
components along the path. In fig. \ref{graphdimension}, nodes 7,8,9
all belong equivalently to a single SCC along the path to node 10.
There is thus a local symmetry transformation that implies hidden
degrees of freedom at this stage of the path.

The tuple coordinates label symmetrical elements within SCCs.  So the
extent of tuple coordinates depend on how many different nodes we can
reach in the strongly connected region along a given path. Movement within
the CCCs behaves like `hidden dimensions' embedded within the larger
spanning set.

\subsubsection{Scale transformations a graph}

Scaling in a graph is not a simple matter of multiplicative
renormalization of measures. Graphs are not self-similar in general,
and thus such a scaling has no meaning.  Figure \ref{scale2} shows
what we mean by scaling in a graph. 
In general, we have seen that the coordinatization of a graph depends
mainly on the edges in between completely connected regions.

At a given scale, we may form a coarse graining of the graph by
constructing the strongly connected components up to a fixed number of
nodes $n$, or horizon.  One ends up with new single elements in place
of CCCs, linked by a new matroid, bounded by a given number of hops.
This is a renormalization of the graph of level $n$.

Repeating the process, one may perform further renormalizations,
until the graph is a forest. This forest yields the large scale 
connectivity of the graph.

\subsubsection{Derivatives and vectors}

A vector on a graph is simply a pair of vertices $(i,j)$, where $i,j \in V$.
It is natural, though not necessary, to restrict vectors to
only directly connected vertices, so that valid vectors are
those for which there is a non-zero element in the adjacency matrix.
If we consider vectors linking nodes that are not directly connected,
implying a path or a route composed of multiple hops, a vector might
not exist or if it does it might not be unique.

The partial derivative of a function defined on the tuple of vertices,
may be defined along nearest neighbour edge $k$, at node $i$, as 
\beq
d \vec f_k \Big|_i = f_{i+} - f_i,
\eeq
where $+$ is the succession operator along $k$.

\subsubsection{Lattices from irregular graphs}

Our familiar notions of spacetime involve tiled lattices, so when we
think of graphs as spatial, they should scale up they have to lead to
familiar topologies.

Like a lattice, we can expect there to be a connection between
neighbouring points in the tuple space spanning a general graph. In a
lattice we generally expect that (1,0,0) and (2,0,0) are adjacent and
connected unless there is a boundary.  In a graph, we cannot assume
that the neighbouring point exists to a uniform value, as a particular
direction can suddenly end in a cul de sac.
However, (1,0,0) and (0,1,0) might actually be the same point. Paths
are not orthogonal.

Question: What requirements need to be imposed to ensure large scale lattice structure from local graphs?

\subsection{Symbolic grammars as spatial models}\label{grammars}

Information sciences are principally about the recognition and
manipulation of discrete sequential string-like patterns called {\em
  languages}.  Languages are formed from alphabets of symbols.  A
symbol is a spatial pattern, or may be representative of one. This
allows us to compress spatial representations, define the notions of
semantic index and code book (see section \ref{codebook}).

The {\em syntax} of any language is the set of all allowed
strings of symbols. It can be modelled by a general
theory of its structure, called a {\em grammar}.  Grammatical methods
assume that arriving data form a sequence of digital symbols (called
an alphabet) and have a structure that describes an essentially
hierarchical coding stream.  The meaning of the data is understood by
{\em parsing} this structure to determine what information is being
conveyed. The leads us to the well-known Chomsky hierarchy of
transformational grammars (see, for instance, \cite{lewis1}).

Language is clearly something closely associated with semantics, so a
pseudo symbolic structure for space is of considerable significance to
our interpretation of its meaning. Documents are nothing if not spaces
of structured symbols which are assigned explicit meaning.

Because of their regularity and conduciveness to formalization,
computer science has seized upon the idea of grammars and automata to
describe processes. Symbolic logics are
used to describe everything from computer programs and language
(\cite{logic}) to biological mechanisms that describe processes like
the vertebrate immune response (\cite{jerne1}). Readers are referred
to texts like
\cite{lewis1,watt1} for authoritative discussions.  For a cultural
discussion of some depth see \cite{hofstadter1}.

\subsubsection{Dimensionality and topology in languages}

Languages are usually written explicitly in a one-dimensional
representation, in the sense that they are viewed as sequences of symbols.
However, context-free grammars support the embedding of parenthetic
sections. However, we are familiar with the idea that a sequential book
can embed structures like chapters and footnotes. Such `digressions' belong
in a different dimension to the main thread of a storyline.

One only has to think about the storage of information on a computer
to understand how dimensionality is an issue of interpretation.
Imagine a two dimensional table, encoded as a one dimensional stream
of bytes read from a three dimensional array of disks. What appears to
be a connected stream is generally stored in very disjoint locations
within a storage array.  What information science refers to as a {\em
  virtual view} of data is simply a redefinition of spacetime semantics.

\subsubsection{Automata as classifiers of grammatical structure}

A grammar is a multi-dimensional structure encoded as a linear stream
of symbols taken from an alphabet $\Sigma$.
The complexity of patterns in a language may be classified according to the level of
sophistication required to compute their structures.  Chomsky defined
a four-level hierarchy of languages called the Chomsky hierarchy of
transformational grammars that corresponds precisely to four classes
of automata capable of parsing them. Each level in the hierarchy
incorporates the lower levels: that is, anything that can be computed
by a machine at the lowest level can also be computed by a machine at
the next highest level.

\begin{center}
\begin{tabular}{ll}
State machine & Language class\\
\hline
Finite Automata & Regular Languages\\
Push-down Automata & Context-free Languages\\
Non-deterministic Linear Bounded Automata & Context-sensitive Languages\\
Turing Machines &  Recursively Enumerable Languages\\
\end{tabular}
\end{center}
State machines are therefore important ways of recognizing input, and
thus play an essential part in human-computer systems.

Time progresses in an automaton or state machine by the arrival of
each new symbol in a language string, even a blank one.  Thus the
automaton that parses a language is its own clock.
Space and time are explicity matched one to one.

\subsubsection{Naming of elements in a grammar}

A grammatical structure may be used to span a space, such as a
document.  Its regions form a matroid, (like a spanning tree in the
case of a hierarchical grammar) of strictly non-overlapping elements.

The atomic elements of such a space are the symbols of the language
alphabet themselves, and but there may be larger regions of repeated
patterns that form regions or coordinate patches, and could be
aggregated in a `rescaling'. Indeed, during data compression it is the
replacement of extended patterns of atomic symbols with single
`meta-symbols' that reduces the size of the space. Take the following string of
symbols from the alphabet: $\Sigma = \{ A,B,G,N,Y,Z,),(\}$.
\begin{alltt}
XXX(YZA(BG))ANXXX(A)...
\end{alltt}
We notice a repeated pattern `XXX' which could be aggregated into a region as a rescaled element, analogous
to an SCC, or an independent set.

Regular expressions (or patterns of an embedded regular language) can,
for example, represent larger strings or `words' in a language which
repeat and are attached significance. This is a common way of parsing
documents in information technology.

Another parenthetic language which used extended symbols for begin and end parenthesis
is the Hypertext Markup Language used in the World Wide Web. The semantics of each region
are encoded into the name for the relevant parenthesis.
\begin{alltt}
 <html>
  <body>
    <h1>My title</h1>

    This is some body of text, where the context has no interpretation
    to the HTML language, it is a human language embedded in a machine
    language.  
  </body> 
 </html>
\end{alltt}

\subsection{Bigraphs: nested structures}

Graphs have many important qualities, but they are elementary
structures.  Let us now consider a generalization of graphs with
intentional semantics, using some concepts from category theory. This
work was pioneers by Milner in an attempt to seek a language for
describing spatial arrangements in computer
science\cite{milnerbigraph}.

Bigraphs are a composite model of space that add to graph theory a
specific notion of semantics in the arrangements of things.  Rather
than discussing boundaries, it talks about interfaces, imagining that
arrangements or fragments of bigraphs will be composed into larger
ones. Bigraphs, are thus a model for machinery.

Milner used a topological categorical model of discrete containers to
mark out regions of agency. This is a non-Euclidean view.  Bigraphs describe places (somewhat
analogous to the primitive points in a topological space) and the
links between them. The difference is that Milner's points may contain
internal structure, including holes to be filled in with other
bigraphs, like pluggable modules.  This assumes the notion of an
embedding of the structures described within some kind of topological
space.

\subsubsection{Bare graphs, forests and links}

\begin{figure}[ht]
\begin{center}
\includegraphics[width=9.5cm]{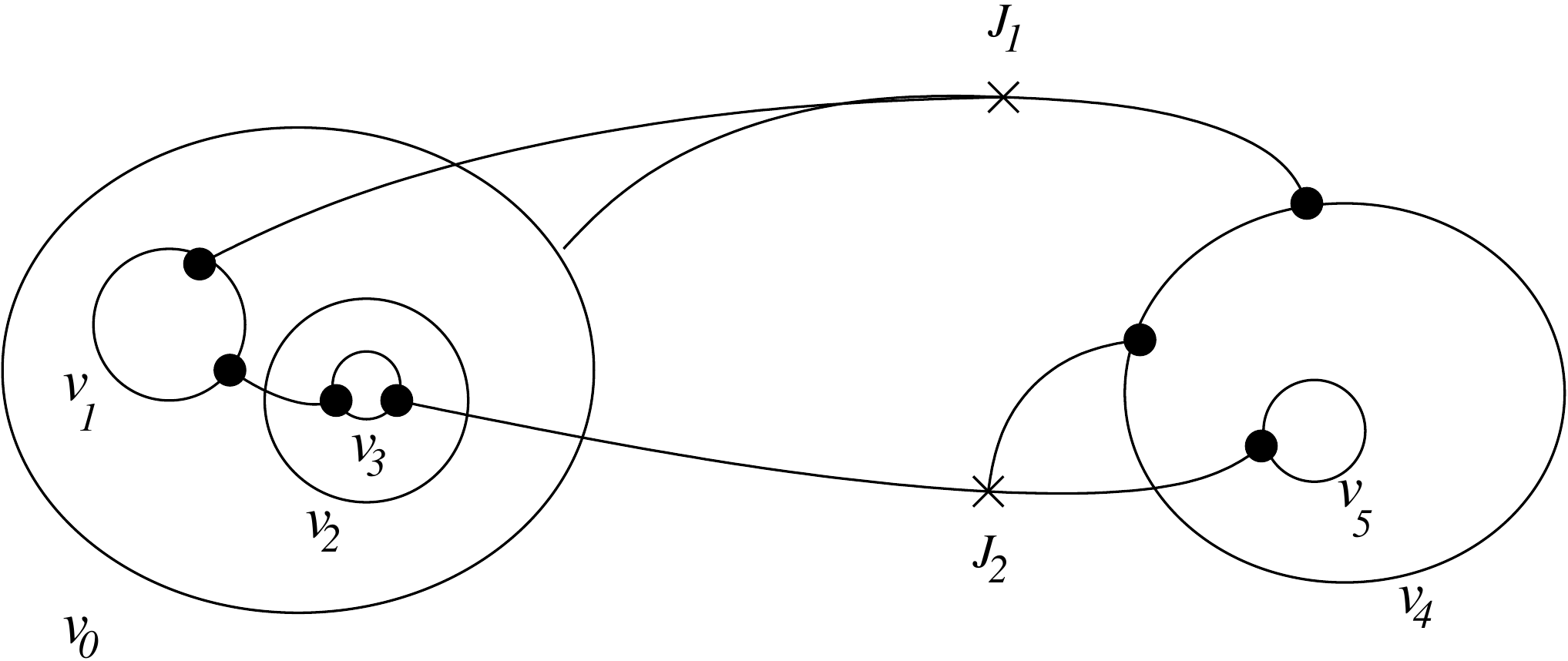}
\caption{A bare bigraph\label{bigraph1}}
\end{center}
\end{figure}

Bigraphs merge several points of view in order to add interpretation to a regular graph.
A `bare' bigraph $\breve G$ may be seen in fig \ref{bigraph1}. It contains
the basic relationships between elements.
vertices $v_i$, some of which are inside one another, and edges $e_i$.
The edges are not labelled in the figure. In standard graph theory, edges are always
point to point adjacencies; however Milner follows his own convention here and treats
edges as `communication buses' that can have junctions. Formally this makes bigraphs bipartite graphs.
For clarity, I shall mark the junctions $J_\ell$ with an $X$ symbol
on the edge, to indicate the presence of
a junction agent (see fig. \ref{bigraph1}).

\begin{figure}[ht]
\begin{center}
\includegraphics[width=4.5cm]{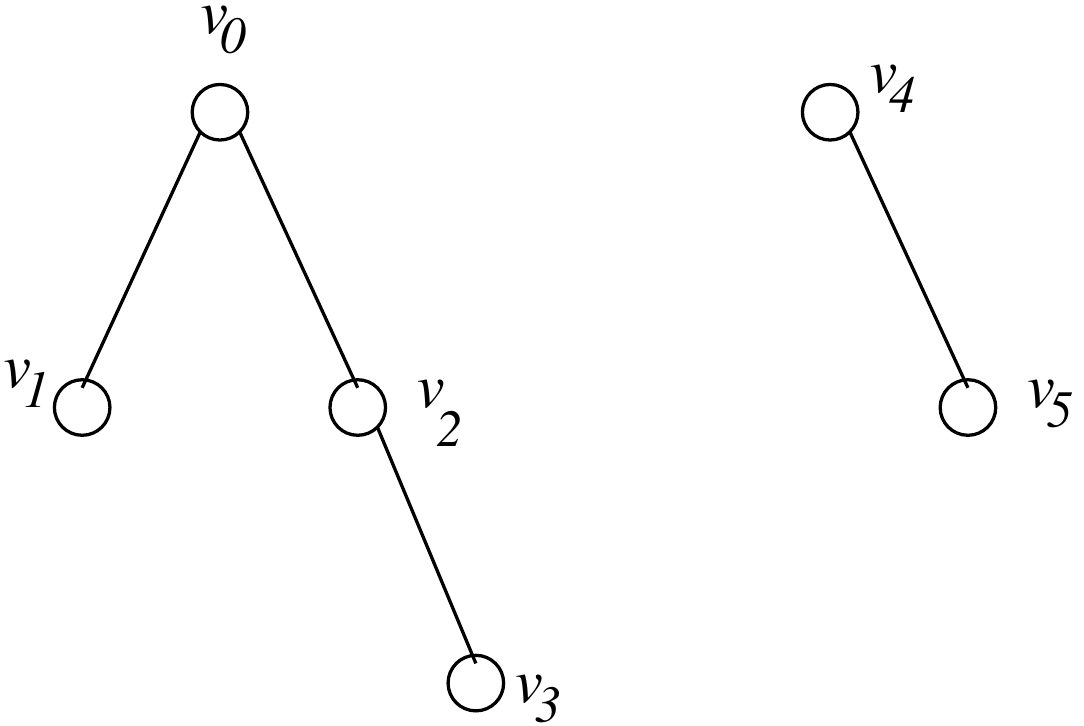}
\caption{A forest view of the bigraph\label{forrest}}
\end{center}
\end{figure}

From the single picture in fig. \ref{bigraph1}, we may infer two views. The first is
called the {\em forest graph} of $\breve G$ (see fig \ref{forrest}).
A forest is a treelike structure without the constraint of connectedness.
The forest view shows the hierarchical relationships between the
vertices, from the largest at the top to the smallest at the bottom.
It indicates the boundaries of containment membranes through which a
link must pass in order to connect vertices.  Here large implies the
ability to contain small (though size is implicit, since bigraphs have
no internal notion of size). It is typically disjoint unless all vertices are encompassed by a
single root. The hierarchy represents a notion of boundary, as in
Gauss' law: that which can emanate from a vertex depends on that which
is inside it.

The second derived view is the {\em link graph}, which illustrates the channels
of connectivity (perhaps communication) between the vertices.
\begin{figure}[ht]
\begin{center}
\includegraphics[width=4.5cm]{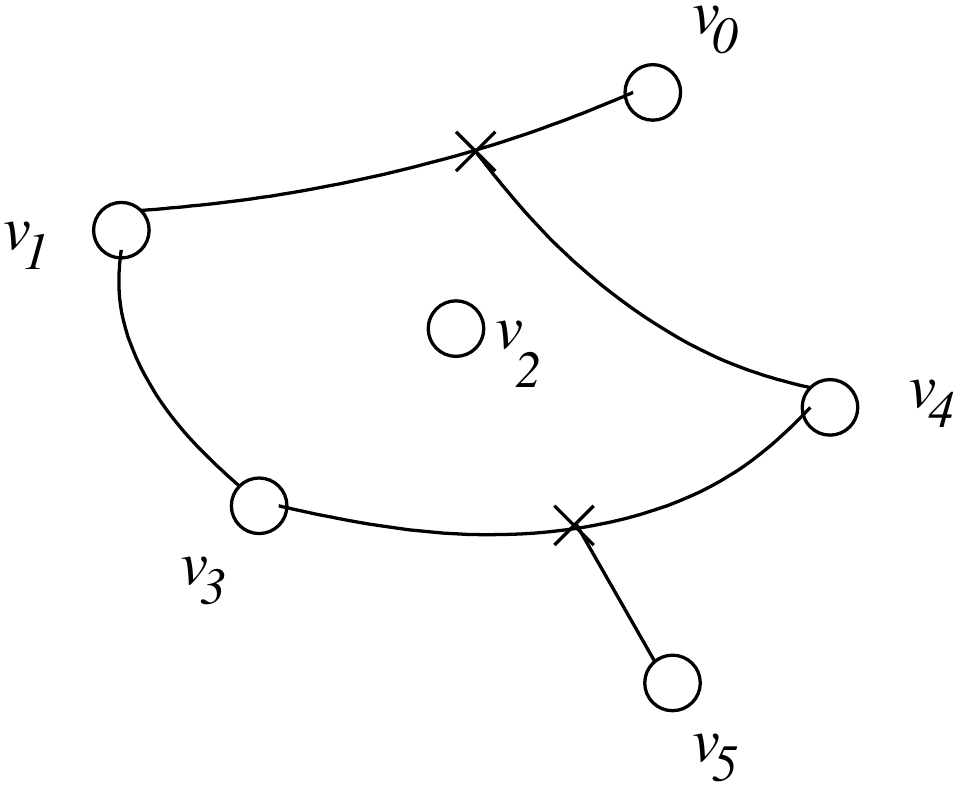}
\caption{A link view of the bigraph\label{link}}
\end{center}
\end{figure}
Note that without the junction markers $J_1, J_2$, these link graphs look
unlike traditional graphs of graph theory.

A bigraph `face' is thus associated with two different sets of description
$G = \langle N, \{ X\}\rangle$, where $N$ is a natural number representing the countable vertices,
and $X$ is a set of links.

\subsubsection{Boundaries and interfaces on a bigraph}

The forest graph indicates containment relationships, while the link
graph indicates connectedness. It is also possible to denote open
ended channels as `doorways' in and out of the graph as interfaces (or
simply `faces' in Milner's language). This will allow bigraphs to be
thought of as representing the structure of pieces of active
machinery, like mathematical operators, and further be composed as fragments.
This will be analogous to having matrices in a vector space.

\begin{figure}[ht]
\begin{center}
\includegraphics[width=5.5cm]{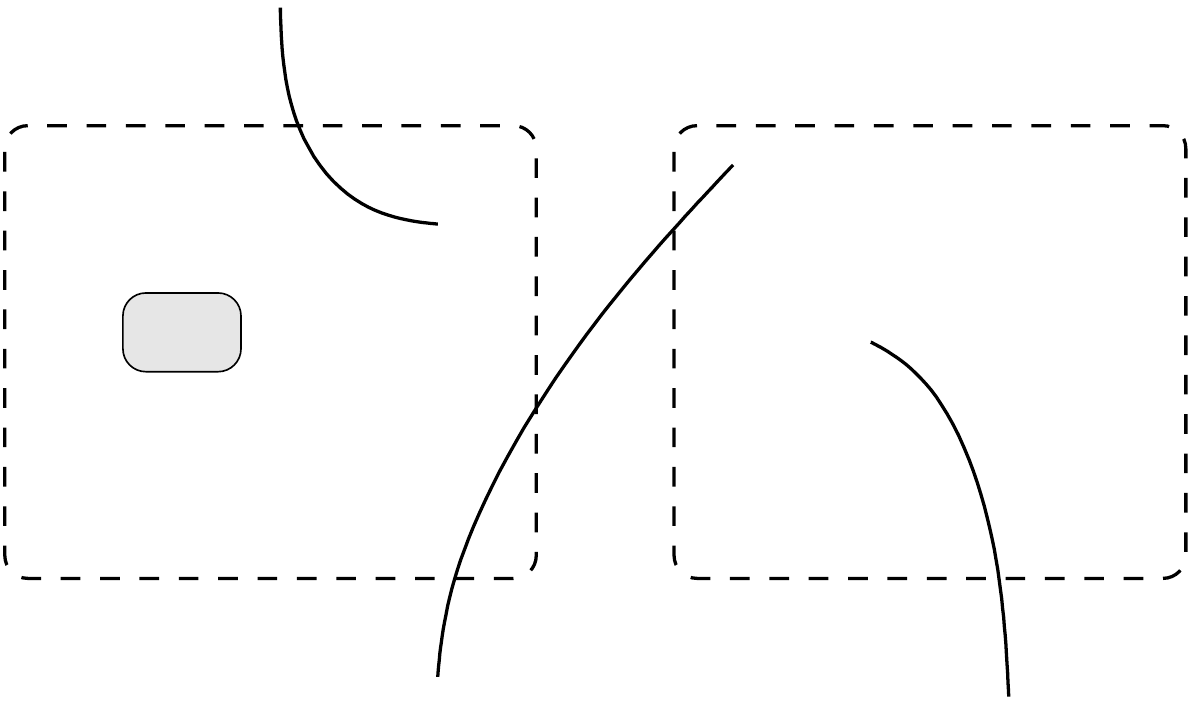}
\caption{A interface mapping one interior site to two regions, and
two input links to one outgoing: $\langle 1, 2 \rangle \rightarrow \langle 2, 1\rangle$\label{faces}}
\end{center}
\end{figure}

A complete bigraph interface is a mapping from an {\em inner} or input
face to an {\em outer} or output face. We write the face in angle brackets
as a transformation from inner to outer faces:
\beq
{\rm inner} &\rightarrow& {\rm outer}\\
\langle {\rm sites}, {\rm inlinks} \rangle &\rightarrow& \langle {\rm regions}, {\rm outlinks} \rangle
\eeq
For example, in fig. \ref{faces}, we see a mapping from one inner site to two
outer facing regions, and two incoming links mapping to a single outgoing link:
\beq
\langle 1, 2 \rangle \rightarrow \langle 2, 1\rangle
\eeq

\begin{figure}[ht]
\begin{center}
\includegraphics[width=5.5cm]{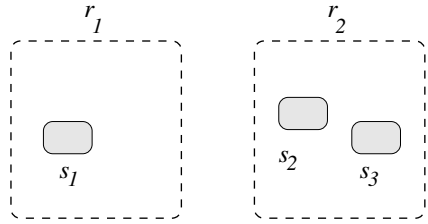}
\caption{A place-oriented interface, with regions and sites\label{placeface}}
\end{center}
\end{figure}

Site-region interfaces are drawn as meta-locations around the bare
bigraph, using boxes (see fig \ref{placeface}). The outer boxes are
regions (sometimes called roots) of the graph are labelled $r_i$, and
represent distinct places. The sites, labelled $s_i$, are like holes
into which other regions could be plugged. Sites therefore take on the
guise of receptors, as in biology, or sockets as in engineering.
Regions are thus the outer face of elements, while sites are the inner
face.  They are the closest representation of Euclidean space in this
description.

\begin{figure}[ht]
\begin{center}
\includegraphics[width=5.5cm]{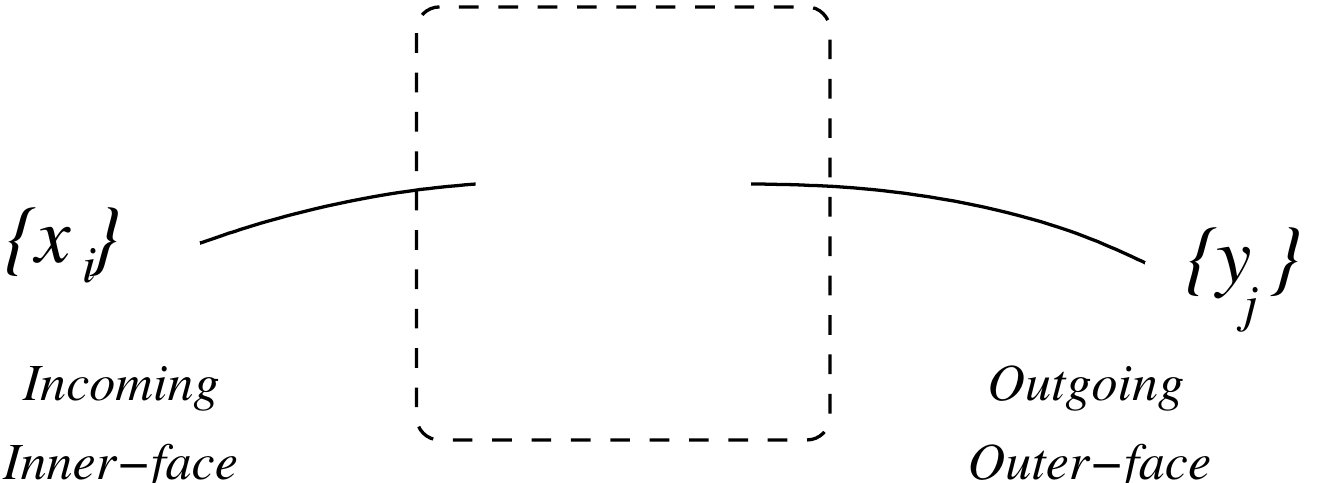}
\caption{A link oriented interface\label{linkface}}
\end{center}
\end{figure}

Links can also form interfaces, ready to connect with the connectors
at the boundary of other places.
Link interfaces are edges that extrude from the outmost region of a bigraph
and `wait to be completed' (like an operator hungry to act on its
operand).  Conventionally one draws incoming interfaces beneath a
bigraph (as inputs) labelled $y_j$ and outgoing interfaces (outputs)
above the graph, labelled $x_i$ (see figure \ref{linkface}).

\begin{figure}[ht]
\begin{center}
\includegraphics[width=10.5cm]{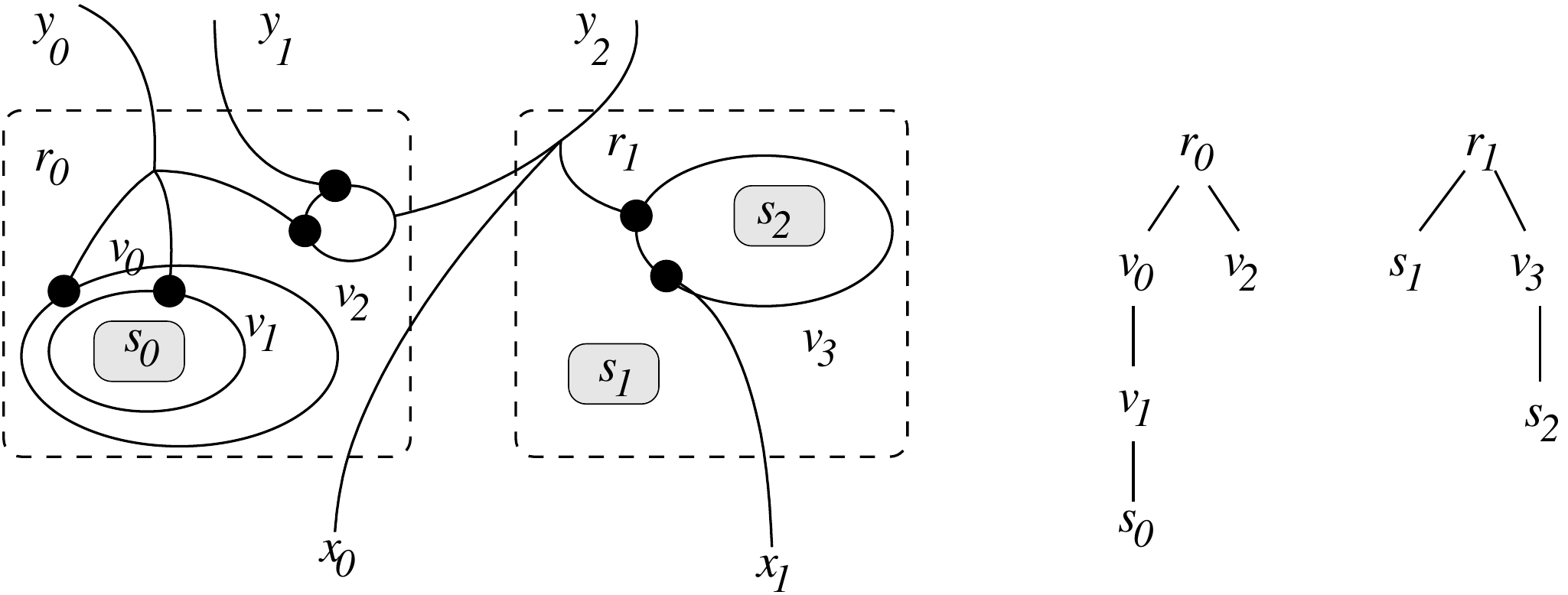}
\caption{A fully featured bigraph, mapping inner to outer faces:
$\langle 3, \{x_0, x_1\}\rangle \rightarrow \langle 2, \{ y_0, y_1, y_2\}\rangle$\label{bigraph2}}
\end{center}
\end{figure}

Combining all the features discussed leads to a general bigraph, such as that
shown in fig. \ref{bigraph2}.

\subsubsection{Signatures and sorts for semantic typing of spatial structures}

The addition of {\em sorts} for bigraph vertices is one way of using an
algebraic concept to endow them with elementary semantics. Type and
arity with respect to links, for each vertex, provides an operator
view of vertices, making them functional or agent-like, i.e. classifying
 them with named roles. We'll see the
same notion of typed agency in promise theory.

The typing of vertices, including their `arity' with respect to edges
specified a functional signature for the bigraph machinery.
In order to arrive at a signature, we have to classify or name the nodes
by role. Milner refers to these names as {\em controls}.
\beq
{\rm Signature} = \left\{ K:2, L:0, M:1 \right\}
\eeq
\begin{figure}[ht]
\begin{center}
\includegraphics[width=5.5cm]{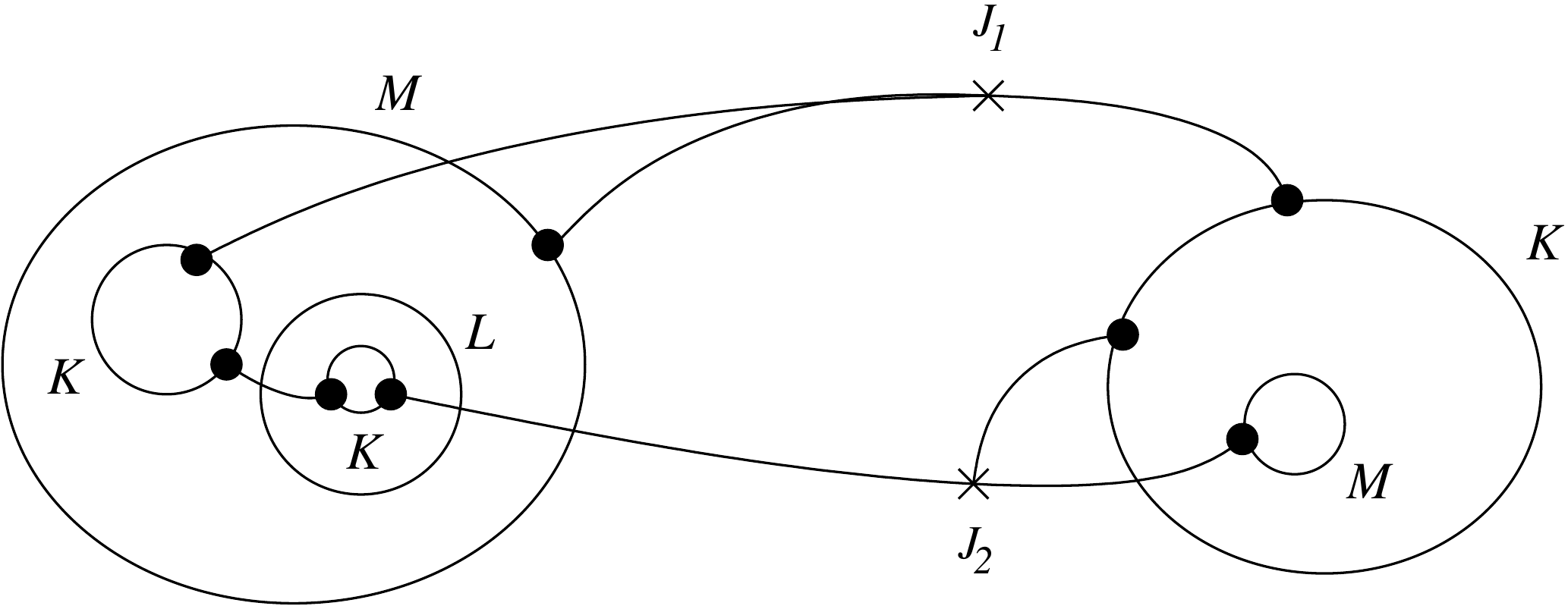}
\caption{A link oriented interface\label{signature}}
\end{center}
\end{figure}
For example, the type $K$ might represent a building, and $M$ might represent a fenced area.
$L$ might represent a table. Or, $K$ might be an oscillator, $L$ an amplifier, and $M$ a coil.

If there is a sufficiently detailed typology of controls and their
arities, then a signature acts like a kind of role inventory for a
bigraph: a description of the basic functions of its machinery.
It will not normally be sufficient to tell the purpose of the machinery,
since that depends on how it is combined with other bigraphs, however
it offers a table of elements for a chemistry of composition, at a molecular level.

\subsubsection{Composition of bigraph operators}

The composition of bigraphs can be carried out in a number of different ways.
The simplest composition is to simply place two graphs alongside one another without
connection. This is called juxtaposition, and is denoted $G_1 \otimes G_2$. It is considered
to be a tensor product.

A proper inner product (see equation \ref{innercategory}), analogous to a
matrix product in linear algebra may be carried out if the outgoing
face of one bigraph matches the incoming face of the other, making
them compatible.  A bigraph of arity 2 is like a matrix, and arity 1 is like a tuple.
For example, suppose we have faces:
\beq
I = \langle a, \{x\}\rangle\\
J = \langle b,\{y\}\rangle\\
K = \langle c,\{z\}\rangle
\eeq
and bigraphs
\beq
G_1 = A: I \rightarrow J\\
G_2 = B: J \rightarrow K
\eeq
Then we may write
\beq
C = B\circ A = G_2 \circ G_1 : I\rightarrow K,
\eeq
which is the analogue of matrix multiplication $C_{ik} = \sum_j A_{ij}B_{jk}$.
Two bigraphs are said to be incompatible if they cannot be composed,  written $G_1 \# G_2$.
This process is illustrated in figure \ref{FoG}, where we see how the regions of
$G_1$ are slotted into the sites of $G_2$, and the outgoing links match up to the incoming 
links (as labelled):
\begin{figure}[ht]
\begin{center}
\includegraphics[width=8.5cm]{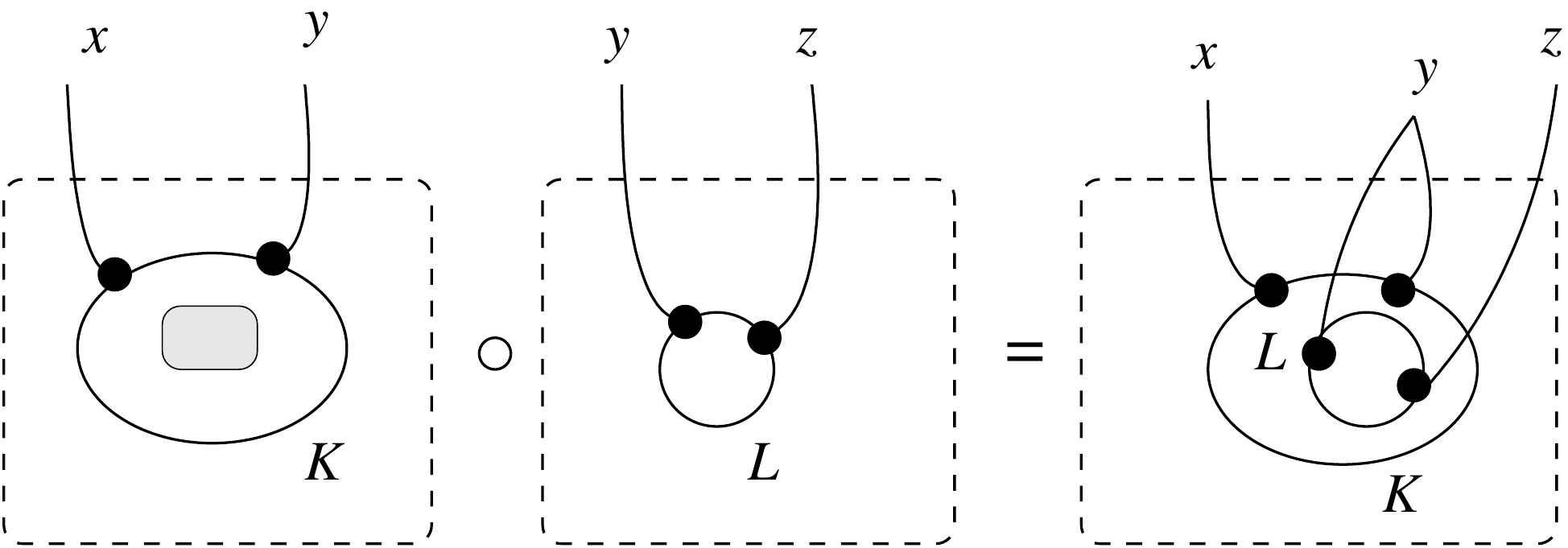}
\caption{The composition of two bigraphs by proper inner product (nesting)\label{FoG}}
\end{center}
\end{figure}

\begin{figure}[ht]
\begin{center}
\includegraphics[width=8.5cm]{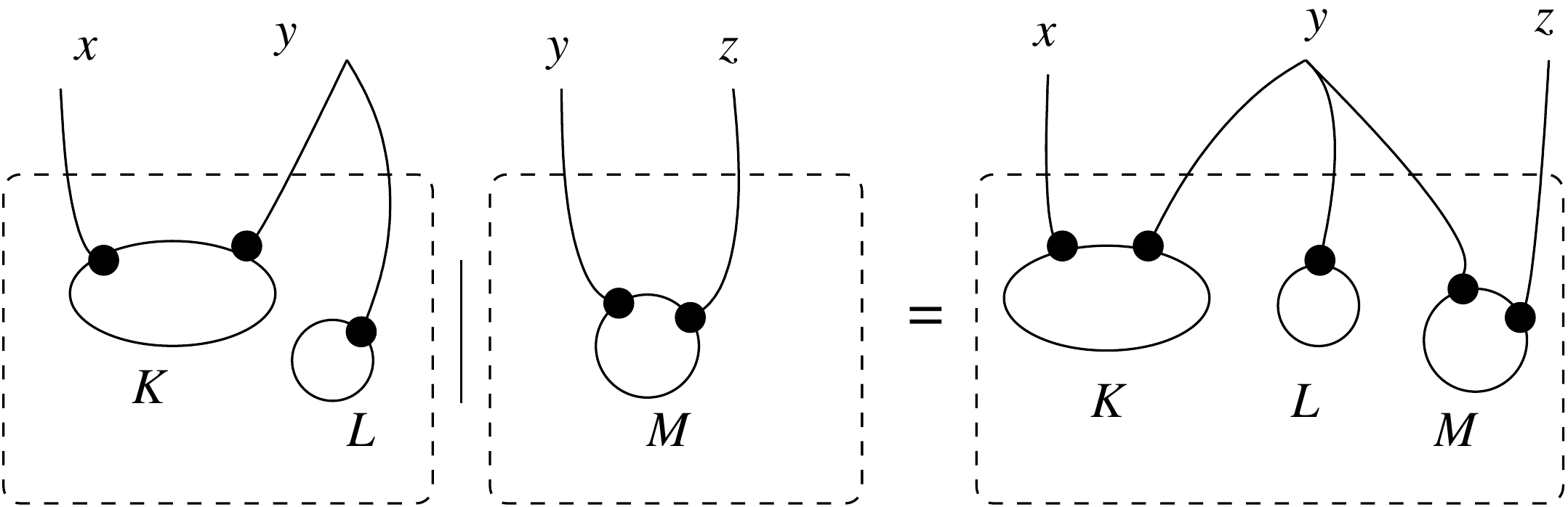}
\caption{A link composition of two bigraphs merging places, by semi-proper inner product\label{FIG}}
\end{center}
\end{figure}

Other kinds of combination are possible, where there is an absence of sites.
In fig. \ref{FIG}, one defines $G_2 | G_1$ to mean the connection of links and the merging
of the regions from the two graphs into a single region, proving a `semi-proper' inner
product.

\begin{figure}[ht]
\begin{center}
\includegraphics[width=8.5cm]{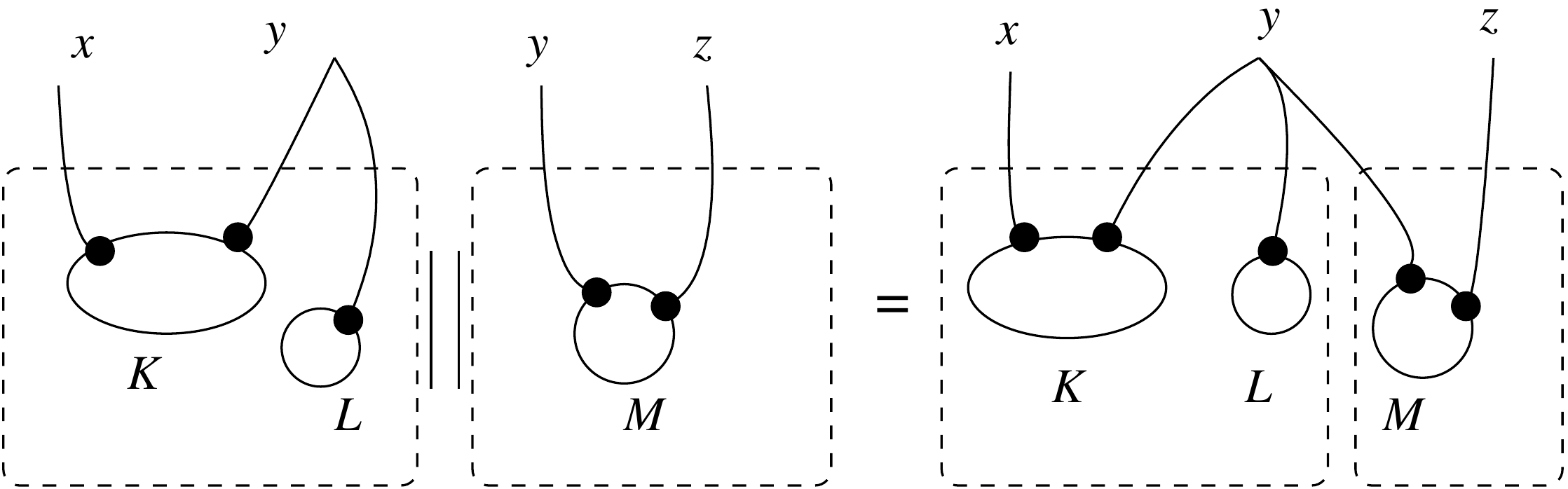}
\caption{A link composition of bigraphs without merging places, by improper inner-product\label{FIIG}}
\end{center}
\end{figure}

Similarly, the same operation can be carried out without merging the regions
of the graph, as shown in fig \ref{FIIG}, in an `improper' inner product..

\subsubsection{Time and motion in bigraphs}

We have established from the nature of composition that a bigraph is
analogous to a matrix in linear algebra, i.e. it transforms other
compatible bigraphs that it operates on with its faces.  Such vectors
are not described by Milner, but clearly they exist as bigraphs with
only incoming faces (sinks), or transverse vectors with only out-going
faces (sources).

In linear algebra, motion of vectors by translation and rotation
occurs by acting on vectors with the matrices, hence there is an
analogous notion of change of motion enacted by the interpretation of
bigraphs as machines that enact change.

Milner's own idea of bigraphs is that the layout of this machinery
itself is what changes over time, and such changes require a separate
kind of transformation known as a {\em reaction rule}.  The analogy
here to linear algebra is that of a group transformation of a matrix
operator.

Time is enacted by reaction rules in Milner's view, but it may also be
marked by changes carried out by the machinery (analogous to matrix
multiplication) as the number of stages of transformation, or number
of intermediate faces (co-domains) through which the machinery passes.
His language and formulation seem anchored to interaction machinery -
biological models, and electrical circuits.  Spatial containers have
interfaces, somewhat like matrices/tensors, that express dimensionality
at the joins between regions.

Bigraphs are important because the theoretical formulation is rigorous
and well developed, in spite if its very particular vision.  In
particular, its notion of containment semantics as a dual viewpoint to
continuity and translational invariance, is an inspired break from the
dynamical traditions of physics, where symmetry and continuity are the
starting points.

The weakness of bigraphs lies in the fact that all changes that can be
represented are semantic changes. This limits what we can say about a
environment where dynamics would be a more appropriate formulation of
change.

\subsubsection{Derivatives and vectors}

A derivative is a rather strange object for a bigraph. It can still be
defined as the difference between two bigraphs, though it is unclear
to what end one would need this concept. Bi-graphs are not a
dynamical model, they are a semantic model. What might a rate of
change mean under these circumstances?  The answer is likely connected
to the idea of versions and proper time, to be discussed later in these notes.

\subsection{Processes algebras, time and motion}

In computer science there is a description of finite system behaviour
expressed in terms of labelled state transitions, or process algebras.
A process consists of agents which undergo transitions.

A transition that happens externally might trigger an input of a value $x$ into input $i$, then the process
continues according to the definition of $P$
\beq
i(x).P
\eeq
Conversely, a process agent $P$ outputs a value on port $o()$ $y$ and continues according to the definition of $P$.
\beq
\overline o(y).P
\eeq
Transitions can also occur internally or non-deterministically (outside the scope of modelled process)
like `noise' on a Shannon channel:
\beq
\tau.P
\eeq
Algebras can be quite expressive at rendering scope and transitions (time).

The iteration implied in the equations is solved, much like a propagation
solved for a differential equation, by tracing the intermediate states.

\section{Spacetime semantics}

A brief summary of some of the lessons learnt from the foregoing
descriptions of spacetime is warranted.
The foregoing reviews
can be summarized in a few points.
\begin{itemize}
\item Space is a set of adjacencies between basic elements.
\item Time is counted by distinguishable changes of state.
\item Spacetime plays the role of a measuring system for transitions that can be represented
in two ways: 
\begin{itemize}
\item As coordinates (names).
\item As transitions (change).
\end{itemize}
\item A set of independent transition matrices form a basis for the
  allowed changes (i.e. vectors) in a space. The superposition of
  these generators of allowed change leads to a global transition
  matrix, or adjacency graph, which represents regular symmetries
  across patches of space.

\item Distance, extent and change are somehow equivalent things, all related to transitions.
Our common notions of distance are based on a very special case of homogeneous lattices.

\item Dimension of a space is a semantic issue: the naming of
  locations (coordinatization) and the association of a tuple of $D$
  base elements with a single location is a matter of convention,
  independent of the connectivity. For a graph, we can define a
  heuristic topological dimension at every point $D = \2 k_{\rm out}$,
  but tuple dimensionality of vertex sets leads to `strangely'
  connected coordinate spaces.

\item Spaces may be composed with inner ($\circ$) and outer ($\otimes$) products, for
joining and exfoliating respectively.

\item Different scales in space imply different levels of coarse graining of the base
elements. Only regular lattices have scale invariant multiplicative renormalizations.

\item Time is measured by change in the states available to an
  observer. Proper time encompasses all states and is therefore not
  independent from space. If we partition a system so that only part
  of it represents the clock for measuring time, complicated
  relativity issues ensue: change can occur and be observed in one
  part, while time stands still in another.

\item The structural compositions, or grammars of spacetimes are algorithmic
in their representations and complexity.

\item In several models there are two classes of object. Just as in physics one
has spacetime and the matter one puts into it, so in bi-graphs there are places
and agents with their connections that one puts into them. By contrast graphs
normally only have a single class of vertex to represent both. We shall return to this
in section \ref{promisemotion}.

\item Finally, from the appendix, we see that the total amount of
information in a spatial structure is the same no matter now many
dimensions we choose for the spanning representation.
\end{itemize}

Our received notion of dimension is closely tied to the idea of
tuple coordinates, in which one has a unique name for each component
value in the tuple in each direction.  Coordinate names are typically
numbered values when we deal with lattices, but they could also be
labels like Internet Protocol addresses, or proper names in an
arrangement of books such as a library, or a semantic network.

Of the foregoing cases, graph theory seems to stand alone in placing
all of the issues on the table plainly. It becomes a natural starting
point to explain the semantics of spacetime. Unlike completely regular
lattices, general graphs force us to see the difference between issues
we can take for granted in manifolds. The percolation of large scale
motion through graphs ought to be compatible with a three dimensional
Euclidean view of the world at large distances.

Category theory haunts the fringes of these descriptions, offering us
a raw functional viewpoint, but is has few insights into the
structural nature of spacetime that are not better described in terms
of specific transition systems (semi-groups and greater). Its main
purpose is to state the regularity of the structures and patterns in a
rigorous way, but it is a viewpoint that offers little in the way of
insight into semantics.

Bigraphs show a contrasting view of space, somewhat like a grammar,
with a relatively significant amount of semantic content.  What is
missing from bigraphs is the semantics of autonomy.  Is a bigraph an
autonomous device that can be a source of its own behaviour, or is it
merely a cog in a machine?  The signature typing suggests that the
former is the case, but the composition of bigraphs offers no notion
of whether one bigraph would be able to `allow' or `disallow' another
to be composed with it, or whether it would reject it. Thus the
concepts of autonomous decision-making and a local observer view seem
to be missing. This is a weakness, as there is no room for reasoning
or individual agency.

%%%%%%%%%%%%%%%%%%%%%%%%%%%%%%%%%%%%%%%%%%%%%%%%%%%%%%%%%%%%%%%%%%%%%%%%%%%%%%%%%%%%%%%

\section{Promise theory - autonomous local observer semantics}

Let us now turn to the main purpose
of this essay, which is to understand spacetime from the viewpoint of
autonomous agents. The principal motivation for this is technological,
but it is also fascinating to reflect on what the conclusions might
mean for the natural world\footnote{In particular, it seems to be
impossible to reproduce ideas of spacetime without the existence of
fundamental global (non-local) interactions that maintain symmetries.
These are often seen as an embarrassment in physics, where Einsteinian relativity
suggests that they cannot exist. However, it could be that Einsteinian relativity
is merely the effective face of those remaining observables that
manifest themselves on our observable large scale.}.

The need to represent artificial and conceptual spaces, especially in
technology, motivates the study of non-uniform observer
interpretations about the relationships between elements of space. Rather than
this being less fundamental than the foregoing models of natural
spacetime, such artificial constructions turn out to have more
features, and thus it requires more work to recreate aspects that are
taken for granted in physics. Such a study might offer some insight
into the assumptions we take for granted about physical spacetime, as
well as provide a framework for more fully understanding artificial
constructions within the information sciences.

Promise theory builds relational structures that look superficially
like graphs or categories.  However, a promise is not a function, nor
a transport channel, so it cannot be computed or traversed.  A promise
{\em does} however require the notion of communication, and hence
adjacency.  Promises allow us to discuss and label the nature of relationships
between elements in a space in diverse ways, as well as address concepts such as
{\em permission} to move from one to another. We shall see that
the notion of autonomy makes the idea of motion harder to understand
altogether.

\subsection{Basics}

One begins with the idea of autonomous agents that interact through
the promises they make to one
another\cite{burgessdsom2005,promisebook}\footnote{Physicists might
  feel confused by the notion of spacetime elements with arbitrary
  semantics, represented as a facsimile of human promises, but this is
  no more or less mysterious than different flavours of charge we
  usually attribute to point-like objects, which are good examples of
  promises made by (or equivalently the semantics of) elementary
  objects in the natural world, at least according to mainstream
  thinking. Physics has inconsistently turned up its nose at the
  admission of semantics in world descriptions. Ascribing intent to
  inanimate objects in generally frowned upon; thus one would not say
  `A promises to move towards B'.  However, inconsistently we do say
  `A attracts B' rather than `A pulls B towards it'.  Both have
  technically `intentional' semantics (merely de-personalizing the
  choice of words does not change that), so the arbitrary line of
  acceptability might just as well be removed.}.  \emph{Agent} is the
term used for the fundamental elements in Promise Theory; these will
be the elements of spacetime too.  Agents have no observable internal
structure, as with the foregoing models, however, we allow them to
exhibit agency or intent, either fundamentally or by proxy; this means
that an observer would interpret behaviour as being intentional, as
observers always impose semantics on what they observe.

An agent is {\em autonomous} in the sense that it controls its own
behaviour, it promises are made by itself, and it cannot be forced to
comply with an imposition; however, it can promise to comply
voluntarily with impositions from external sources.

Promise Theory\cite{burgessdsom2005,promisebook} is an elemental
framework for expressing intended behaviour graphically between
autonomous parts. This makes it analogous to atomic theory, as a
framework of elemental building blocks with which to recreate systems
of greater sophistication. Observability cannot be taken for granted
in promise theory; like statistical and quantum theories, it is a
theory of incomplete information.

We write a promise from $\rm Promiser$
to $\rm Promisee$, with body $b$ as follows:
\beq
{\rm Promiser} \promise{~b~} {\rm Promisee}.\nonumber
\eeq
and we denote an imposition by
\beq
{\rm Imposer} \imposition{b} {\rm Imposee}.\nonumber
\eeq
Promises come in two polarities, denoted with a $\pm$ signs, as below.
The $+$ sign gives assertion semantics:
\beq
x_1 &\promise{+b}& x_2 ~~ ({\rm I~will~give}~ b)
\eeq
while the $-$ sign gives projection semantics:
\beq
x_1 &\promise{-b}& x_2 ~~ ({\rm I~will~accept}~ b)\label{pmdef} 
\eeq
where $x_i$ denote autonomous agents.  A promise to give or provide a
behavior $b$ is denoted by a body $+b$; a promise to accept something
is denoted $-b$ (or sometimes $U(b)$, meaning use-$b$). Similarly, an
imposition on an agent to give something would have body $+b$, while
an imposition to accept something has a body $-b$.  In general, intent
is not transmitted from one agent to another unless it is both $+$
promised and accepted with a $-$. Such neutral bindings are the
exchange symmetry.

A promise model thus consists of a graph of vertices ({\em agents}), and
edges (either {\em promises} or {\em impositions}) used to communicate
intentions. Agents publish their intentions and agents in scope of
those promises may or may not choose to pay attention.  In that sense,
it forms a chemistry of intent \cite{burgess_search_2013}, with no
particular manifesto, other than to decompose systems into the set of
necessary and sufficient promises to model intended behavior.

A promise binding defines a voluntary constraint on agents.
The perceived strength of that binding is a value
judgement made by each individual agent in scope of the promises.  If
an agent offers $b_1$ and another agent accepts $b_2$, the possible
overlap $b_1 \intersect b_2$ is called the effective action of the
promise.

For example, $A$ promises $B$ `to give an apple'. This does not imply
that $B$ will accept the apple. $B$ might then promise $A$ to `accept
an apple'.  Now both are in a position to conclude that there is a
non-zero probability that an apple will be transferred from $A$ to $B$
at some time in the future. If the promise is to continuously transfer
apples, then the timing is less ambiguous.  Thus a promise binding is
the basis for interaction, and this must also include adjacency.

The constraints implied by the scope of observability for agents
complicates this.  Consider an exchange of promised behaviour, in
which one agent offers an amount $b_1$ of something, and the recipient
promises in return to accept an amount $b_2$ of the promised offer.

\beq
\pi_1: x_1 &\scopepromise{+b_1}{\sigma_1}& x_2\\
\pi_2: x_2 &\scopepromise{-b_2}{\sigma_2}& x_1
\eeq
Then any agent in scope $\sigma_1$ of promise $\pi_1$, will perceive
that the level of promised cooperation between $x_1$ and $x_2$ is likely $b_1$.
An agent in scope $\sigma_2$ of promise $\pi_2$, will perceive
that the level of promised cooperation between $x_1$ and $x_2$ is likely $b_2$.
Finally, an agent in scope $\sigma_1\intersection \sigma_2$ of both
promises $\pi_1$ and $\pi_2$, will perceive
that the level of promised cooperation between $x_1$ and $x_2$ is likely $b_1\intersection b_2$.
The relativity of observations can lead to peculiar behaviours, contrary to expectation.
Ultimately every agent makes decisions based on the information it has.

If a promise with body $S$ is provided subject to the provision of
a pre-requisite promise $\pi$, then the provision of the pre-requisite by
an assistant is acceptable if and only if the principal promiser also promises
to acquire the service $\pi$ from an assistant (promise labelled $-X$):
\beq
x_T \promise{+b(\pi)} x_1,
\left.\begin{array}{c}
x_1 \promise{S|b(\pi)} x_2 \\
x_1 \promise{-b(\pi)} x_2
\end{array}
\right\rbrace
\sim x_T \promise{+b(\pi)} x_1, x_1 \promise{S} x_2
\eeq

The relativity of observers will be key to understanding the semantics
of spacetimes.  Intent, being an interpretation offered by an
observer, brings with it a variety of anthropomorphisms, like trust and
level of belief which are equally important to science (witness the
Bayesian interpretation of statistical observations for instance).
This should not be considered a problem; it is merely the reflection
of a received interpretation by local observers. 
Similarly, promise theory, like statistics and quantum mechanics, is a
theory of incomplete information.
The promise formalism
is described in\cite{promisebook}.

\subsection{Agent names, identifiers and namespaces}

Agents may or may not be identifiable to one another. In order to be
identified by another agent, each agent has to promise to be visible
or identifiable. The property of observability might be an
interpretation of the property of reflecting light, or of sending
messages.  This is an example of what is meant by semantics.

\begin{enumerate}
\item Agents may or may not promise their name or identity. If no name
  is promised, then only nearest (adjacent) neighbours can attribute
  information to them, by virtue the agent's own labelling.

\item Observers may promise to accept, and hence associate a unique identity
  with each agent they can observe directly or indirectly.
That identity is local to the observer.
\end{enumerate}
Promise theory makes the concept of a namespace unambiguously something that
belongs to an observer alone.

\begin{figure}[ht]
\begin{center}
\includegraphics[width=6.5cm]{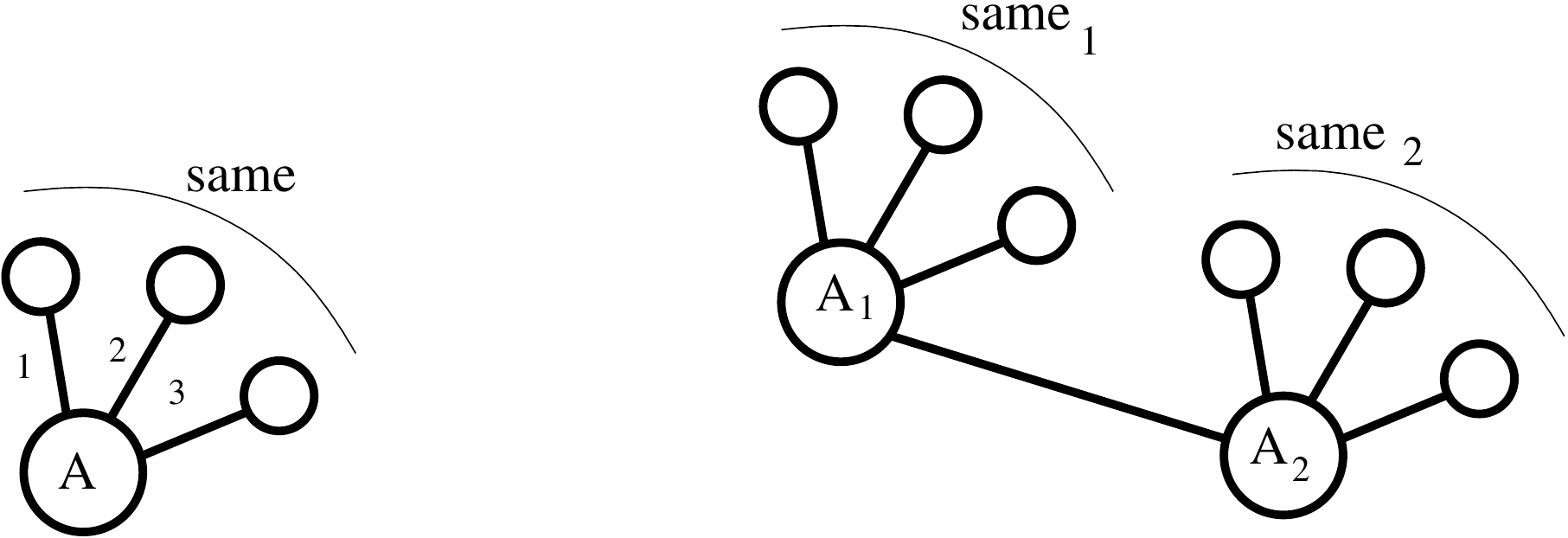}
\caption{Agents may be distinguished by having distinguishable
promises, by observer labelling, or by knowing the path
between them and the observer (if trusted).\label{agentunique}}
\end{center}
\end{figure}

Assume that an agent $x_0$ can distinguish its nearest neighbours
$x_N$ by labelling its adjacencies internally $N = 1 \ldots n$ (see
fig.  \ref{agentunique}). If neighbouring agents are completely
interchangeable, with respect to what can be observed by $x_0$, then
$x_0$ can label the adjacency channels but cannot know if agents
exchange places along those channels.
If agents are not directly adjacent, but make similar promises in
different locations, they are potentially distinguishable by
information about the path along which information passes. This in
turn assumes that information about the path is passed along the path
by neighbouring agents and that there is a chain of trust.  Trust in
information is usually taken for granted in natural science (with
exceptions in quantum mechanics), but it is by no means assured in a
techno-social network\footnote{A difference here with the foregoing models of spacetime, is that the
naming of agents cannot be assumed constant or even unique over an
extended region.  We might wish to say something like: every element
in a set must have a unique name.  In
Promise Theory however this is an imposition or {\em obligation}, and
cannot be assumed.}. 

As autonomous elements, agents might not have agreed to coordinate
with every other to ensure unique distinguishable identities (a
non-local property).  Even if they have, these might not be a promises
they are able to keep.  We must renormalize our notions of what can be
taken for granted.

\subsection{Reconstructing adjacency, local and non-local space}

Promising implies the ability to communicate messages.  We assume that
adjacency and the ability to exchange information are synonymous.
After all, space is merely a reflection of the ability to observe and
transport information\footnote{See also the theorem of consistent
  knowledge propagation\cite{promisebook}.}. This might seem peculiar
from the viewpoint of absolute spacetime, but distance is just one of
many possible associations between agents that change with perception,
circumstance and individual capability. We must begin by rekindling the
notion of distance from the more primitive concept of adjacency.

\begin{assumption}[Promised adjacency]
  A promise of adjacency can be made by any agent to any other,
  allowing them the promiser to offer information to promisee.
\end{assumption}
This is the same assumption by which we build topologies from sets.
To complete any kind of information exchange, we need a match an
imposition (+) or a promise to send with a promise to use (-).

Not all promises are about spacetime structure: being blue or having
positive electric charge does not qualify as an spatial promise, for
instance, as it doesn't tell us about neighbouring points. Many
promises offer membership in some group of similar agents, thus could
be non-local properties (like charge and mass), but do not explain any
relativity or connectivity between them. 

Being close to, being able to see or hear a neighbour, being able to
point to, or even being attracted to, are examples of adjacency, as
they name a specific target.
Thus an adjacency promise is more than mere
continuity\footnote{Database normalization rules (first normal form)\cite{date1,burgess04analytical}
  are promises of regularity of form (internal structure on spacetime
  elements that are tables) but not just any promise tells us about
  how to traverse from one place to another.}.
\begin{definition}[Adjacency promise]
  An adjacency promise is a promise that relates an agent $x_i$ to
  another specific unique agent $x_j$ ($i\not= j$), and may give
  a local interpretation to a relative orientation i.e. direction between the two.
\end{definition}

Agents making an adjacency promises to more than one agent cannot
simply be exchanged for one another without changing the linkage.
Thus adjacency is a form of {\em order} (see section \ref{lro}).
Let us now examine how many primitive promises are needed to 
bind adjacent points in a spacetime.

\begin{definition}[Adjacency promise binding]
A bundle of bilateral promises, analogous to a contract, binds
an agent $x_n$ with another agent $x_{n+1}$, promising a channel between them. 

$x_n$ promises that $x_{n+1}$ may transmit (+) directed influence to it.
$x_{n+1}$ promises to use (-) $x_n$'s offer.
$x_{n+1}$ promises that $x_n$ may transmit (+) directed influence to it. 
$x_n$ promises to use $x_{n+1}$'s offer (-).
\beq
x_n &\promise{+{\rm accept\;msg}}& x_{n+1}\\
x_{n+1} &\promise{+{\rm accept\;msg}}& x_n\\
x_n &\promise{-{\rm accept\;msg}}& x_{n+1}\\
x_{n+1} &\promise{-{\rm accept\;msg}}& x_n
\eeq
\end{definition}
Notice that Newton's third law is not automatically guaranteed in
Promise Theory: that which is given is not necessarily received; hence
conservation of promised properties is not guaranteed, it must be
documented with explicit promises just like charge. In this respect,
familiar dynamical concepts of the continuum are puzzling from a discrete
information perspective. Neither mass nor velocity are easy to incorporate.

By the duality rules in promise theory\cite{promisebook}, we can
interpret the acceptance of an accept-message promise like a promise
to send messages (though the anthropomorphism makes that
sound stronger than necessary for a natural spacetime). 
Accepting a message is the same as accepting the presence of the agent.
We might say that the agent promises
to avail itself of the opportunity to send messages directly to its
neighbour, cementing this relationship. The symmetry between $A$ and
$B$ makes the adjacency relationship into an undirected graph.

The scope of any promise defaults to the two agents involved in the
adjacency relationship, but it could also extend beyond them, allowing
others to observe the relative positioning of points, allowing in turn
the coordination of distributed behaviours. In semantic structures
like swarms (flocks of birds, or shoals of fish), nearest neighbour
observations are sufficient to maintain the coherence of the emergent
cohesion, suggesting that a spacetime formed from autonomous agents
with nearest neighbour interactions could be sufficiently stable
without long range interactions.

A promise network may thus be partitioned into two parts: the graph of
promise bindings referring to adjacency and communication between the
agents, and all other promises that use the former to expand their
scope. If we think of the classical concept of matter living within
spacetime, then we reduce this to bundles of `scalar' promises (see
below) that associate with a location by virtue of being anchored to
an agent that has position as a result it its adjacency promises. Thus
matter is simply spacetime with special properties, explaining how
matter (quite literally) occupies space.

\subsection{Continuity, direction, and bases}

Nearest neighbour adjacency is a difficult enough concept to understand
in an autonomous framework; continuity across regions is even harder.
Why, for example, would agents align themselves with a reduced symmetry
such as a lattice?

\begin{definition}[Spatial continuity]
  Continuity is understood to mean that, if a direction exists at a
  certain location $x_i$, it continues to exist in the local neighbourhood
  around it.
\end{definition}

Suppose we wanted a concept of travelling North. How can this be
understood from an agent perspective? The concept of North-ness is
non-local, and uniform over a wide region. In order to image
continuing in the same direction, we also need to know about the
continuity of directionality.  Direction too is thus a non-local
concept. When we speak of direction, we mean something that goes
beyond who are the agents closest to us.
Any agent can promise to bind to a certain number of other neighbours,
calling its adjacencies to them with the same name (say North, South,
etc), but why would the next agent continue this behaviour?
How does each agent calibrate these in a standard way?

As established for graphs, membership in a basis set is a semantic
convention used by observers. It cannot be imposed. A point in a space
need not promise its role in a coordinate basis, because that
information is only meaningful to an observer, and could simply be
ignored by the observer.

An agent can promise to be adjacent to another agent, but to propose
its own classification as a member of some basis would be to impose
information onto others from a different viewpoint.  By autonomy, each
agent is free to classify another agent as a member of an independent
set within a matroid that spans the world it can observe.

The dimensionality of spacetime, perceived by any observer belongs to 
the rank of the matroid it chooses to apply to the agents
it can observe. The consequence of this is that spacetime can have any
dimension that is compatible with the adjacencies of the observer.
Indeed, the notion of dimensionality experienced by the elements of a
promised space is different for every agent, at every point.  The
observer with $n$ outgoing adjacencies may regard each independent
adjacency as a potential basis vector or direction.

\begin{assumption}[Matroids are observables]
Every autonomous agent decides its own set of independent sets to span
a space. Hence direction is a local observer view, as noted for graph theory.
\end{assumption}

Consider an ordered sequence of agents $x_i$ that are mutually adjacent.
An agent $x_i$ recognizes a direction $\mu$ if it promises adjacency ($+{\rm adj}_\mu$)
along a locally understood direction $\mu$ to a subsequent neighbour $x_{i+1}$, {\em and}
it promises to accept adjacency ($-{\rm adj}_\mu$) with a previous neighbour $x_{i-1}$:
\beq
x_i &\promise{+{\rm adj}_\mu}& x_{i+1}\\
x_i &\promise{-{\rm adj}_\mu}& x_{i-1}\\
x_i &\promise{+{\rm adj}_\mu}& x_{i-1}\\
x_i &\promise{-{\rm adj}_\mu}& x_{i+1}
\eeq
for all $x_i$.
Or in shorthand:
\beq
x_i &\promise{\pm{\rm adj}_\mu}& x_{i\pm 1}, \forall x_i.
\eeq
We shall need to say what happens at edges where we run out
of $x_i$ (see section \ref{promiseboundary}).
These promises are local but require long range 
homogeneity between the agents, i.e. the condition $\forall x_i$ is
a non-local constraint. It is equivalent to promises by every
agent to conform to these promises:
\beq
x_i &\promise{C({\rm adj}_\mu)}& x_{j}, \forall i,j
\eeq
The issue is not the numbering $i = 1\ldots N$ of the agents, as this
may be freely redefined. Any local agent will bind to another
exclusively and the ordering can easily emerge by self-organization,
however, the notion that all of the agents or spacetime points would
coordinate with long range order in keeping these promises does beg an
explanation.

A multi-dimensional interpretation of the different spanning sets,
does not really add further difficulties, but
emphasizes further the non-local cooperation in terms of promise
homogeneity. If we choose a 3 dimensional basis with coordinate names $(x,y,z)$,
\beq
(x_i, y_j, z_k) \promise{\pm{\rm adj}_\mu} (x_{i\pm 1}, y_j, z_k)\\
(x_i, y_j, z_k) \promise{\pm{\rm adj}_\mu} (x_{i}, y_{j\pm 1}, z_k)\\
(x_i, y_j, z_k) \promise{\pm{\rm adj}_\mu} (x_i, y_j, z_{k\pm 1})
\eeq
The directional names belong to the local agent's coordinate basis,
and the non-local cooperation is assumed homogeneous. This leads
to the possibility of a collection or misaligned, non-oriented agents
self-organizing into a crystal lattice. In this way, a space can
acquire {\rm long-range order} with only local, autonomous promises, provided
they are homogeneous over a sufficient region.

\subsection{Symmetry, short and long range order}\label{lro}

Adjacency (vector) promise bindings are exactly analogous to chemical
bonds, with the addition of semantic types.  A graph of autonomous
agents with only without adjacency bindings has a state of maximal
symmetry; we may call it `gaseous' or `disordered', by analogy. If
agents promise adjacency promises in a uniform and homogeneous way,
one may speak of long-range order as in a `solid'. It is natural to
describe such a space with a latttice coordinate system (e.g.
Cartesian), like that of a crystal. For something we take for granted,
this is quite non-trivial.

If we are considering spacetimes of technological origin, e.g. computing
infrastructure, then we readily see a mixture of these two states
in everyday life. Mobile phones, pads and computers migrate without
fixed adjacency on a background of more permanent fixtures: servers, 
disks, network switches, and other `boxes'. Thus, we should be prepared
to view spacetimes of mixed phase within a single picture.

The self-organization of autonomous agents is quite analogous to a phase
transition by local interaction in matter. Indeed, it should be clear
by now that in a discrete spacetime, elements of space must behave and
interact according to the same essential mechanisms as elements of matter,
causing one to wonder if there should be a distinction at all.

A spacetime with a graph structure thus exists in phases analogous to
gaseous (disordered) with short-lived adjacencies, or crystalline with
long-lived adjacencies (ordered). The disordered state is said to be a
symmetrical state, and a crystalline order is a broken or reduced
symmetry, which is usually related to breakdown to arbitrary symmetry
operations, by selection of a subset that appears to bring long range
order. A lattice exhibits long range order, while a fluid, replete
with vortices and flows, might exhibit only short range (nearest
neighbour) order, for instance.

For a physical world, the remarkable point here is this: if regular
spacetime structure requires long range order with underlying global
symmetry, what is the origin of such order? Why should all points make the
same promises? Perhaps they don't.

These notions are usually ready-built into most descriptions of
spacetime, and as such taken for granted, but in a graphical view of
the world we are forced to confront how this arises, as we cannot
separate symmetry breaking from the structure of spacetime itself.
How one breaks the autonomous symmetry of spacetime is thus the
ultimate question for understanding its structure.
The Internet does not possess a natural long-range order, for example,
which is a hindrance to the creation of a global network addressing
scheme.

Long range order can exist for many kinds of promises, not just
adjacency promises, but only adjacency promises lead to connectivity.
Similarly, there are many ways of breaking a symmetry. One is to build
a function on top of spacetime with a monotonic gradient which serves
as the generator of a notion of consistent direction.  This is how
chemotaxis works in cellular biology for example. If one takes cells
to be autonomous agents and considers bio-space, one could form a
directional basis for self-organizing adjacency provided each cell
could homogeneously promise to measure the functional gradient, as in
foetal morphology.

As much as one tries to build the concepts of regular space from a local
observer perspective, one never quite escapes the notion of non-local
symmetries at work in creating long range structure. 
Like it or not, in a discrete spacetime, we are forced to confront the
idea of phases and the possibility of transitions between them.
In promise theory,
this amounts to understanding why more than a single agent with no
direct adjacency should make the same promises. In this respect, the model
is like that of a cellular automation with an undetermined topology.

\subsection{Material (scalar) properties as singularities and cliques}\label{properties}

Consider now how to represent point-like properties in agent space.
There are two possible local representations of the assertion 
that agent $A_1$ is blue. In the first case:
\beq
A_1 &\promise{+{\rm I~am~Blue}}& A_2\\
A_2 &\promise{-{\rm I~am~Blue}}& A_1
\eeq
$A_1$ merely asserts a property of itself to an
observer $A_2$.  
The observer $A_2$ can take or leave the promise of blueness, thus it must
promise to use or accept the assertion, (it
might be colour blind, for instance, and hence effectively not promise to use the
information). Notice here that the conceptual world of blueness lives entirely within the body
of the promise, and does not affect the type of objects between
which promises are made. In this representation,  concepts thus live in a parallel world that 
does not intersect with the space formed by the agents $A_i$ themselves.

Any agent might make the same promise, and a priori there is
be no objective calibrated standard for blueness. Different agents
might interpret this promise differently too.

For the second representation, consider the merger of the physical and conceptual 
worlds, by introducing special agents for material properties concerned:
\beq
 A_{\rm Blue} &\promise{+{\rm Blue}}& A_1\\
A_1 &\promise{-{\rm Blue}}& A_{\rm Blue}
\eeq
A special kind of agent, whose function it is to label things as blue, now
promises this quality as if providing the property as a single point of service.
Association with this source of blueness is what gives $A_1$ the
property. To be blue, all $A_1$ has to do is promise to promise to use the
service.  

\begin{figure}[ht]
\begin{center}
\includegraphics[width=8.5cm]{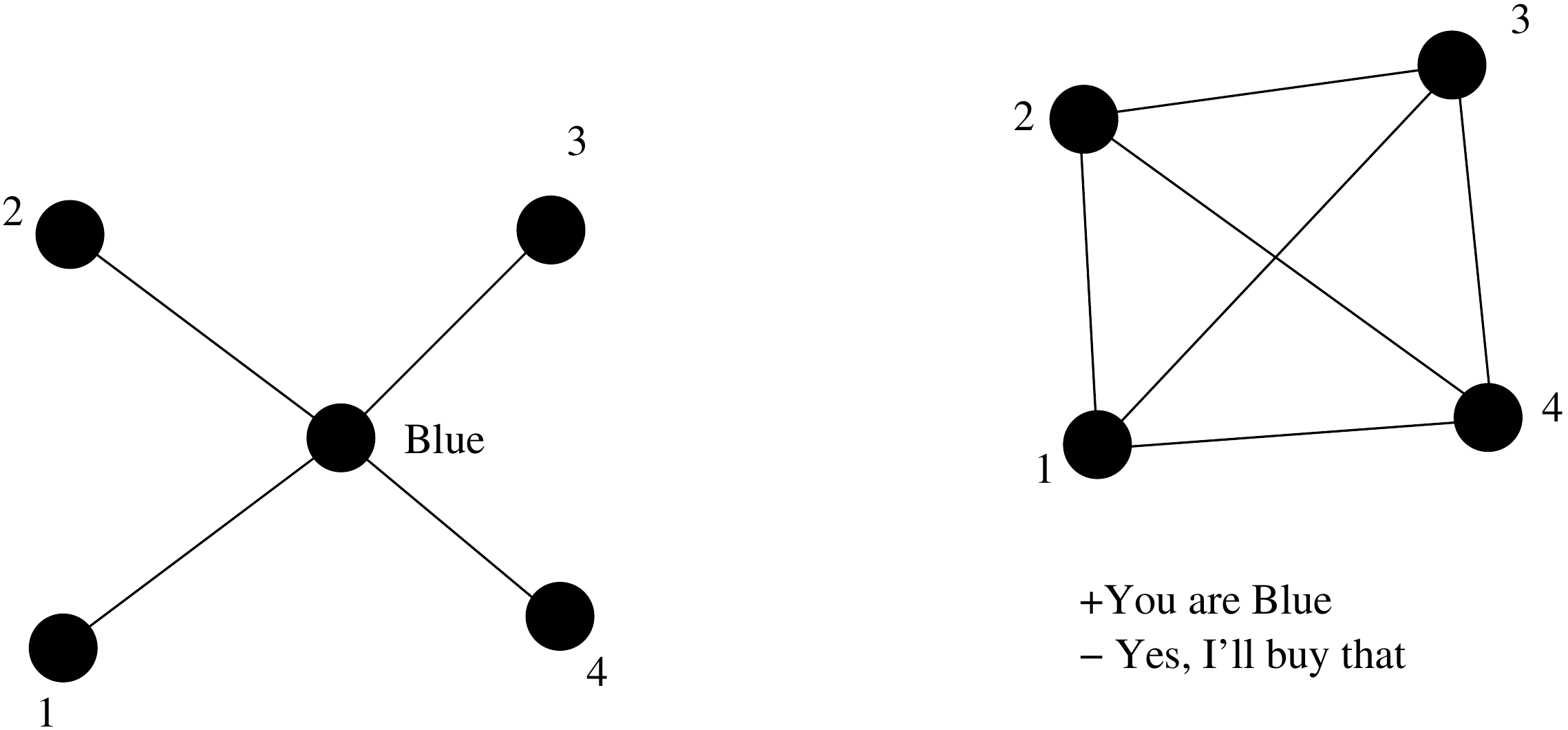}
\caption{Global symmetries - calibrating a property, either by equilibrating with
a single source (singularity), or everyone individually (clique).\label{calibration}}
\end{center}
\end{figure}

Multiple agents may now in two ways: either by coordinating their
definitions individually in a peer-to-peer clique,
\beq
\{A_i \} \promise{\pm{\rm blue}, C({\rm blue})} \{A_j \}, \forall i,j
\eeq 
or by using a
definitive calibration source, 
\beq
 A_{\rm Blue} &\promise{+{\rm Blue}}& \{ A_i\}\\
\{ A_i\} &\promise{-{\rm Blue}}& A_{\rm Blue}
\eeq
which then acts as a kind of hub for
the property of blueness (see fig.  \ref{calibration}).

\begin{figure}[ht]
\begin{center}
\includegraphics[width=8.5cm]{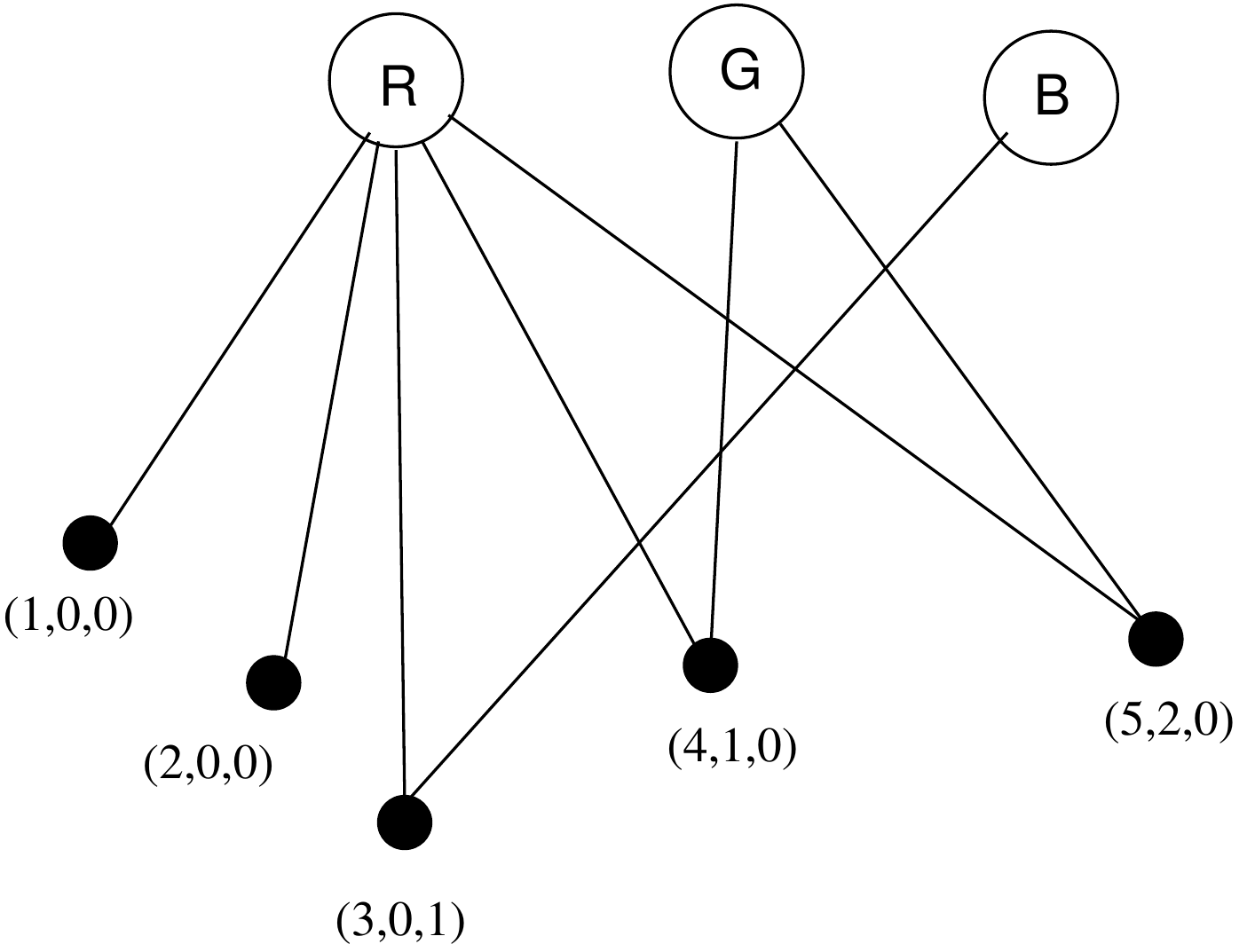}
\caption{Matroid basis for global properties with three property hubs
adding three components to the coordinate tuples.\label{coordkm}}
\end{center}
\end{figure}

Notice that, since each property has at least one single vertex
associated uniquely with it (the source for that property), the set of
links emanating from it is automatically an {\em independent set}.  Each property or
type of promise that refers to an intrinsic quality is in fact a {\em
  basis} vector, belonging to a matroid (see fig. \ref{coordkm}). What is interesting is that,
unlike a vector space, this vector ends at a singularity, like a
charge radiating lines of force.

The scalar promises bear an obvious resemblance to the use of `tags'
or `keywords' to label information documents and inventory items in
databases. They act as orthogonal dimensions to the matroid that spans
agents in the ordered phase of a spacetime.

\subsection{Spacetime (vector) promises and quasi-transitivity}\label{3types}

Scalar promises imbue elements of spacetime with intrinsic properties;
vector promises describe cumulative relationships between
them\footnote{In the language of chemistry, scalar promises are like
  atomic isotopes, while vectors are their interatomic bonds, forming molecules
and crystals.}.  Vector
promises are, in principle, interpretable as one of the following
cases:
\begin{itemize}
\item $A_1$ can influence $A_2$ (causation)
\item $A_1$ is connected to $A_2$ (topology)
\item $A_1$ is part of $A_2$ (containment)
\end{itemize}
e.g.
\beq
A_1 &\promise{\rm Causes}& A_2\\
A_1 &\promise{\rm Precedes/Follows}& A_2\\
A_1 &\promise{\rm Affects}& A_2\\
A_1 &\promise{\rm Is~a~special~case~of}& A_2\\
A_1 &\promise{\rm Generalizes}& A_2
\eeq

Adjacency too may be considered quasi-transitive, for while $A$ next
to $B$ next to $C$ does not imply that $A$ is next to $C$, if we
reinterpret adjacency only slightly as connectivity, we can make it
true, e.g. $\pi_L$: $A$ is to the left of $B$ is to the left of $C$.
\beq
\pi_\adj \rightarrow \left\{  
\begin{array}{l}
\pi_{\rm connected}\\
\pi_{L,R}\\
\pi_{N,S,E,W}\\
\pi_{\pm\mu}
\end{array}
\right.
\eeq

Clearly if $A$ causes $B$ and $B$ causes $C$, there is a sense in
which one might (at least in some circumstances) interpret that $A$
causes $C$, hence there is a kind of transitivity.  Mathematically,
there is exists a generator of a translational symmetry which can be
repeated more than once to bring about a sense of continuity of
motion.  In functional terms, these relational maps have arity 2 and
span a single coordinate direction. Such relationships generate a
spacetime, analogous to a vector space.

Next there are container models, analogous to bi-graphical positioning:
\beq
A_1 &\promise{\rm Is~contained~by}& A_2\\
A_1 &\promise{\rm Is~found~within}& A_2\\
A_1 &\promise{\rm Is~part~of}& A_2\\
A_1 &\promise{\rm Is~eaten~by}& A_2
\eeq
These generate forest graph relations. Thus we have a way of incorporating
both types of spatial semantics in the promise framework, and
we can link translation and containment through their quasi-transitive nature.
These promise types provide a notion of {\em spatial continuity}.

Promises that cannot be made into a succession of symmetrical
translations belong to the singular properties discussed in section
\ref{properties}.  They represent promises about self, rather than
about relationships to others (in functional terms they have arity 1).
Such expressions may be formulated and interpreted in two ways,
depending on who or what are the recipients of the relationship: as
promises, or as general associations between topics in a topic map.

\subsection{Fields and potentials}

The classical continuum notion of a field or potential is now seen to be a
functional representation of the split between a common underlay of vector
adjacency promises (spacetime), and a set of local material promises (the potential)
at each spacetime element. 

\begin{figure}[ht]
\begin{center}
\includegraphics[width=12cm]{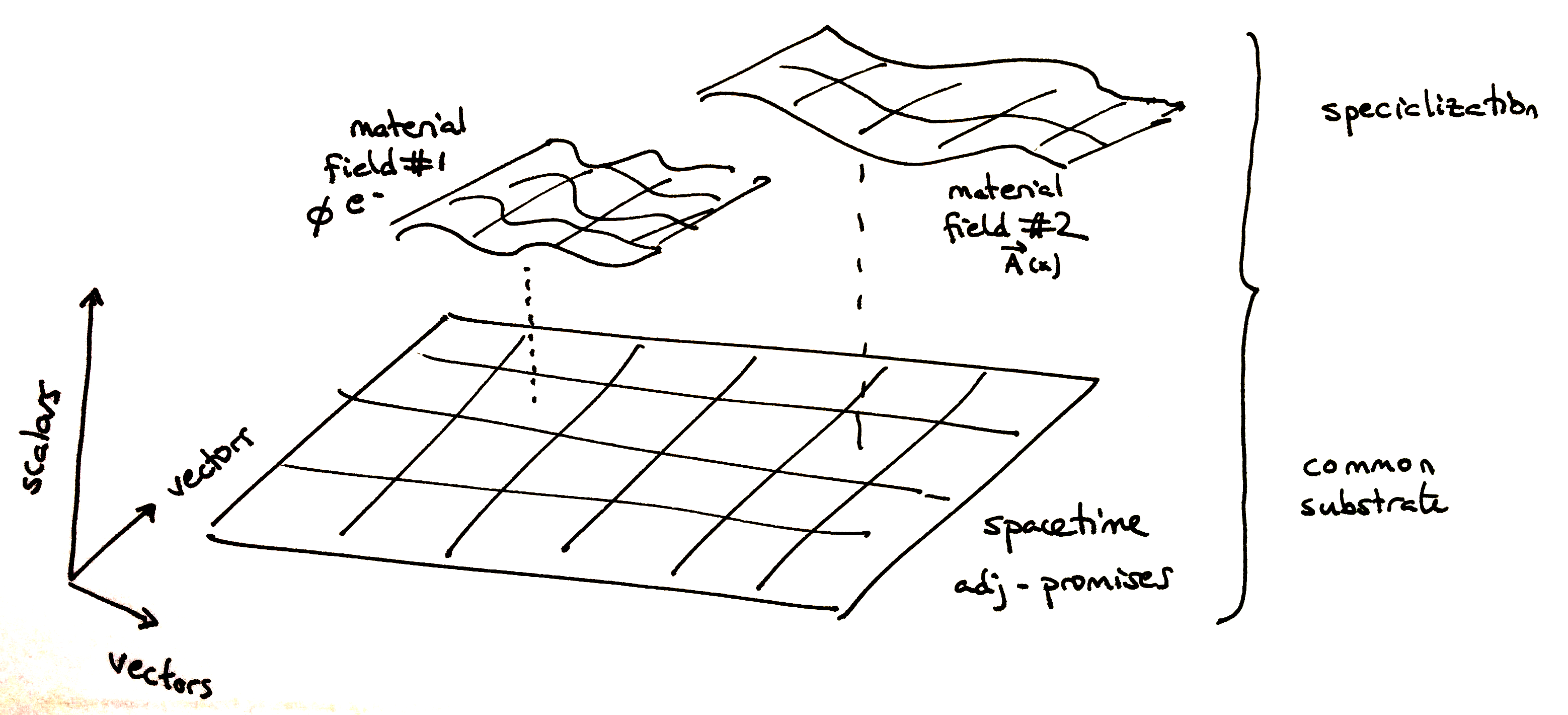}
\caption{In a continuum approximation, the idea of a potential field
  $\phi(x)$ (as in electrodynamics) whose value $\phi$ is specified at each location on top
  of a spacetime $x$, appears as a separation of
  scalar promises on top of the common denominator of adjacency
  promises.\label{qfield}}
\end{center}
\end{figure}

This construction is very similar to Schwinger's formulation of
quantum fields (see figure \ref{qfield}), except that we have
autonomous agents for spacetime elements, and the promise of field or
material properties, at some location, belongs to each spacetime
agent.  The promise to measure `particle events' might be kept in a
particular location (to be assessed by an observer). This shows how
promises capture the field idea in a discrete way, and emphasize the
symbolic semantics of particles alongside dynamical
properties\cite{schwingerfield1,schwingerfield2}.

\subsection{Boundaries and holes}\label{promiseboundary}

The interruption of vector continuity is what we mean by a boundary. Boundaries
explicitly break symmetries and seed the formation of structure by
anchoring symmetry generators to some fixed points.

The simplest notion of a boundary is the absence of a promise of adjacency.
For autonomous agents, this can have two possible directions.
Suppose one names agents in some sequence; 
at position $n$, 
\beq
x_n \promise{\emptyset} x_{n+1}
\eeq 
An agent might be ready to accept messages from a neighbour, but there is no neighbour
to quench it, 
\beq
x_{n+1} &\promise{+{\rm accept\;presence}}& \emptyset\\
\eeq
or no promise given to keep such a promise.
\beq
x_{n+1} &\promise{+{\rm accept\;presence}}& x_n\\
x_n &\promise{-{\emptyset}}& x_{n+1}
\eeq
We can summarize different cases:

\begin{definition}[Continuity boundary]
If an agent $x_i$ does not promise $+ {\rm adj_\mu}$ to any other agent
may be said to be part of a $\mu$-transmission boundary.
\end{definition}

\begin{definition}[Observation boundary/event horizon]
If an agent $x_i$ does not promise $- {\rm adj_\mu}$ to any other agent
may be said to be part of a $\mu$-observation boundary, or event horizon.
\end{definition}

\begin{figure}[ht]
\begin{center}
\includegraphics[width=4.5cm]{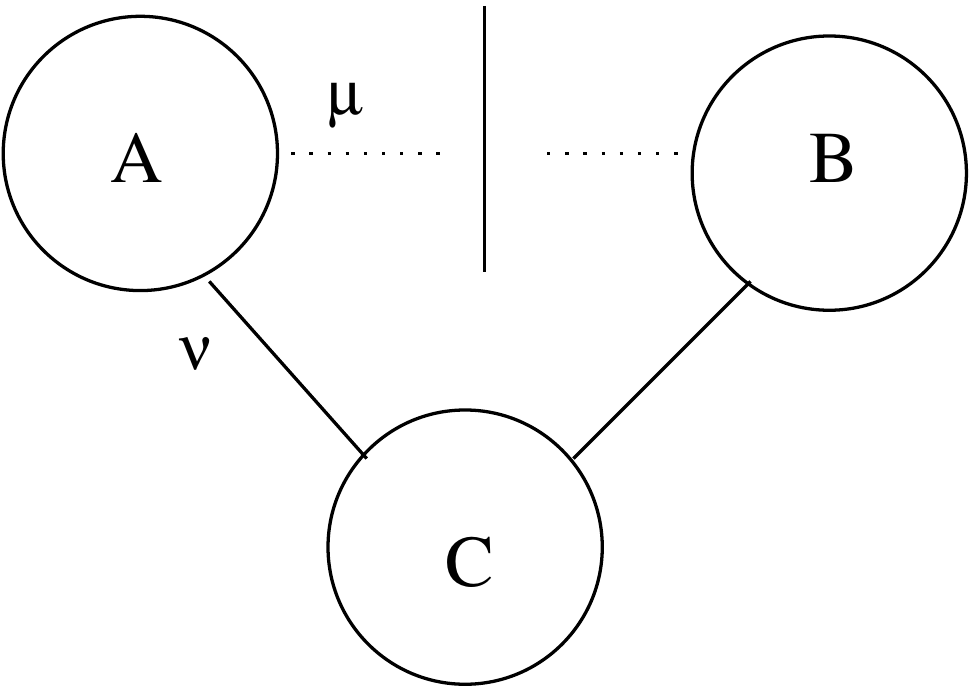}
\caption{A boundary or a partition in a graph may be the absence of an adjacency.\label{boundary}}
\end{center}
\end{figure}

Boundaries can thus be semi-permeable membranes. These are quite
common in biology.  Boundaries can be localized or extended (see fig.
\ref{boundary}). Their perceived extent depends on observer semantics,
or coordinatization. The absence of an adjacency along a direction
labelled $\mu$ between agents $A$ and $B$ may be called a
$\mu$-boundary, even though there is still a path from $A$ to $B$ via
$C$, in a direction $\nu$.

Boundaries are usually considered to be the discontinuation of a certain 
degree of freedom or direction. The default state in a network of autonomous
agents is boundary.

\begin{definition}[Edge boundary]
If an agent $x_i$ does promises $\pm {\rm adj_\mu}$ to an agent that does not exist,
we may say that $x_i$ belongs to the $\mu$ edge of the space.
\end{definition}

\begin{definition}[Material boundary]
The edge of a vector region consisting of those agents that uniformly promises material property $X$.
\end{definition}

Thus we attribute a variety of semantics to boundaries: 
\begin{itemize}
\item The absence of an agent adjacency (an edge of space or a crystal vacancy).
\item An agent that selectively refuses a promise to one or more agents (a semantic barrier, e.g. a firewall, passport control, etc).
\end{itemize}
For example:
\begin{itemize}
\item The edge of space itself
\item The edge of a property e.g. blue, table
\item The edge of an organization e.g. firewall `DMZ', freemason
\end{itemize}

\subsection{Containment within regions and boundaries}\label{containment}

How shall we represent the idea of one object being inside another in a world of
autonomous agents? Agents are atomic, and one atom cannot be inside another.
The clue to this lies is viewing containment as a bulk material property.
We start by defining membership in regions or cliques.

\begin{figure}[ht]
\begin{center}
\includegraphics[width=8.5cm]{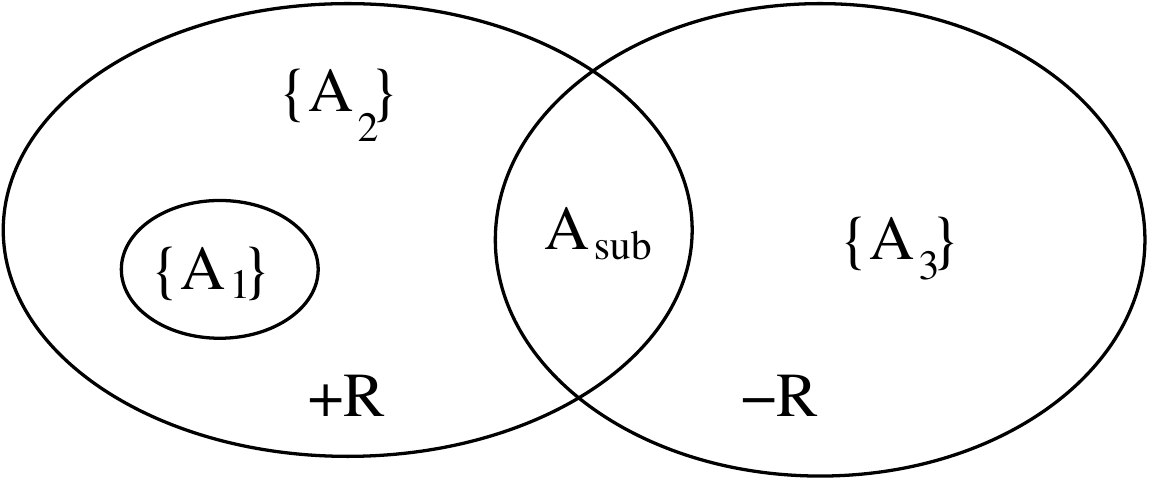}
\caption{Motion of the first kind: extrinsic motion in an untyped spacetime\label{contain}}
\end{center}
\end{figure}

A compound agent, denoted $\{ A_i \}$, with role $R$ is the set of
agents that mutually promise to belong to set $R$.  Membership in a
group, role of property follows the discussion in section
\ref{properties}.
Containment and overlap may now be defined with reference to fig. \ref{contain}

\begin{definition}[Containment promise]
Compound agent $\{ A_1\}$ is $R$-inside (or $R$-contained by) compound agent $\{ A_2\}$ iff 
\beq
\{ A_1\} \promise{-R} \{ A_2\}\\
 \{ A_2\} \promise{+R} \{ A_1\}
\eeq
\end{definition}
The promise of $+R$ represents membership in region $R$, which defines a compound or `super-agent',
that is a coarser grain of space than the component agents.

\begin{definition}[Spatial overlap promise]
Compound agent $\{ A_1\}$  $R$-overlaps with compound agent $\{ A_2\}$ iff 
\beq
\{ A_{\rm sub}\} \subseteq \{ A_3\}  \promise{-R} \{ A_2\}\\
 \{ A_2\} \promise{+R} \{ A_{\rm sub}\}
\eeq
\end{definition}

When we say one region is inside another, it is sometimes convenient
to describe the boundary of the region rather than the region's name.
There is a straightforward relationship between these, given that a
region is simply a material boundary in which there exist adjacent
agents, some of which promise $-R$ (inside the region's boundary) and some of
which do not (outside the region).

The same criterion of membership applies. Using the container as a label for the 
type membership is the complement of using the type itself, since defining an edge requires
us to specify edge with respect to which property.

Compound agents inside one another or adjacent to one another can all be represented
as regular tuple coordinates, by defining matroids as indicated here. An incidental example
of this may be seen in the construction of data network addresses, like
IP addresses, VLAN numbers, etc\cite{DBLP:journals/corr/BorrilBCD14}.

\subsection{Local, global, and proper time (What counts as a clock?)}

To the eye of an all-seeing `godlike' observer, time is simply a series of spacelike
hypersurfaces. Any change of the properties of one of these represents a new
time.  We refer to this total picture as {\em proper time}. Each
moment of proper time represents a single version of all the promised
attributes of the universe.  In an artificial system where one has the
ability to observe everything `instantaneously', one could always
define a proper time accordingly. Each universe is its own clock.

Locally, agents do not have access to this complete instantaneous
overview.  They have information horizons, and information travels at
a finite speed. They can only observe what can be transmitted to them,
and what they have promised to receive, up to the limitations of their
own faculties. For local time, agents can choose any convenient set of
states within themselves, to represent their clock.  This is a
convention that might vary to place to place\footnote{Many of our
  assumptions about time come from the doctrine that time exists
  independently of space. In Maxwell-Einstein relativity, the
  constancy of the speed of light in all frames warns of the falsity
  of this assumption, but one never quite lets go of time as an
  independent, yet coupled quantity. On a macroscopic level, it is
  possible to arrange that illusion by partitioning information.
  However, at a microscopic level, one cannot measure or experience
  anything at all without causing change. This information, causation,
  spatial configuration, and time are all tightly interwoven.}.

To keep matters simple, the clocks by which agents measure their own
passage of time are best kept sufficiently local as to not be affected
by the need to cooperate with neighbouring agents, and the finite
speed of communications that entails. In practice, we assume that clocks are
internal and of zero size\footnote{In the special theory of
  relativity, the locality of clocks is glossed over.  What is, in the
  twins paradox, the clocks of both twins were always two halves of a
  single clock, or it was so large as to span the distance travelled
  by one of the twins?}. Another possibility is to build a co-clock
with a neighbour, in the manner of a dialogue or handshake.

A local clock is one whose state counters are all contained internally
by an agent $A$. Thus we exclude adjacency bindings from the
measurement of local time. This is not without controversy since time
for a super-agent necessarily includes the bonds between atomic
agents.  Moreover, as we'll see, the ability to measure derivatives
requires agents to be able to remember at least two or three points,
which could require the cooperation of other agents. Constructing a
shared clock goes beyond the current paper.

Agents must be able to measure space and time in order to cooperate
and evaluate one anothers' keeping of promises.  In a discrete
spacetime, this is a lot more difficult to understand than if space
and time were separate, since both velocity and acceleration are
themselves only understandable from the viewpoint of a continuous
independent spacetime.

\subsubsection{Concurrency, simultaneity, and timelines}

A clock tick is an observable change of state, and a timeline is a
sequential ordering of such changes, observed by some agent. Every
agent is free to form its own timeline, within its own world, because
intentions arise from each agent individually (the intent to see, and
the intent to interpret).  Any agent that accepts intentions from
another can promise to use its intended order, or not. That is the
first level of possible distortion, as it is essentially a `voluntary'
or autonomous act.

If multiple agents communicate, then each neighbour interaction can
add distortions of its own. Moreover, when there are multiple routes
for information to take through the web of adjacencies, different
routes might lead to different distortions. In physics we assume that
information is passed with integrity, only delayed predictably in
time. This is mainly because there is no way to verify if it isn't.
However, this is too strong an assumption for general transmission of
information in intentional structures.

An agent can only tell if one event is intended to follow another if it
is promised the causal sequence in the form of a dependency. This is what
conditional promises do:
\beq
A \promise{b_1|b_2} A'
\eeq
A dependency promise of this kind offers non-local (non-Markov) information.
It starts to build a journal of events, or a {\em timeline}.
Here the agent $A$ promises that the intention described in body $b_1$
must follow the intended outcome of $b_2$. If, on the other hand, $A'$
receives promise or their outcomes in a certain order, without documented
dependency:
\beq
A \promise{b_1} A'\\
A \promise{b_2} A'
\eeq
then $A'$ cannot be certain of their intended order.
Regardless of what order the messages arrive in, without the
conditional promise dependency, $A'$ must view $b_1$ and $b_2$ to be
concurrent (both as outcome events, and promised processes).
This is an important definition of concurrency.

The question of a whether one event happens before another is not
really a meaningful one in any spacetime, because until an event has
been observed there is no causal connection between one part of space
and another; the real mystery is why changes or events happen at all.
If space is gaseous and time is space, it implies that time must also
be non-ordered.  However, if there is a channel for communication,
then there it is possible to place some limits on the order of
transitions, by sharing information on trust\footnote{This is the
  essence of Lamport clocks or vector clocks in the literature of distributed computing.}.

Agents observer limited views of the world around them. The horizon of
the observability defines their clocks, and hence their notion of
time. Often observers measure time based on internal states, which are
only available to them, so every agent must experience time
individually.  This is the essence of relativity. It is well known
that different agents, distributed in space, can experience changes in
different ways, and even disagree about the order of certain events.

The issue of measurement takes on an even greater importance
when considering observer semantics, because an event only acquires
meaning once it has been noticed and interpreted. Signals might lie
latent until observers choose to `process them', e.g. such as in queueing
systems.

If we are interested in intentional behaviour, then the issue of
causation is more about what an observer chooses to see than what
it is potentially capable of seeing. This might come as a slap in the
face to those who think that special relativity (i.e. the effects
brought on by the finite speed of communication) already makes matters hard
enough.

\subsubsection{Shared (non-local) timeline example}

Expecting distributed consensus between individual observers is a weak
strategy for any observer. One should expect diversity, as this is a
promise more likely to be kept. Consider an example.

Suppose four agents $A$, $B$, $C$ and $D$ need to try to come to
promise each other a decision about when to synchronize their
activities. If we think in terms of impositions or command sequences,
the following might happen:
\begin{enumerate}
\item $A$ suggests Wednesday to $B$,$C$,$D$.
\item $D$ and $B$ agree on Tuesday in private.
\item $D$ and $C$ then agree that Thursday is better, also in private.
\item $A$ talks to $B$ and $C$, but cannot reach $D$ to determine which conclusion it reached,
\end{enumerate}
In a classical view of time, resolving this seems like a trivial matter. Each
agent check the times of the various conversations according to some
global clock, and the last decision wins.  That viewpoint view is
problematic on several levels however.

First of all, without $D$ to confirm the order in which its
conversations with $B$ and $C$ occurred, $A$ only has the word of $B$ and $C$
that their clocks were synchronized and that they were telling the truth.

\begin{figure}[ht]
\begin{center}
\includegraphics[width=9cm]{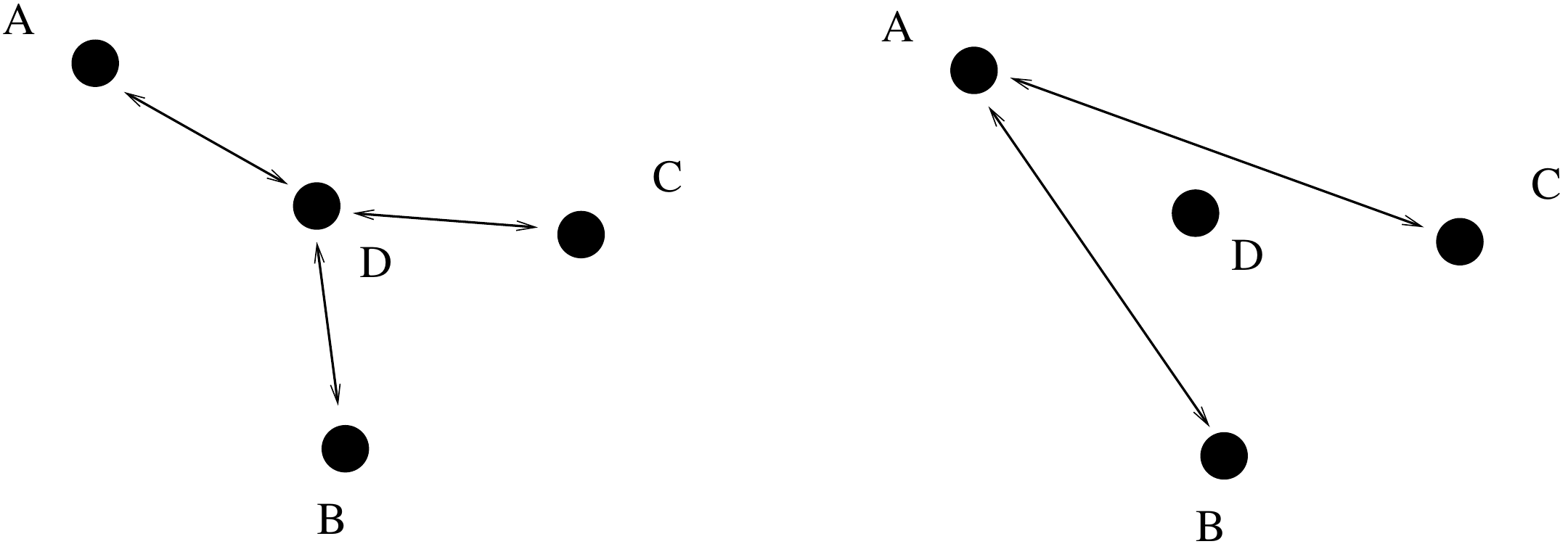}
\caption{Comprehending intended time}
\end{center}
\end{figure}

$A$ is neither able to find out what the others have decided, not able
to agree with them by `signing off' on the proposal. To solve this,
more information is needed by the agents. The promise theory
principles of autonomous agents show where the problem lies very
quickly. To see why, let's flip the perspective from agents telling
one another when to meet to agents telling one another when they can
promise to meet. Now they are only talking about themselves, and what
they know.

\begin{enumerate}
\item $A$ promises it can meet on Wednesday to $B$,$C$,$D$.
\beq
\pi_1: A \promise{{\rm Wed}} \{B,C,D\}
\eeq
\item $D$ and $B$ promise each other than they can both meet on Tuesday.
\beq
\pi_2: \{B,D\} \promise{{\rm Tue}} \{B,D\}
\eeq
\item $D$ and $C$ promise each other that they can both meet on Thursday.
\beq
\pi_2: \{C,D\} \promise{{\rm Thu}} \{C,D\}
\eeq
\item $A$ is made aware of the promises by $B$ and $C$, but cannot reach $D$.
\beq
\pi_2: \{B,D\} &\scopepromise{{\rm Tue}}{A}& \{B,D\}\\
\pi_2: \{C,D\} &\scopepromise{{\rm Thu}}{A}& \{C,D\}
\eeq
\end{enumerate}
Each agent knows only what promises it has been exposed to. So $A$ knows
it has said Wednesday, and $B$,$C$, and $D$ know this too. $A$ also knows that
$B$ has said Tuesday and that $C$ has said Thursday, but doesn't know what
$D$ has promised, because the two promises involving $D$ are concurrent
according to its information.

There is no problem with any of this. Each agent could autonomously
meet at the promised time, and they would keep their promises. The
problem is only that the intention was that they should all synchronize at
the same time.  That intention was a group intention, and somehow it
has to be communicated to each agent individually, which is just as
hard as deciding when to meet for dinner.  Each agent has to promise
($+$) its intent, and accept ($-$) the intent of the others to coordinate.

Let's assume that the promise to meet one another (above) implies that
$A$,$B$,$C$, and $D$ all should meet at the same time, and that each agent
understands this.  If we look again at the promises above, we see that
no one realizes that there is a problem except for $D$. To agree with $B$
and $C$, it had to promise to accept the promise by $B$ to meet on
Tuesday, and accept to meet $C$ on Thursday. These two promises are
incompatible. So $D$ knows there is a problem, and it is responsible to
accepting these (its own actions).  So promise-theoretically, $D$ should
not accept both of these options, and either $B$ or $C$ would not end up
knowing $D$'s promise.

To know whether there is a solution, a God's eye view observation of
the agents (as we the readers are) only need to ask: do the intentions
of the four agents overlap at any time (see the right hand side of
the figure)? We can see that they don't, but the autonomous agents are
not privy that information.  The solution to the problem thus needs
the agents to be less restrictive in their behaviour and to interact
back and forth to exchange the information about suitable overlapping
time. This process amounts to the way one would go about solving a
game-theoretic decision by Nash equilibrium.

With only partial information, progress can still be made if the agents
trust one another to inform about previous communications.
\begin{enumerate}
\item $\pi_1$: $A$ promises it can meet on Wednesday or Friday to $B$,$C$,$D$.
\beq
\pi_1: A \promise{{\rm Wed}} \{B,C,D\}
\eeq
\item $\pi_2$: $D$ and $B$ promise each other than they can both meet on Tuesday or Friday, given knowledge of $\pi_1$.
\beq
\pi_2: \{B,D\} \promise{{\rm Tue,Fri}|\pi_1} \{B,D\}
\eeq
\item $\pi_3$: $D$ and $C$ promise each other that they can both meet on Tuesday or Thursday, given knowledge of 1 and 2.
\beq
\pi_3: \{C,D\} \promise{{\rm Tue,Thu}|\pi_1,\pi_2} \{C,D\}
\eeq
\item $\pi_4$: $A$ is made aware of the promises by $B$ and $C$, but cannot reach $D$.
\beq
\pi_2: \{B,D\} &\scopepromise{{\rm Tue,Fri}|\pi_1}{A}& \{B,D\}\\
\pi_2: \{C,D\} &\scopepromise{{\rm Tue,Thu}|\pi_1,\pi_2}{A}& \{C,D\}
\eeq
\end{enumerate}
Now $A$ does not need to reach $D$ as long as it trusts $C$ to relay
the history of interactions. It can now see that the promise $\pi_2$
was made after $\pi_1$, and that $\pi_3$ was made after $\pi_2$. Thus
$A$ can surmise that the agents have not agreed on a time, but that
$B,C,D$'s promises overlap on Tuesday. It could now alter its promise
to reach a consensus or equilibrium.

The notion of time, in a world of autonomous agents, becomes nothing
more than a emergent post-hoc narrative that summarizes the
dependencies between interacting agencies. To put it another way, we
can draw a flowchart based on what we think happens in a distributed
system, perhaps even in a number of different ways, from different observer
perspectives, and with potentially different interpretations.

\subsection{Motion, speed and acceleration in agent space}\label{promisemotion}

Behaviour consists of exhibiting certain promised attributes; this can
be assessed by an agent acting as an observer. Motion, accordingly,
can be measured as a change in the location of promised
attributes\footnote{Attribute information might well exist without
  being promised, but in that case it is unobservable.}.  This can now
be defined in a few different ways, some of which amount to cellular
automata, in which the agents are cells, and others are re-wirings of
the fabric\footnote{Motion is a change in the relative adjacencies
  between bundles of properties.  This can happen by treating agents
  as containers for the promised attributes and redefining
  free-floating agent relationships (as in a swarm), or by dividing
  agents into fixed sign-post and mobile traveller type agents, i.e.
  building a rigid agent scaffolding on which free floating agents can
  move by binding and re-binding to an infrastructure network (somewhat
  like the way mobile phones attach to different cells in a cellular
  network).}.
In a space of autonomous agents, even the familiar concepts of uniform
motion in a straight line are non-trivial. 

\subsubsection{Foreword: is there a difference between space and matter that fills it?}

A question that becomes relevant when we approach the matter of spacetime
semantics is whether there is a basic difference between empty space
and something material that fills it? 

Imagine a blank storage array, which becomes filled with data. Is the
absence of an intended content fundamentally different from the
promise of an unspecified value?  Is a tissue of stem cells different
from cells that have been given material semantics by expressing
differentiated types? Both have DNA; they merely express different
promises. Is empty space merely an agent without a promise to behave
like matter? There is at least the possibility that matter is simply
the breakdown of indistinguishability in space.

This question arises naturally in the possible descriptions of motion
in a discrete agent-based spacetime.  Three distinct models of motion
make sense from the perspective of promise theory.  These might seem
excessive from a physical viewpoint, but all are in fact common in the
artificial spaces of technology, as well as in material science.

In the first case, there is only a single kind of agent in a gaseous
state.  In the second, there is a two-phase model with a solid spatial
lattice and material properties bonded loosely to them at certain
locations.  In other words, the position of matter is by bonding to an
element of space.  Finally, in the third model, there is only a single
kind of agent, but the physical properties promised as matter can bind
to a specific agent and be transferred from one to another.

Technologically, we have need for all three models of motion.  The
first relates to ad hoc mobile agents, the second to base-station
attachment of mobile devices, and the third to fixed (virtual)
infrastructure.  It is also fascinating to speculate as to the meaning
of these processes in nature. Certainly, these alternatives exist within
material structures. The question remains open as to whether spacetime
itself is constructed in the same way.

\subsubsection{Motion of the first kind (gaseous)}

\begin{figure}[ht]
\begin{center}
\includegraphics[width=4.5cm]{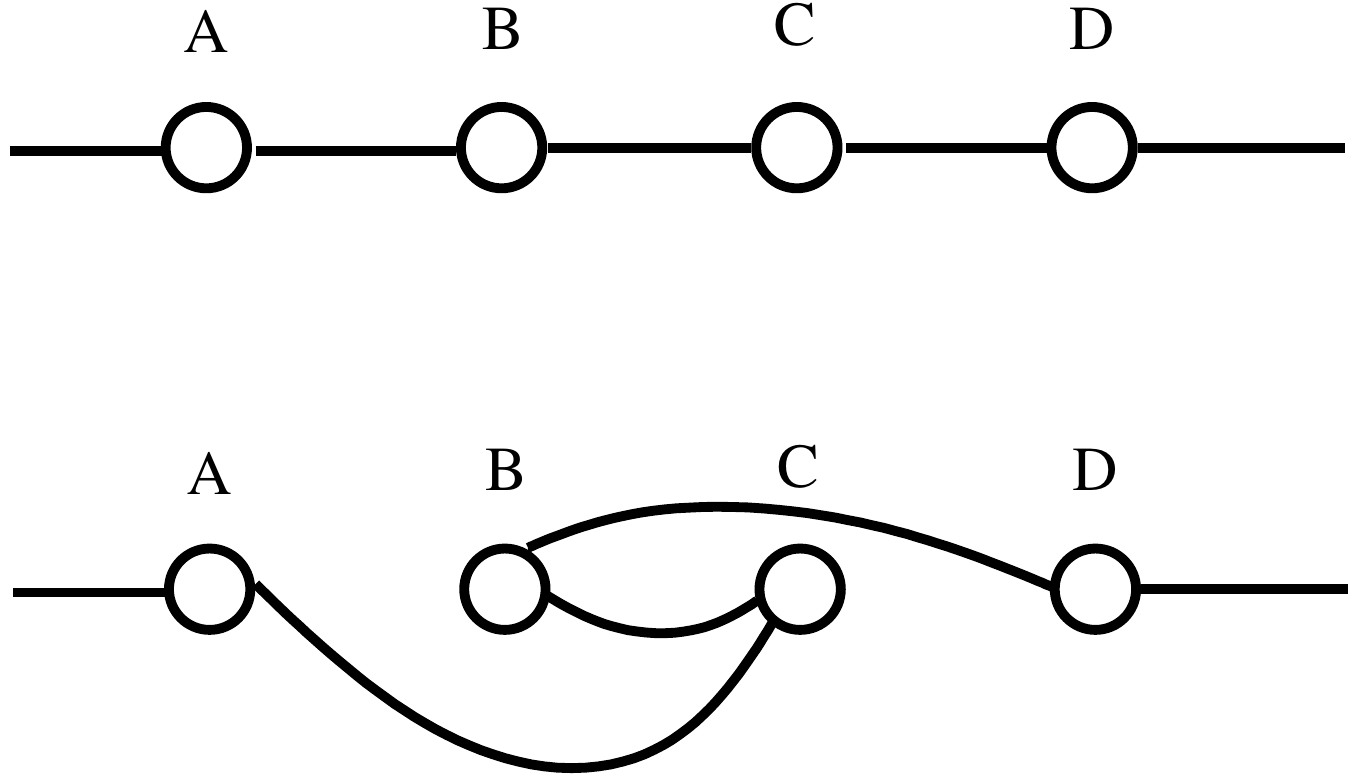}
\caption{Motion of the first kind: intrinsic motion in an untyped spacetime.\label{agentmove1}}
\end{center}
\end{figure}

The first case deals with a homogeneous collection of agents that can move
by swapping places, within an ordered graph of adjacencies (see fig.
\ref{agentmove1}).  This becomes increasingly complex in a multi-dimensional
lattice, so we'll restrict this to one dimension only to understand its
properties.

In order to be able to form a new adjacency, an agent must know about
the existence of the agent to which is wants to bind, and vice versa.
This can be assumed for nearest neighbours only. However, it turns out
that exchanging places\footnote{This is essentially a bubble sort
  algorithm} also requires knowledge of next-nearest neighbours.  This
would seem to involve multiple messages back and forth to discover one
another. This in turn seems to create a bootstrap problem for
spacetime.  How can spacetime form structure without such structure
existing to begin with? However, as long as agents promise to relay
information about their adjacencies to their neighbours, this can be
handled in a purely local manner.

A second, but related problem, is how to form a coordinate system in a
gaseous phase. If one cannot use the sequential, monotonic nature of
integer labels, then coordinate names become ad hoc and lose the
extrapolative power of a pattern.

Consider a simple one dimensional model with adjacencies to the left
and right of each agent (see fig. \ref{agentmove1}). We denote a trial
agent that has a non-zero velocity from left to right by $A_i$ ($A_i =
B$ in the figure). $A_{i+1}$ is to the right of $A_i$, and $A_{i-1}$
is to the left of $A_i$. The agent must be able to distinguish left
from right.

In fig. \ref{agentmove1} we see that, in order for an agent to move one place the right,
four agents' promises need to be coordinated. This is a non-local phenomenon, since
$A_i = B$ and $A_{i+2} = D$ are initially non-adjacent.

We may denote the bodies for adjacency promises by $\pm {\rm adj}_{L,R}$. In addition we denote
the property of momentum to the left or right by $+p_{L,R}$.

We need to introduce a value function on the agents' promises to allow one combination
of promises to be preferred over another $\pi_1 > \pi_2$. Two promises will be considered incompatible 
when we write $\pi_1 \# \pi_2$. If two promises are incompatible and the first is preferred,
i.e.  $\pi_1 >\# \pi_2$ then we may assume a promise transition in which $\pi$ is withdrawn in favour of $\pi_1$.

Motion from left to right occurs with the following promises, for all $i$:
\begin{enumerate}
\item $A_i \promise{+p_R} *$, i.e. we assume an agent has the property of right momentum, which
is promised to all observers.

\item $A_i \promise{\pm A_{i\pm 1}} A_{i\pm 1}$, i.e. all agents mutually promise their neighbours to relay information
about their neighbours to them, both $\pm$.

\item $p_R >\# -{\rm adj_R}$ the acceptance of a right-adjacency (from the left) is incompatible with
the attribute of right momentum. i.e. an agent with right momentum  would immediately
withdraw its promise of right adjacency. This allows $B$ to drop its adjacency to $A$ in the figure.

\item If the adjacency with right neighbour $A_i$ goes away, try the next neighbour $A_{i+1}$:
$$A_{i-1} \promise{+{\rm adj}_R|\neg -{\rm adj}_R} A_{i+1}$$
This allows $A$ to connect with $C$ in the diagram.
Note this requires a memory of the next neighbour's identity, which means that spacetime
has to store more than information about attributes.

\item To join $B$ with $D$ and reverse links $BC$

An agent $A_i$ with right momentum $+p_R$ might prefer to bind to $A_{i+2}$ and relabel
its offer of right adjacency to $A_{i+1}$ as acceptance of a right adjacent $-\adj_R$ from $A_{i+1}$.

\item To drop $CD$, an agent might prefer to use a combined promise of momentum and adjacency from $B$,
than pure adjacency from $C$, i.e. $p_R , +\adj_R >\# +\adj_R$.

\item To flip the use-right-adjacency of $CB$ to a give right adjacency offer $CB$,
this might be preferred if the $-\adj_R$ is no longer received from $D$.

\item Finally, all the left adjacencies also need to be connected along with the right adjacencies
in a similar manner.
\end{enumerate}
This is a lot of work to move an agent one position in a lattice. It
seems unnecessarily Byzantine, in violation of Occam's razor.  The
valuation preference, for instance, favours promises from farther away
than nearby agents. This also seems contrary to physical experience.

This algorithm does the job with only locally passed information but
it has loose ends, as well as the unsatisfactory need for extended
memory. Moreover, in higher dimensions it becomes even messier. It
does not seem viable as a method for motion, though possible.  Loose
ends include adjacency promises that are not withdrawn, leading to
non-simple connectivity.  This can be dealt with by introducing
conditional equilibrium promises: I will make you a promise only if
you accept it.  
\beq
A \promise{+ \adj_R | - \adj_R} A'\\
A \promise{+ \adj_L | - \adj_L} A' 
\eeq 
In this construction, the
promises are invalid unless the recipient is listening so the momentum
can break the symmetry of the equilibrium\cite{burgess_search_2013}.
 To view a `particle' as
moving in this spacetime, one would have a spacetime element promise the particle's
properties. 

In physics, one assumes that motion is an intrinsic behaviour of
bodies, but here we see that it is a cooperative non-local behaviour.
This might be unavoidable in any description. This requires no
non-locally transmitted knowledge (i.e. the bootstrapping is
self-consistent). However, it is far from satisfactory. Perhaps one might also view such multiple
connectivity as a form of tunnelling akin to {\em entanglement}, in
the sense of quantum mechanics\cite{weinberg2012lectures}.

What this exercise in promise theory reveals is the logical need for
{\em memory} of non-local information in order to bootstrap motion.
Whether spacetime is artificial or not, this structural information
conundrum is unavoidable.

\subsubsection{Motion of the second kind (solid state conduction)}

In the first approach to motion, there is only a single kind of agent,
which suggests great simplicity, but this leads to a highly complex
set of behaviours to explain motion. A second kind of motion may be
explained by separating agents into two classes, which we may refer to
as spatial skeleton agents $S_i$, which account for the ordered
structure of spacetime, and a second kind of agents which promise
non-ordered material properties $M_j$. Adjacency $\{ M\} \promise{\pm\adj} \{ S\} $
accounts for the location of `matter' within `space', and matter becomes
effectively a container for material promises.

\begin{figure}[ht]
\begin{center}
\includegraphics[width=4.5cm]{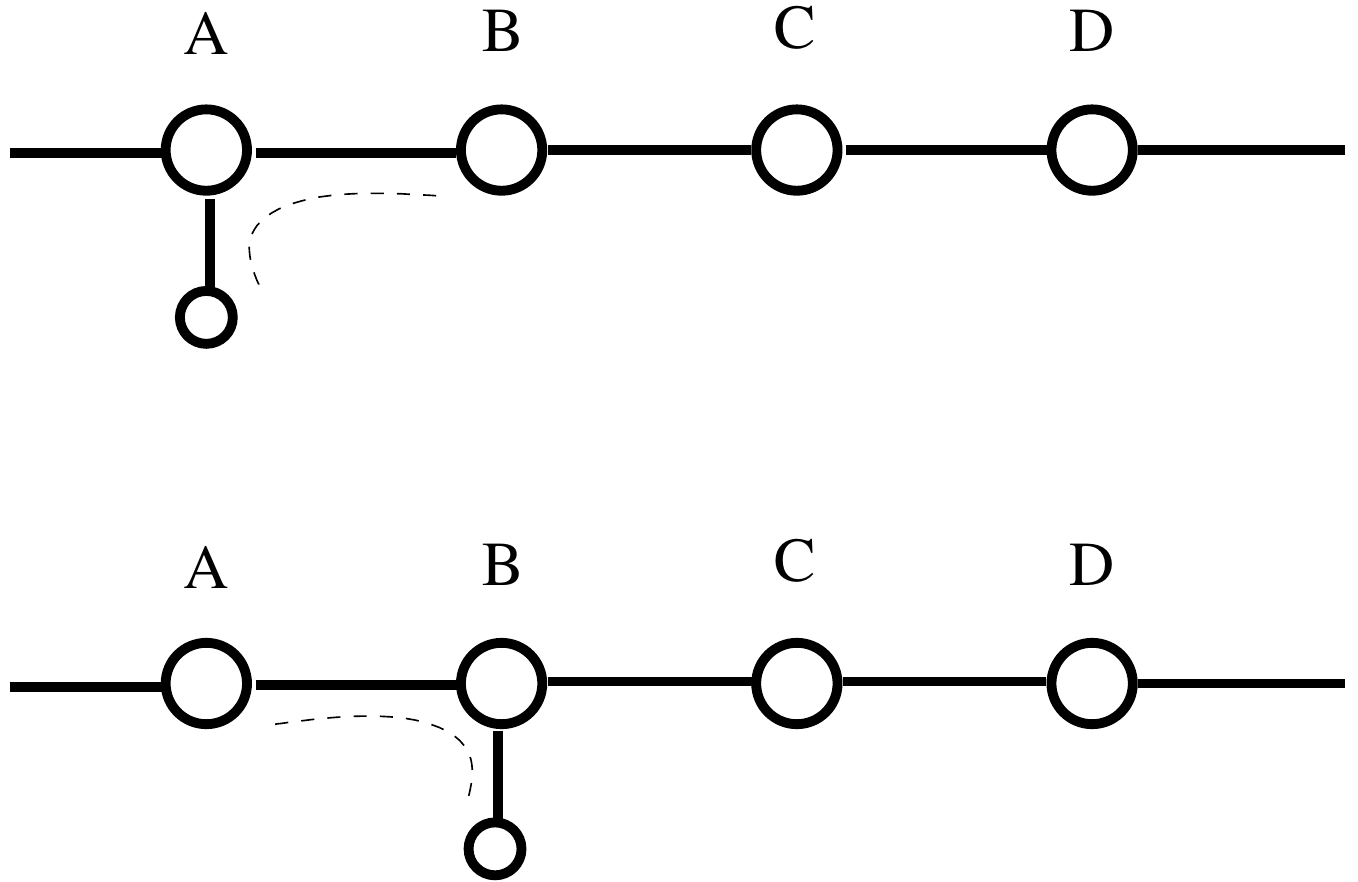}
\caption{Motion of the second kind: extrinsic motion on a typed spacetime.\label{agentmove2}}
\end{center}
\end{figure}

Motion within such a model then consists of re-binding material agents
at new spacetime locations (see fig. \ref{agentmove2}, with $S_1 = A, S_2 =B$). 
\beq
{\rm withdraw} ~~~~ M_1 &\promise{\pm \adj}& S_1\\
{\rm make} ~~~~ M_1     &\promise{\pm \adj}& S_2
\eeq
This is the way mobile phones attach to
different cell base-stations, by re-basing or re-homing of a satellite
agent around a fixed cell location. It is also the model of location
used by bigraphs. Locations are fixed seeds around which material agents
accrete or migrate. Since the property of velocity belongs to the material
agent, it must be an autonomous promise to bind to a new site.

Just as before, agents carrying a momentum need to know of the next
location binding point. That information has to be relayed between the
spatial agents, just as in the previous section.  With two classes of
agent, new questions arise: how many $M_j$ can bind to the same $S_i$?
This discussion goes beyond what there is room for here (in physics this relates to
the particle classes called bosons, fermions and anyons).  The same
issues will persist as we generalize the kinds of agents to many
classes in section \ref{semantic}

In this model, motion is still not a behaviour that is purely
intrinsic to an agent; it is non-local, as the existence of
non-adjacent points is not knowable without information by the
skeletal agents being exchanged between $A$ and $B$.  However, in this
case only two skeletal agents and one mobile agent need to be involved
to transport the mobile agent. This is much simpler than motion of the
first kind.

Although this model recreates the old idea of absolute spacetime, or
`the \ae ther', it seems preferable than the first model.

Notice how other forms of spatial belonging fall into the same category
of binding to a seed. Membership on a person in a club or organization
can be viewed in the same light. Then the adjacency promise takes on the
semantics of `is a member of' or `is employed by', etc. Another way to form a group
or `role' as it is called in promise theory is for all agents to promise
all other agents in a clique to have the same promises. This is a de-localized
alternative to the localized binding envisaged here.

\subsubsection{Motion of the third kind (solid state conduction)}

To remove the notion of an explicit \ae ther, one could avoid
distinguishing a separate class of agent, and keep all agents the
same.  In technology, this is what is known as a peer to peer network.
Movement of promises between agents is the third and perhaps simplest
possibility. It is essentially the same as motion of the second kind,
without a second type of agent but with two classes of promise.

\begin{figure}[ht]
\begin{center}
\includegraphics[width=4.5cm]{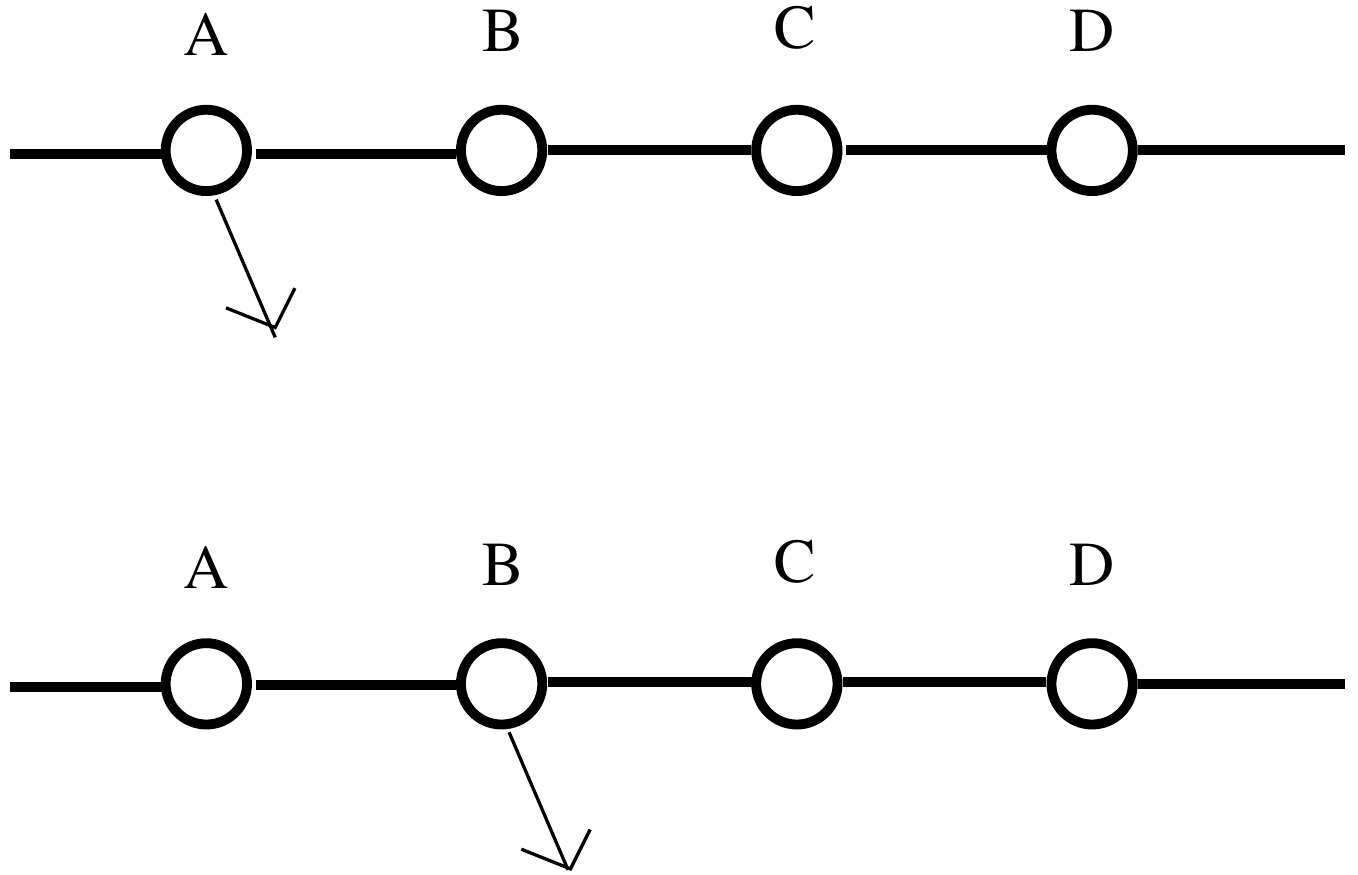}
\caption{Motion of the third kind: extrinsic motion of promises on a non-typed spacetime background.\label{agentmove3}}
\end{center}
\end{figure}

Motion now consists of transferring promises from one location
to the next. This brings new assumptions of global homogeneity: now
every agent in space must be capable of keeping every kind of promise.
There can be no specialization.

The promises to do this are straightforward, and the continuing need for global
cooperation with local information in this picture looks very like Schwinger's
construction of quantum fields\cite{schwingerfield1,schwingerfield2}, and
Feynman's equivalent graphical approach\cite{feynman1,dyson1}.

\subsubsection{Measuring speed, velocity and transport properties}

What is the meaning of speed and acceleration of transitions for a
finite state machine? Speed is defined as the translation of a
measurable quantity over a definite time interval.  While
philosophically interesting in its own right, this has practical
consequences for the transport of material promises from place to
place in a cooperative system, so it is worth spending a few words on.

The three kinds of motion described above do not give us an automatic
notion of speed.  Motion of the second and third kinds are
representative of what we observe classically in materials.  Motion of
the first kind is akin to gaseous collisions, quantum tunnelling and
topological effects. In those phenomena, time is always part of the
independent backdrop. In a space of autonomous agents, that assumption is
invalid.

Consider then how we can define speed and acceleration, the fixtures
of classical mechanics on which we base much of our idea of reality, in
an autonomous agent space.
\begin{itemize}
\item How long does a transition take? No agent can measure a transition
faster than its own clock ticks, so transfer of information is limited
by each participant's local ability to experience time. Events
which happen faster than a clock tick of a receiver appear concurrent
or simultaneous.

\item So how do we even measure the concept of velocity in a state machine
that makes transitions, where those transitions {\em are} the very definition of time?
Clearly it is not possible without promises and trust in the version of reality
they express.
\end{itemize}

Suppose now we try to measure speed and acceleration, as one would in
an experiment at our macroscopic scale. If we can measure speed, then
we can also measure acceleration, by measuring successive speeds and
comparing for a change.  In continuous spacetime, we can take many
things for granted; as we'll see, in a discrete network, we cannot.
It's not important which model of motion we choose. For the sake of
argument, let's take motion of the third kind in which promises move
(like electrons skipping through an atomic lattice).

The conclusion is quickly seen by the following argument: clocks to
measure time have to be built from space, so a change in spatial
configuration implies a change in time. However, a change in time from
other states which do not alter position in a particular direction, so
a change in time does not imply a change in space. This implies that a 
change in speed must always be less than or equal to some maximum value:
\beq
\Delta x &\Rightarrow& \Delta t\\
\Delta t &\not\Rightarrow& \Delta x\\
\Rightarrow \frac{\Delta x}{\Delta t} &\leq& v_{\rm max}.
\eeq
\begin{lemma}[Universal speed limit]
  In any discrete transition system that measures its own time, there
  is a maximum speed (which we can define as unity).
\end{lemma}

Consider just two nearest neighbours in isolation. Our only measure of
distance is in adjacency hops, so the distance between neighbours is
always 1 unit.  Similarly, the time it takes for a transition can
always be defined to be 1, since an observation leads to a change of
state and hence a tick of the clock. Actually, astute readers will
note that the agents cannot measure the `time' in any other sense than
this. There is no information about when a signal started that is
available. Transitions are considered instantaneous. However, the
arrival of the event leads to a change, which in turn counts as a tick
of the clock so that change must be 1.

Regardless of what any external observer might see or think, the local
observers always see information travel at a constant speed of 
\beq
v_{\rm neighbour} = \frac{1}{1} ~~~{\rm hops/transition} 
\eeq 
This is
straightforward because the observer and the locations are the same.
By this reasoning, an observer would never be able to observe any
transport at a speed different from 1\footnote{This is analogous to
  the constancy of the speed of light in electrodynamics: speed
  depends only on universal constants, in the frame of the observer.
  However, this wave equation result implicitly assumes the uniformity of
  spacetime.}.
It suggests that all signals must travel at the same speed, and that
there can never be any acceleration. However, if we extend to paths
that go beyond nearest neighbours, then the state space grows and it
becomes possible to measure differences of {\em effective speed} on
average, because of the incompleteness of information available to
observers.

\begin{figure}[ht]
\begin{center}
\includegraphics[width=4.5cm]{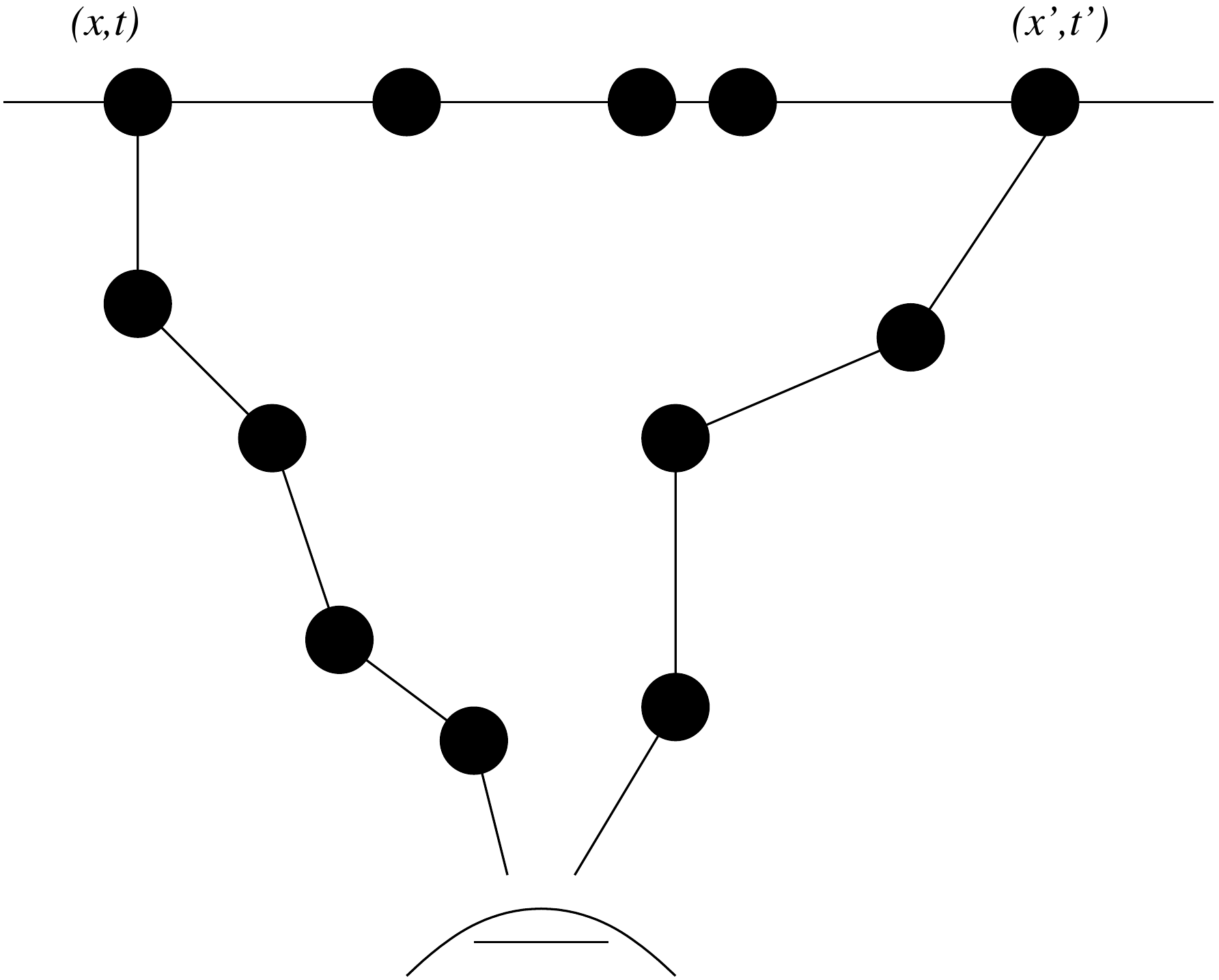}
\caption{Motion observed by a third party.\label{measure}}
\end{center}
\end{figure}

To extend the model to a greater distance, look at figure \ref{measure}.
Suppose an observer $O$ observes a promised property $\pi$ at
$(x_1,t_1)_O$ according to its own measurements, and later observes
that $\pi$ has moved to $(x_2,t_2)_O$. Then, according to our ordinary understanding
of speed, it can compute the speed as the ratio of distance travelled per unit time:
by: 
\beq v_O = \frac{x_2-x_1}{t_2-t_1}.  
\eeq 
Since there are more agents involved now, it is possible that other interactions
might have caused $O's$ clock to advance while $\pi$ is travelling, so we are freed
from the idea that the time for a transition must be a single tick.

This assumes that we are numbering the coordinates and times in an
ordered sequence. This assumes either non-local cooperation between agents, or swarm-like
cooperation.
If the coordinatization of space and time by $O$ is not monotonic,
this arithmetic difference of values means nothing. Thus the semantics
must be policed by $O$ alone. In a
gaseous state, this is a risky assumption, but let's imagine that the
relative positions of the agents are frozen for now to keep things as
simple as possible.

Assuming that the transfer takes place, and $O$ makes its
measurements, the first issue is that $O$ has no direct knowledge of
how many intermediary agents might lie between $x_1$ and $x_2$,
without foreknowledge. This is hard to imagine given that we normally
assume a regular, predictable Cartesian coordinate lattice. Without
that safety net, simply knowing the names of the end points of an
interval does not uniquely determine a path or history. This is
analogous to the matter of geodesics in curved spacetimes.  The only
measure of distance we have is nearest neighbour adjacency: `hops'.

So $O$ has no knowledge that $\pi$ travels monotonically from
$x_1$ to $x_2$: it might take a non-direct path, or it might actually
reverse course sometimes and even oscillate between two agents for a while
before continuing. All the while, $O$'s restricted clock might be
ticking (or even reversing). What speed would be inferred then?

\begin{lemma}[Speed can only be inferred]
  The speed of transport of an observable property cannot be measured with
  certainty by any observer, since this would require it to impose a
  promise to report on behalf of all agents along the path taken by
  the observable property. This would violate the autonomy of the agents.
\end{lemma}

Even if none of this possible weirdness happens, the promise that $\pi$ moved
from $x_1$ and has reached $x_2$ has to be communicated by passing
through a number of intermediate agents, with all the same concerns
above. Thus there is ample opportunity to count additional ticks of
$O$'s clock. Since the act of measurement opens it to state which
becomes part of its experiential clock, this suggests that $O$'s
time can never run backwards as long as it is interacting with other agents.

So, $O$'s clock is ticking at a rate which it experiences as fixed.
And we seem to have discovered that the speed of any object observed by $O$
\beq
 v_O \le 1.  
\eeq
Now that speed is no longer fixed, there is a possibility to measure
accelerations too. The question of how this relates to forces must be
left for another time (on the reader's clock).

The ability to carry out this experiment actually requires us to
assume a knowledge of the structure of spacetime in advance.  Clearly,
this is already a problem.  What makes this uncertain is that an agent
(or compound super-agent) is basically free to measure time according
to any clock it wants to. If the clock is not entirely internal, then it
has to deal with the fact that speed is an illusion of the local passage of
time. There are only transitions, which take unit time (by definition) to complete.
The details go beyond the scope of these notes.

The bottom line is that an agent can only rely on what it observes
directly.  Promises made by other agents help it to form a model of
the external world, but that is subject to uncertainty, and requires
extensive cooperation between agents to be able to trust\footnote{In
  distributed knowledge models, like Paxos and Raft, etc, the so-called
  guarantees of consensus about local knowledge between agents rests
  entirely on the non-local assumption that all agents follow the same
  set of rules without error.}.

\subsection{Growth and death of agent based spacetime}

Usually, in a spacetime, one considers the number of places to be
constant.  Cosmology, biology and technology are exceptions to this.
In a discrete spacetime, as in a cell colony, there can be spawning of
new agents (mitosis, meiosis), and both intentional termination
(apoptosis) and unintended termination (necrosis).

In technology, the cosmos of infrastructure is constantly being upturned,
growing and shrinking, like a rich ecosystem.
Injection of agents, representing either matter or space, has to be handled
in a symmetrical way: both involve either an increase or decrease in
the number of discrete agents. How agents are imbued with properties
is beyond the scope of this work.  Semantically, this is only
important if the agents make promises to other agents. The number of
agents in a `noble gas' of non-interacting agents is actually quite
uninteresting, except perhaps from a global resource
perspective\footnote{Sources and sinks are requires to explain the
  addition of information into a closed space; however, the entire
  question of conservation of quantities is beyond the scope of this
  discussion about spacetime, and it does not seem to be simply
  related to the idea of semantics.}.

The expansion and contraction of spacetime, through the birth and death of agents, is
mostly about the keeping or not keeping of promises.
\begin{itemize}
\item Is there a cost to introducing/removing a promise?
\item Is a new agent addressable (recognition)? 
\item Is it unique (naming/entropy promise)?
\end{itemize}

%%%%%%%%%%%%%%%%%%%%%%%%%%%%%%%%%%%%%%%%%%%%%%%%%%%%%%%%%%%%%%%%%%%%%%%%%%%%%%%%%%%%%%%

\section{Semantic (Knowledge) spaces}\label{semantic}

The final chapter of these notes addresses the goal of unifying
dynamics and semantics into a single description at the spacetime
level.  The motivation for this begins with technology, but the
implications are much wider. Indeed, I've already used examples from
chemistry, biology, and physics. The goal for this section is only
to sketch out the mathematical preliminaries, using promise theory to
unify the semantics and dynamics.

A few simple semantic aspects of spacetimes have already shown up in
this discussion so far.  A further compelling reason to study the
spacetimes with more sophisticated semantics is to model databases,
semantic webs, or knowledge maps, in which relationships and
adjacencies have diverse interpretations. Databases are ubiquitous and
knowledge maps are used for artificial reasoning (e.g.  Bayesian and
neural networks) as well as the creation of expert systems; even
`smart' public infrastructure (computing clusters and clouds) and
human crowds (e.g. a typing pool, or a classroom) may be considered
phases of a semantic space, along side the materials of the physical
world. The utility of this is that is brings a unified framework of
description that can cross interfaces and disciplines, allowing
us to understand the scaling behaviour of semantics.

Semantics are called `meta-data' in information technology.  Another
term for a knowledge space is an `index' for another space.
Confusingly, an entire spacetime can be a semantic representation of
something, and that semantic space might also contain its own index,
or semantic sub-space. This nesting is how we can begin to understand scaling.

Paths through semantic spaces form what we think of as processes, stories or
narratives. The concept of a resource, or valuable thing, is a
semantic one. Clearly, semantics are so ordinary that we neglect their
proper consideration, taking them for granted.

Knowledge spaces extend the foregoing un-categorized spaces by not
merely having structural adjacency and dimensionality, like direction
etc, but also types or `flavours' of adjacency.  By examining the
semantics of such generalized spacetimes, there is an opportunity to
better understand the challenges and potential solutions that present
themselves in technology, and perhaps even in fundamental physics too.

We must ask: what kind of a space is a knowledge space? Conversely: in
what sense can any other kind of space share the characteristics of
and be interpretable as a knowledge map? Would this help in the design
of shops, warehouses or museums, designed for organizing things by the
promises they make?

\subsection{Modelling concepts and their relationships}

Knowledge modelling introduces abstractions, such as the notion of
{\em concepts}, and the relationships between them. Such ideas are
mysterious without a framework like promise theory to help formalize
them. Using it, we can de-personalize these building blocks, and
understand the issues without too much ado\footnote{Clearly,
  linguistics and psychology have definitions for these things, but
  they are not satisfactory from the viewpoint of formalization for
  engineering purpose.}. There are two main questions to answer:
\begin{itemize}
\item How do we encode a concept within a spacetime?
\item How do we relate concepts to one another, using adjacencies?
\end{itemize}
Ironically, to understand knowledge better, we need to understand its
local scaling properties. Approximation, i.e. de-sensitization to
specific location into regions of space, is one of the most
significant precepts for modelling at the conceptual level. Concepts,
after all, are always described relative to a context of other
concepts, things and ideas.
This suggests that the application of the compound super-agent to
encoding semantics will be important. 

\begin{definition}[Context]
  Promise theoretically, the context of an agent is the collection of
  agents in its neighbourhood, that influence its semantics from the viewpoint
  of an observer.
\end{definition}
This rough description will be the model of context for explaining how
names become concepts.  From it, we can obtain trains of thought,
patterns of usage, and so on, by linking agents together.

\begin{definition}[Concept]
  A concept is an agent, in a semantic space, which has been labelled
  by an observer to have a non-numerical coordinate name, and a
  non-empty {\em context}.
\end{definition}
A concept agent may be adjacent to a number of exemplary agents, in
which case the concept agent links together a class of other agents.
A concept is therefore a source or sink, in the graph theoretical
sense, that binds together exemplars or occurrences into a
super-agent, i.e. it mediates a {\em containment} promise (see section
\ref{containment}), aggregating agents into a generalization or
umbrella class.
\begin{figure}[ht]
\begin{center}
\includegraphics[width=5cm]{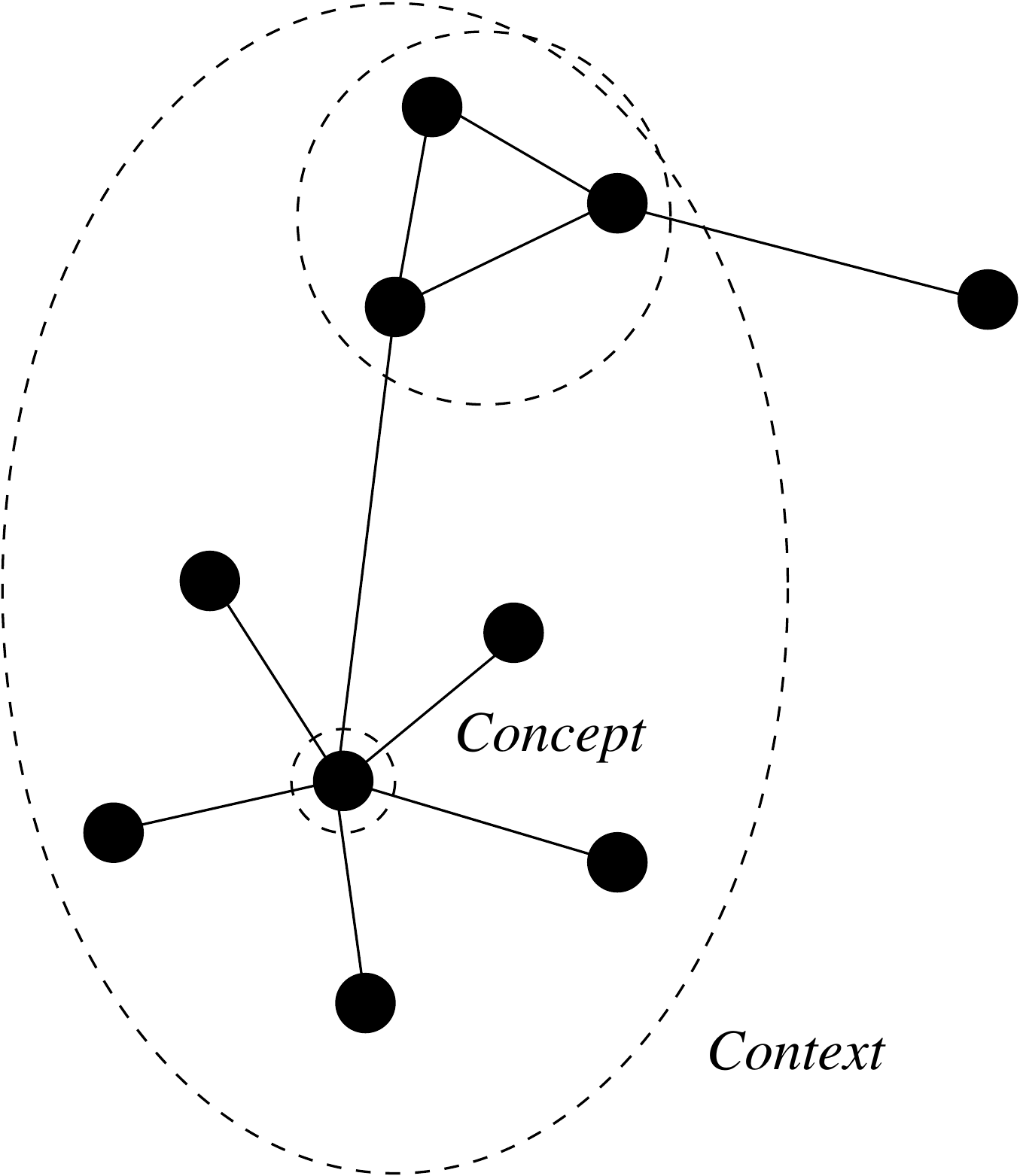}
\caption{A concept's significance comes from its name and its context together.
Without a context, a concept makes an ambiguous promise.\label{concept}}
\end{center}
\end{figure}

A similar definition was given in
\cite{semanticspace2,burgessaims2009}.  When does a name or coordinate
label become a context?  The aim of this section is to see how this
comes about from the promise theory of the foregoing sections.

\subsection{Coordinate systems for knowledge spaces}

One of the important problems in knowledge management is how to find
concepts within a repository space or map.  How we coordinatize a
space, i.e. name its locations, is the key to the addressability of its
resources: to finding things and reasoning about them.

The lack of an intuitive coordinate system for concepts and their
relationships makes finding objects into a brute force search; though
indexing is a strategy that can help to reduce the cost of a search.
Usually the burden is placed on the observer to have significant
knowledge of the global topology of a knowledge space in order to
locate elements. For humans, this learning takes a lifetime.

The traditional approach to knowledge structures was to model them as
tree-like taxonomies, refining names into ever smaller categories.
This was particularly popular during the reductionist phase of science
of the 1800s.  Database theory later came up with the notion of
relational tables, in which one has many similar data records with a
fixed template that are interconnected via abstract references.
Later still, hyper-linked structures like Topic Maps and RDF were
invented to describe information webs. All of these structures can be
accounted for within the scope of promise modelling.  Spacetime points
now become data records or some other kind of aggregated data, linked
by adjacencies that are both promised by the agents themselves and possibly
reinterpreted by observers\footnote{Recently the idea that the brain
  itself might organize itself physically as a semantic space have
  been proposed\cite{Huth20121210}, though the evidence for this is
  weak. I speculate that the reason is that we do no know how to code
  concepts in the brain in terms of neuron coordinates (indeed we do
  not know how the brain coordinatizes concepts at all).}.
Even a mundane data network can be considered to
be a semantic web, in which devices are linked together by physical
and logical connections. Elements of space can promise storage,
computing capabilities, memory, and so on.

If we coordinatize something, we encode {\em memory} of the adjacencies in a
non-local region of spacetime, at the same time as encoding structural semantics.
A space is not a Markov process, unless it is in the gaseous state.

\begin{itemize}
\item Tree structures represent branching processes, which suggest
  refinement of an idea, or the reasoning about an idea. The tree
  forms a history, tracing the roots of particular outcomes.  If the
  tree flairs out, we move to more possibilities of ever lower
  significance (deduction).  If the branches converge to a root, it
  represents the identification of greater meaning from hints of
  evidence (induction or abduction).

\item Cartesian tuples suggest a regular lattice structure, based of
  the generators of a translational symmetry, or transitivity.

\item Networks have a hub and spoke architecture which suggests
  radial, or graph-like adjacency. This is a web or ecosystem,
  emphasizing cooperative relational significance.
\end{itemize}

Knowledge space agents have two roles, in the promise theory sense, to
explain their significance: the {\em concepts} or tokens of meaning, and the
{\em exemplars} or {\em occurrences} of those concepts.  In one common
interpretation of knowledge maps, the conceptual map is considered to be an
associative index of the exemplars.

Since a concept might be composed of a number of others, a scaled
understanding, in which the local coding of concepts is separated from
associations between them, emerges naturally from the autonomous agent model.
The link between these two parts is a set of associations called an
index.
Moreover, because concepts (e.g. animal) are singular generalizations
for exemplars, which are multitudinous, we would not expect concept
space to have anything like a simple lattice structure.  More likely, it
would be a sparse, {\em ad hoc} graph.  Occurrences (e.g. dog, cat), are many and
share promised (material) characteristics in common.

\subsection{Semantic distance}\label{coordskm}

The value of semantics lies in the ability to aggregate similar
things, more than to decompose things into separate classes.
Branching creates diversity, but aggregation creates uniqueness (see section \ref{hierass}).
Branching into more categories is a common modelling viewpoint, but it
leads to intentional inhomogeneity, whereas aggregation of
observations into categories is natural empirically, and leads to
continuity and homogeneous meanings.

It is natural to keep related things close together in some sense; or,
conversely, we might attribute closeness to mean relatedness. This
duality is why the concept of a semantic space proves useful.

Conceptual closeness (aggregation) leads to increased stability of
semantics or spacetime regions by membership. Thus the formation of
concepts, as agents which bind several exemplary agents, is more
valuable than the individual exemplars, because the concept is more
likely to be `remembered' through the bindings to others. The association
with neuron connections is surley not accidental.

This vaue in aggregation also suggests one way of localizing or
coordinatizing knowledge agents, around sign-posts called concepts.
Localization itself has semantics, as distance can be promised in
different ways.  Two obvious candidates are:
\begin{itemize}
\item Semantic distance: the abstract distance between concepts, as measured by the
number of transformations of hops to get from one concept to another in 
some representation of the knowledge.

\item Occurrence distance: the sum of adjacencies from one occurrence of knowledge
to another, as measured in a spacetime that contains both.

\end{itemize}
Ultimately, it is up to observers to evaluate the closeness of knowledge
elements in their own model.
In section \ref{grammars} we looked at document spaces, where chapter
and section numbers could be used to span a document as vector
references.  The alphabetical index pointing to a linear page
numbering is also a simple (scalar) standard that is helpful in
locating strings in a document.  It has been replaced by hyperlinks in
groups of non-linear documents.  But what about the space of all
documents?  

Imagine looking for a book in a shop or library.  If you happen to
know the coordinate system by which the books are stored then finding
a book will be quicker than searching through all the rows (as long as
you promise to use the same coordinate categorization as the provider
of the index).  Libraries invented the Dewey decimal system of
numbering for books (later the universal decimal system) as a
coordinatization of book categories.  It tries to make similar things
close together, and separate dissimilar things, by mapping a subject
ontology to a relative positive in space.

A typed or categorized matroid to cover such a graph must have
duplicate coverings for each type of link relationship. A numbering of
things or topics within `subject categories', `promise roles' or `contexts'
depending on nomenclature of choice assigns a number to agents , e.g.
occurrences of a word or phrase in a text, or a particular kind of
device in a network.
We know such objects by the term `index'. The coordinates are
alphabetically ordered, and map to linear page references.

Using the semantic spatial elements, 
\begin{definition}[Semantic coordinate system]
  A tuple numbering of elements in a material network, spanned by a
  matroid of the form $({\rm scalars, vectors},...)$
\end{definition}
The vector components are covariant in the tensor sense. The scalar
elements are obviously invariants\footnote{Notice how there is a simple correspondence between semantic elements
with their attached scalar promise networks and the notion of tables in relational
databases. Note also that the homogenization of scalar spacelike components
amounts to normalization of a database in first normal form.}.

The final matter context for concepts has to be dealt with, as
semantics are context dependent. Intuitively we expect that
observer semantics must be related to use (-) promises.

Recall the example of blueness in section \ref{properties}.
What if there were multiple occurrences of a property or topic, e.g. the name `blue' in
different contexts and with different meanings (homonyms)? Suppose we
were then searching for `blue', how could we distinguish different
meanings, represented by different sources?  The classical answer in
taxonomy is that each blue occurs in a different type or category, as
part of some tree classification of containers for concepts. However,
we know that matroids are not non-overlapping taxonomies in general,
they can overlap. A more general answer could be that each one has a
different independent set for each different source (see section
\ref{coordskm}).  This generates a different tuple member. Thus we can
always form a non-trivial tuple space from a set of singular properties.
This means that we may unify spacelike properties and singular properties
of point-like locations in a regular coordinate system.

\subsection{The autonomous agent view of a knowledge map}

\begin{figure}[ht]
\begin{center}
\includegraphics[width=8cm]{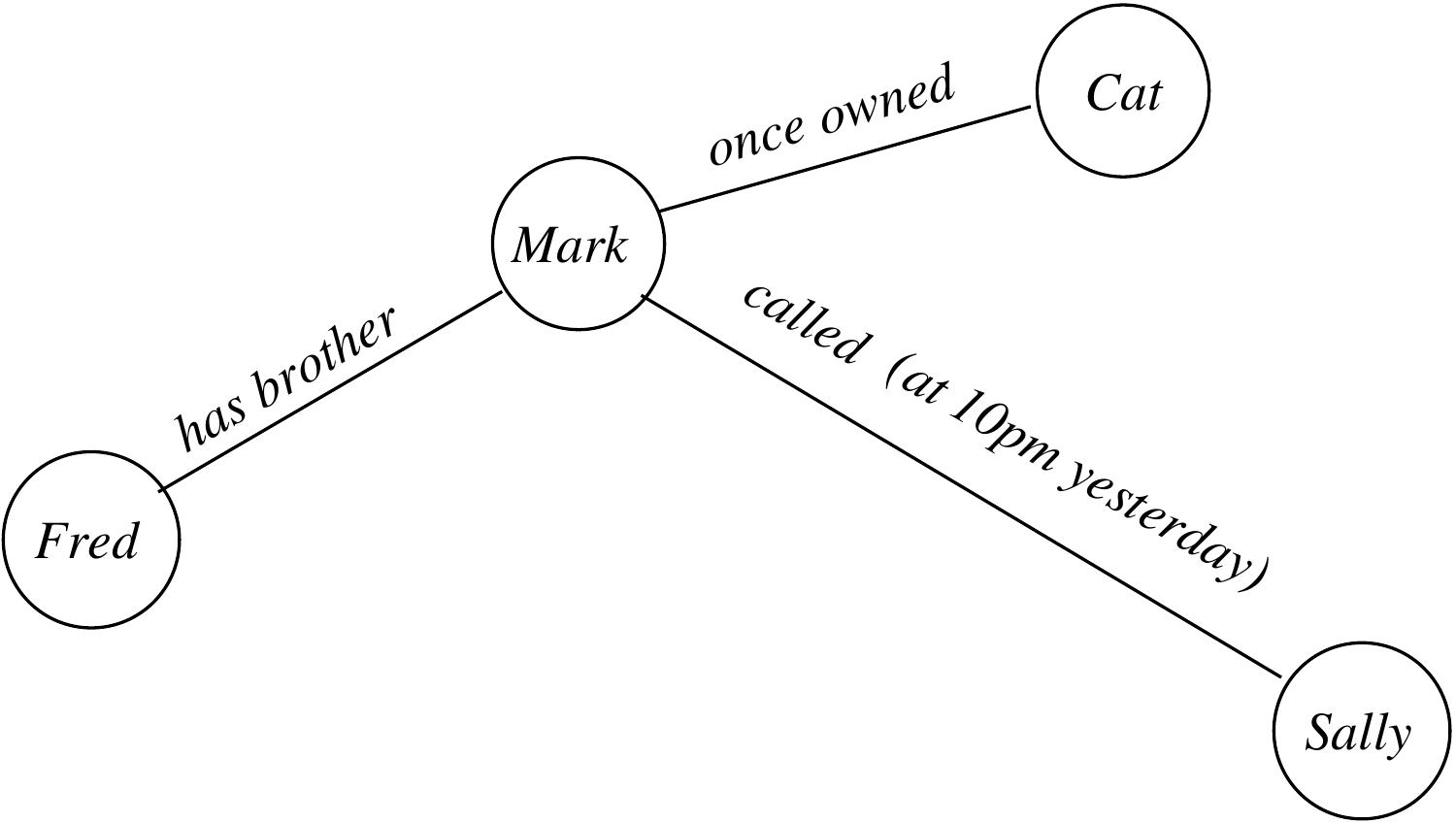}
\caption{Knowledge maps or semantic webs express typed relationships.
Notice how this form mixes an idea of time into the relationships,
so this cannot be a pure representation of spacetime.\label{knowledgemap}}
\end{center}
\end{figure}

Let's invoke promise theory to explore some of the basic properties of a
knowledge map (see fig. \ref{knowledgemap})\footnote{An overview of
  knowledge maps and their challenges was given in reference
  \cite{burgesskm}. In that review, the cost of globally obliged order
  was evaluated against that of a purely local approach built up from
  promises.  }.
From the introduction, we have the idea that a knowledge map has to solve
two main problems:
\begin{enumerate}
\item How to represent and encode concepts, as well as their exemplars, in spacetime structures.
\item How to relate concepts together to form context and narratives, through adjacency.
\end{enumerate}
A knowledge map is often an abstraction built by an observer as a
representation of a real world, but one may also interpret a real
world as a knowledge space in its own right. One could consider a
virtual reality facsimile of a city to be a knowledge map of it.
Alternatively, one might consider an encyclop\ae dia index to be a
knowledge map.  The important feature is that the entire map has a
maintainer, and each maintainer may create their own map as a valid
representation.

\begin{definition}[Knowledge map and maintainer]
  A knowledge map is a collection of agencies, called topics or {\em things},
  which each promise identifying names, and possibly a number of other
  semantic inter-relationships.  The promised links, or adjacencies,
  are called {\em associations}. The map belongs in its entirety to an
  independent agency, called the {\em maintainer}.
\end{definition}

Because topics or things have intended meanings, we may view them as intentional
agents, in the sense of promise theory. Each associative link asserts
an intended property in the manner of an obligation or imposition to
accept; each observer is free to re-interpret this according to its own
convention or world view. 

One of the interpretations of autonomous agents is that they exhibit
voluntary cooperation.  There is no real sense in which knowledge
relationships can be considered `voluntary' in a knowledge structure;
however, we can still think of the relationships as being autonomous,
signalling their independence and local significance.  Topics in a
pre-designed knowledge map can't refuse to bind with one
another\footnote{This is a curious oversight in data modelling, since
  all real physical and programmatic systems exhibit the properties of
  access control, and the right to refuse an assertion made by
  another.  We choose instead to assume that such matters are
  irrelevant to the semantics.  This is another example of classic
  assumption of absolute or necessary space.}, they are imposed by the
authority of a designer.  However, as knowledge is passed on and
assimilated by merging the maps of different observers into shared
map, one still has a choice about whether to connect two things that
assert their connectedness.

\subsection{Semantic agencies: things}

In promise theory, intentional agents only differ in the promises they
make. A priori, they are similar elements, like stem cells, or empty
boxes waiting to be filled. They are both the promisers and the observers of
characteristics.  This models reality better than the assumption of
authoritative design, encompassing it as an option.

Let's consider first how some
simple semantic relationships can be modelled as promises, and then
try to contrast this with a database viewpoint.  Imagine an agent
describing its properties to an observer. The promising agent's
probably semantic type is hinted in parentheses:
\beq
A_1 ({\rm pixel?}) &\promise{\rm~Has~colour~RGB}& A_2\\
A_1 ({\rm Swimmer?}) &\promise{\rm~Swims~breast~stroke}& A_2\\
A_1 ({\rm Book?}) &\promise{\rm~Is~a~work~of~fiction}& A_2\\
A_1 ({\rm Resistor?}) &\promise{\rm~Has~resistance~100 \Omega~\pm 5\%}& A_2\\
A_1 ({\rm Girder?}) &\promise{\rm~Is~made~of~ high~ grade~ steel}& A_2 
\eeq 
The words in parentheses are possible interpretations of the agent
$A_1$ in each case.  Nothing demands that we interpret the agents
$A_1$ in this way. Indeed, on receiving promises, it is the purview of
$A_2$ to decide the nature of $A_1$ by observation alone.  It is as
if the agents are playing a game of charades, and $A_1$ asks: this is
what I promise, what am I?

So who or what is $A_2$? Here, it is some other agent with no
particular type.  $A_1$ and $A_2$ are symmetrical comparable
things in a promise model.  To use an biological analogy, agents are like
generic stem-cells whose characters are differentiated only by the
promises they make.

Now compare this promise structure to knowledge relationships
in a database of normalized objects. For example, a hypertext web 
has the form:
\beq
\thing(A_1) &\assoc{\rm~Has~colour}& \thing(RGB)\\
\thing(A_1) &\assoc{\rm~Swims}& \thing({\rm breast~stroke})\\
\thing(A_1) &\assoc{\rm~Is~a~work~of}& \thing({\rm fiction})\\
\thing(A_1) &\assoc{\rm~Has~resistance}& \thing(100 \Omega \pm 5 \%)\\
\thing(A_1) &\assoc{\rm~Is~made~of}& \thing({\rm high~ grade~
  steel}) 
\eeq 
The `things' on the right hand side of these arrows are primitive
in comparison to intentional
agents; they are simply abstract entities, like information records in
a database.  The left and right hand sides of the relationships here
are very different kinds of objects, so traversing the knowledge
relationship is not like a simple adjacency, one ends up in a
completely different world: from say a world of pixels to a world of
colours. Once we arrive there, if we kept going, where might we end
up?  What is related to colour? The answer is clearly very many
things, so knowledge relationships tend to be nexuses of connectivity
rather than mere adjacency.

\subsection{Promising knowledge as a typed, attributed, or material space}

\begin{figure}[ht]
\begin{center}
\includegraphics[width=8.5cm]{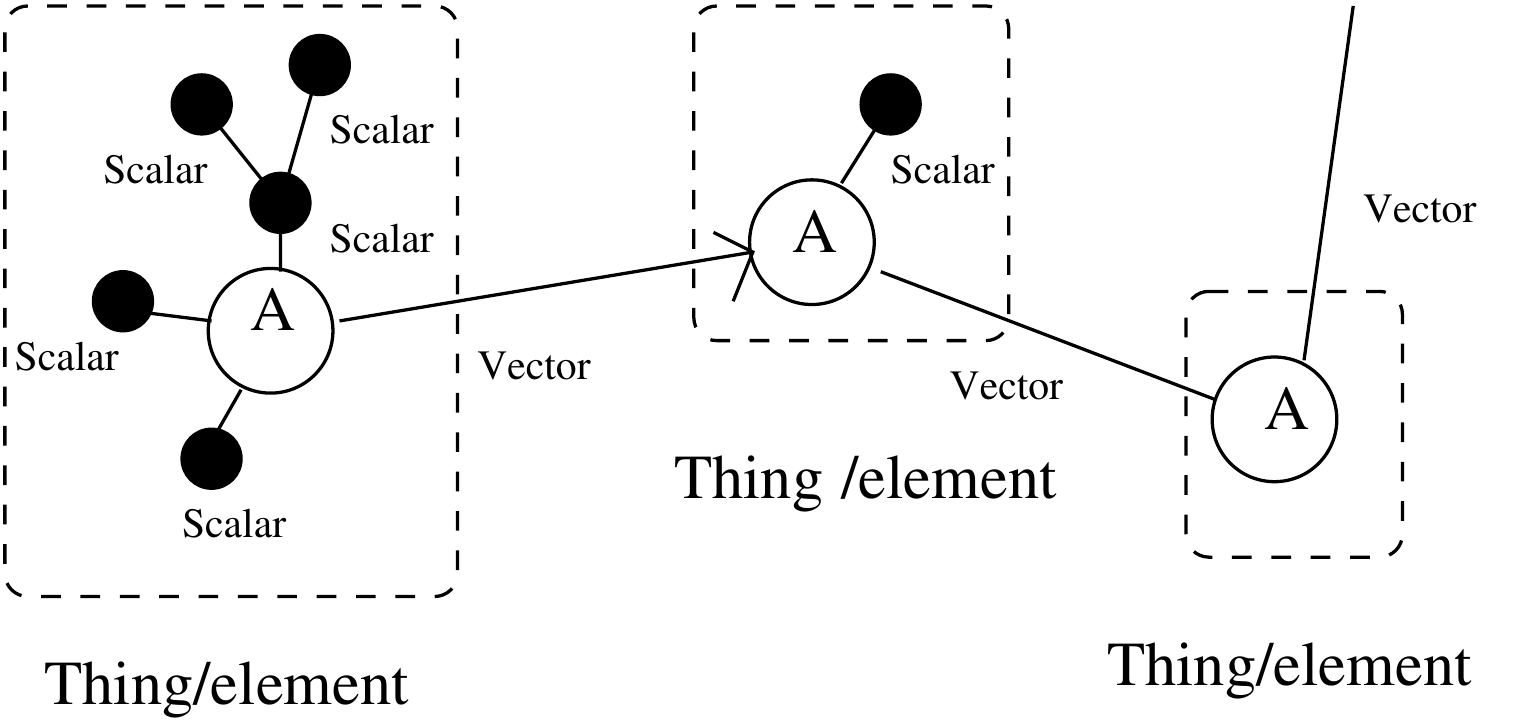}
\caption{Elements of semantic space may be viewed as autonomous agents
surrounded by a cloud of scalar promises, joined into a semantic web
by vector promise bindings. Compare this to figure \ref{qfield}.\label{kspace}}
\end{center}
\end{figure}

To make a spacetime, with both scalar (material) attributes and vector (adjacency)
bindings, from the atomic building blocks of autonomous agents, it is
a simple issue of defining the basic elements in a semantic space:

\begin{definition}[Semantic element (topic)]
  A semantic element or topic $T$ tuple $\langle A_i, \{ \pi_{\rm
    scalar~j}, \ldots \}\rangle$ consisting of a single autonomous
  agent, and an optional number of scalar material promises.
\end{definition}
A semantic element is thus an autonomous agent from promise theory,
surrounded by a halo of scalar promises that imbue it with
certain semantics (see fig. \ref{kspace}). 
This can be compared to the field concept in figure \ref{qfield}.

Familiarity is a valuation ranking that an agent can promise to encode
Bayesian-style learning into a knowledge map.  This indicates the
general utility of the promise model.

For example, to use the biological analogy, the autonomous agent acts
as a stem cell, while the scalar promises correspond to a genome that
is expressed by the cell and interpreted semantically, as well as a
ripeness level that ranks its familiarity dynamically. This indicates
that we may view a biological organism (or indeed any functional
environment) as a semantic network too.

The associations between the semantic elements may also be defined in terms
of promise bindings (I assume that there is no need to model permission to associate).
\begin{definition}[Semantic association]
An uni-directional association between 
semantic elements $T_1$, $T_2$ is a function $f(\pm b)$, in a promise binding between the
agent component of $T$, denoted $A(T)$
\beq
A(T_1) \promise{+f(b)} A(T_2)\\
A(T_2) \promise{-f(b)} A(T_1)
\eeq
The inverse association is
\beq
A(T_2) \promise{+f(b)^{-1}} A(T_1)\\
A(T_1) \promise{-f(b)^{-1}} A(T_2)
\eeq
where $b$ is one of the three types of promise body described in section \ref{3types} (causation, topology and containment). e.g. $f(b)$ is `eats', and $f(b)^{-1}$ is `is eaten by'.
\end{definition}
 A vector promise which can be mapped to a $\pm$ vector promise binding,
whose body is a function of

Semantic typing initially seems to destroy any notion of simple translational
symmetry. A topic map could not have long range order, and no obvious
notion of continuity, because it does not deal with instances
(occurrences, exemplars) of the concepts directly. Instead it ties them
together through the aggregate concepts. This is a centralized hub
network model.

Thus gives semantic knowledge maps like taxonomies, topic maps and RDF
their bi-partite, hierarchical structure. Locally, one is motivated to
restore some translational symmetry however, as this makes reasoning
simpler. In the work with Alva Couch, we proposed re-interpretation of 
associations for causal inference as one approach (see section \ref{3types}).

\subsection{Indices, meta-data, and code-books for semantic spaces}\label{codebook}

An index is a map of space, organized semantically. It associates
locations within a knowledge space, with a short identifying name
representing the semantics associated with those locations. It may be
viewed as a collection of additional promises made about bulk regions
(like pages instead of chapters) so that low resolution skimming can
be used to speed up brute force a search.

An index can only work if there is already a
coordinate system by which locations can be referenced, ordered in
some predictable monotonic lattice.
\begin{definition}[Index map]
  An index is a semantically structured map. It associates knowledge
  items with coordinates of elements in a knowledge space. Any
  structure that maps knowledge items to locations may be considered
  an index. 
\end{definition}
Although we think of an index as a physically represented code book,
an index could also be an algorithm, which implements the mapping
by computation. Hashing and tree-sorting algorithms satisfy these qualities.

In a book, topics are (usually) listed alphabetically for quick
lookup.  Alphabetization is a cheap {\em hashing function} that allows the user
to skip obviously irrelevant entries, and search for a matching name.
Coordinates are then provided by page number.
In electronic indices, references can be hypertext references to
specific documents or tagged regions within a document model.

A code book index is usually encoded in a small region of a spacetime.
Finding something in a book with an index is efficient only if we can
ignore what is written on the pages as we flick through them in order
to find the right page. It is tempting to think that one could simply
jump directly to a location if one knows its reference, rather than
traversing all the intermediate locations, but that depends on the
assumption of whatever underlying coordinate system has been
implemented to traverse it.  In other words, we can perform a code
book compression of the full space into a single coordinate (like
G\"odel numbering), but this does not in itself tell us about the
efficiency of finding the coordinate.

An efficient coordinate system, from an indexing perspective, is
one that minimizes the number of points one has to traverse
in order to identify the destination.

\begin{definition}[Index compression]
  An index may be called efficient if the cost of using or locating an
  item in the index added to the cost of finding the location from the
  coordinates is less than the time needed to locate the knowledge
  item directly in the knowledge space.
\end{definition}

Categorization of information, such as a taxonomy, could be considered
a form of indexing, but conceptually, taxonomies are often based on
the idea of sub-dividing categories so as to separate things into
exclusive containers. (As discussed in section \ref{hierass}, this is more likely to be a
successful strategy for associations than concepts.)
This leads to exponential growth in the number
of outcomes.  Aggregation of objects (e.g. alphabetically or by
conceptual generalization) is usually a more efficient way of ending up with
fewer categories to search.

Coordinate systems that exploit the natural structure to navigate a space can offer
short cuts to locating elements of the space, but only if we can
reduce the information content being addressed. A discussion of caching in semantic
networks is also forthcoming, but beyond the current scope\footnote{To discsuss
a cache as a kind of index, one must also deal with semantic replication of indentity, see section \ref{hierass}.}.

\subsection{World lines, processes, stories, events, and reasoning}

Much of human thinking revolves around our perception of timelines,
i.e. narratives, in which we also sew together events and attribute the
ordered sequence itself a meaning, as a collective entity.

\begin{figure}[ht]
\begin{center}
\includegraphics[width=10cm]{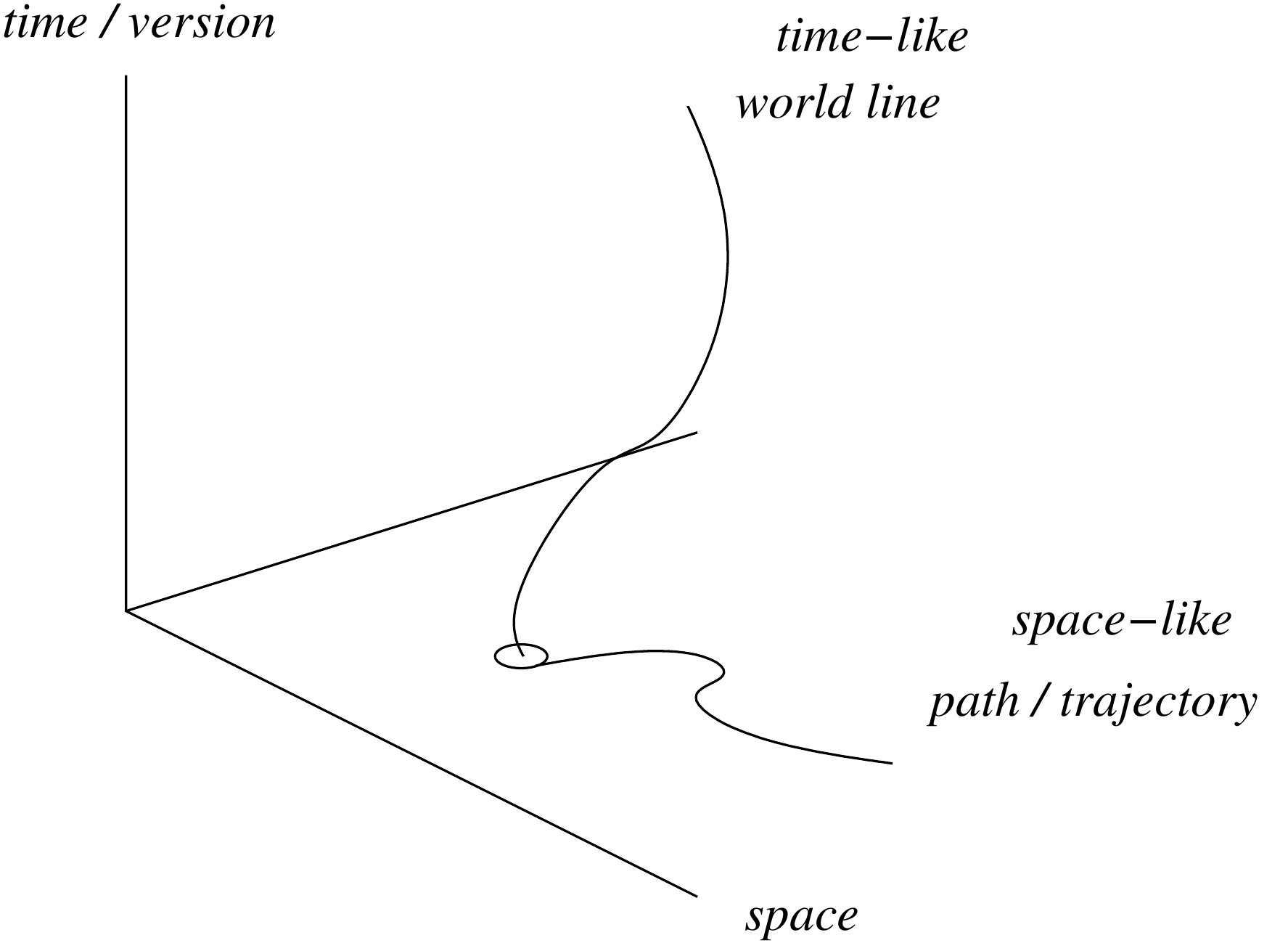}
\caption{World lines may be space-like (stories) or time-like (histories).\label{worldline}}
\end{center}
\end{figure}

We need not speculate why we have this predilection, but it plays a
large role in the way we organize semantics. In Einsteinian
relativity, we have come to refer to the story of a material body as
it moves through spacetime as a {\em world line} (see figure
\ref{worldline}). It represents the journal or history

For example, when we design human jobs, experiences, and even spaces
for human enjoyment, we try to craft a storyline or narrative that is
either practical or emotionally appealing.  Often a sequence fits this
requirement, because of our fondness of stories, and the desire to
know how they end.  

There might be a sequence of interactions between a mobile (gaseous)
agent, like a customer, and a number of other fixed agents that alter
the state of the mobile agent, by interaction with the services
provided by the fixed stations. This is the arrangement in an assembly
line, or the converse of a bee pollenating flowers.  Semantics and narratives
are at the root of our basic attitudes to information.

For example, when we check into an airport: the airport staff follow
you from the outer airport zone check-in desks, and then follow you
through to the boarding gate, where they see you on your way.  The
same staff carry out both stages, as the information from both
contexts is shared, and the experience is humanly satisfying. They
hand over to security personnel for a screening, and to baggage
personnel for luggage transport. There are thus three agents working
alongside from start to finish: boarding, baggage and security. The
same staff follow the passengers throughout their whole process.

Maintaining constant agency throughout an entire process has the advantage of
continuity, not requiring information to be communicated to another
agent thus wasting time. It is also satisfying for both customer and
staff.  One could organize things differently, however. Different
teams could handle the different specializations. Inside the baggage
super-agent, there is a chain of intermediaries. The security is
focused at a single entry point, and walled off by a boundary.

\begin{figure}[ht]
\begin{center}
\includegraphics[width=10cm]{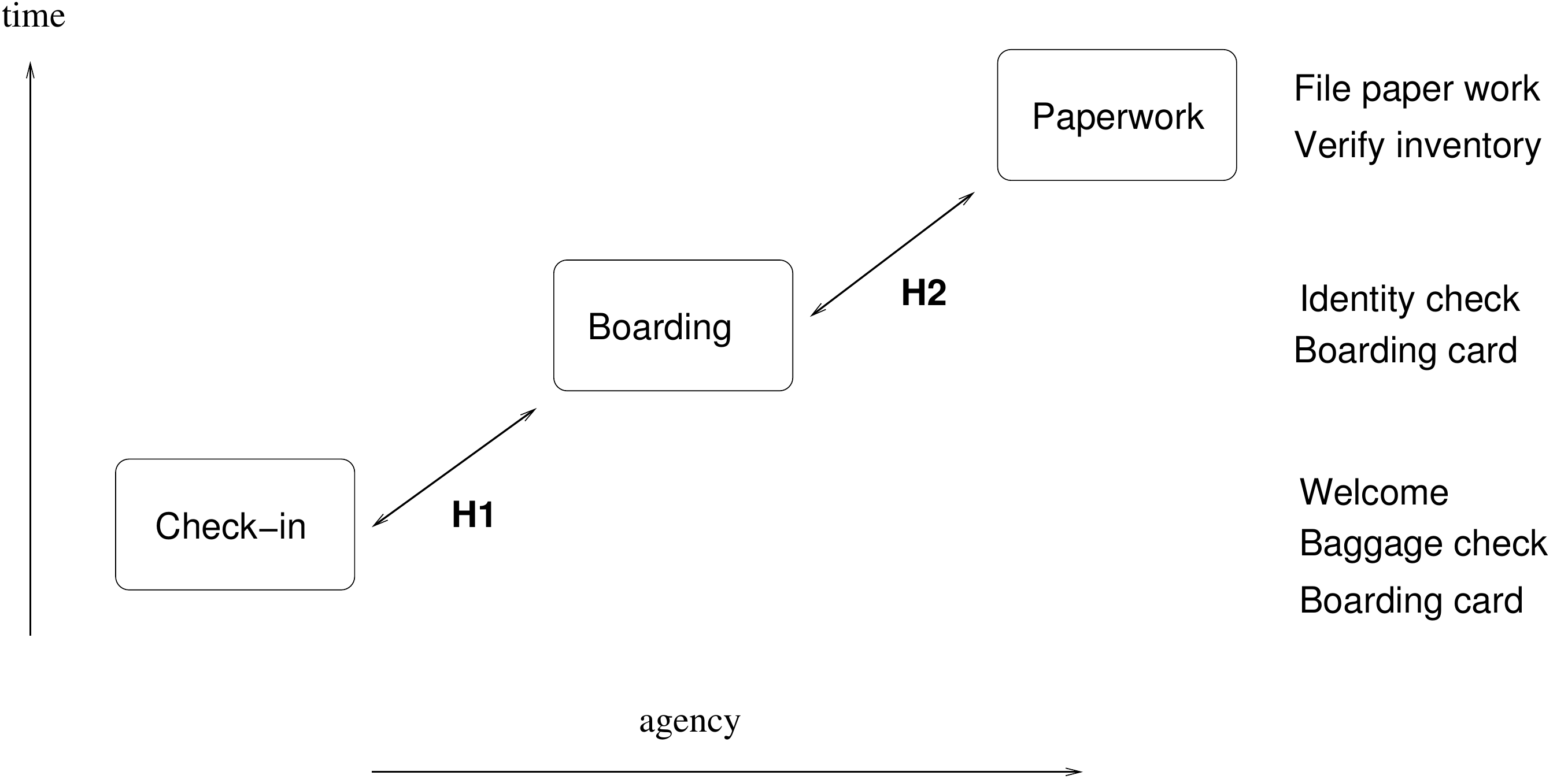}
\caption{Following a job narrative {\em thread} along the timeline (longitudinally)
  or {\em staging} component-wise (transversely), with hand-offs to
  different agents H1 and H2.\label{tl1}}
\end{center}
\end{figure}

\begin{figure}[ht]
\begin{center}
\includegraphics[width=10cm]{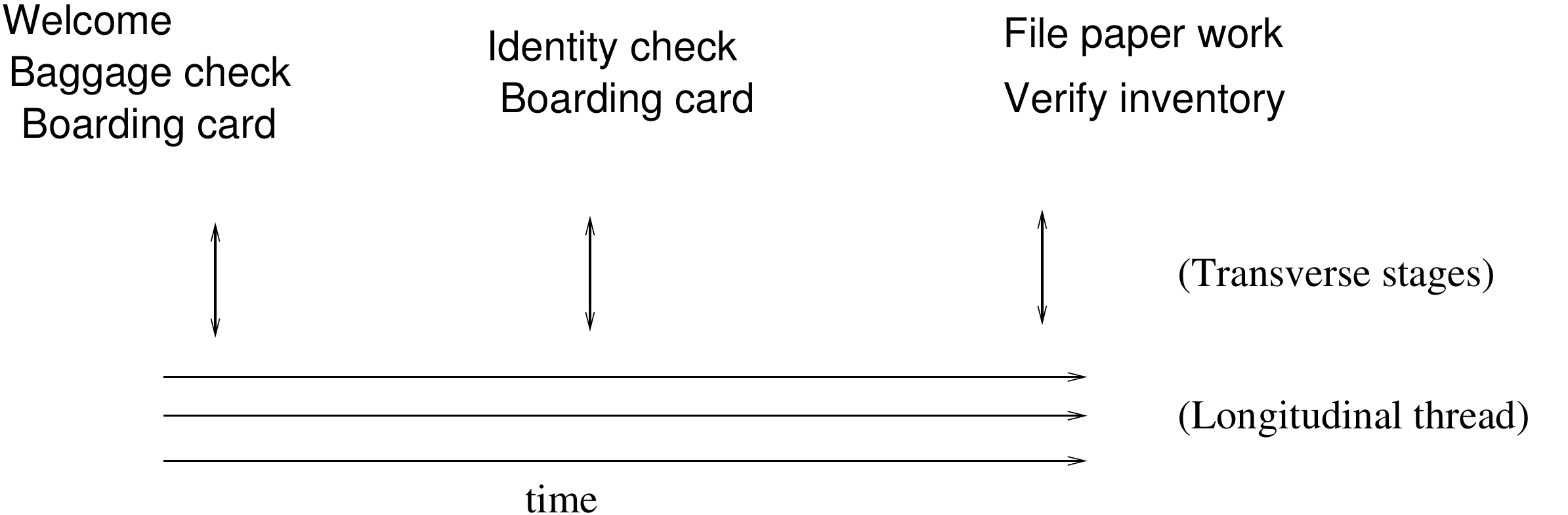}
\caption{Three agents could work longitudinally (threading), or transversely (staging).\label{tl2}}
\end{center}
\end{figure}

Check-in and boarding are two different skills and could also be
handled by different agents, but then both stages would have to
promise to collaborate and pass on information.
These two decompositions of a timeline could be referred to as follows (see figures \ref{tl1} and \ref{tl2}):
\begin{itemize}
\item Continuous {\em threading} involves a single agent whose role adjusts to enact the different
phases of the whole process from start to finish, forming a time-like world-line. 
\item Discrete {\em staging} involves breaking a story into different
  specialized agents for different phases of the whole, forming a space-like story.
\end{itemize}

If we stage a timeline, by handing off to different agents, we need
{\em interfaces} between them, as each agent really has its own world
and timeline with local information\footnote{In programming, one
  speaks of APIs or Application Programmer Interfaces in between the
  stages. These are like function calls.}. This is the approach
modelled by bi-graphs. Reasoning semantics are
exposed in staging, and hidden in threading.  In order to unravel
reasoning as a storyline, we use the rewriting logic of section
\ref{3types} to create a quasi-transitive
chain\cite{stories,burgessaims2009,burgesskm}.
\begin{definition}[Story or narrative]
An ordered collection of elements belonging to a connected semantic space, joined
by associations with a quasi-transitive interpretation.
\end{definition}
This, in turn, can be converted into a sequence of promises.
Consider the following story example:
\begin{quote}
\sc
(Computer X) {\bf signals} (error 13)\\
(error 13) {\bf stands for}  (disk fault)\\
(disk fault) {\bf can be caused by}  (loss of power)\\
(loss of power) {\bf can be caused by}  (diesel generator)\\
(diesel generator) {\bf is manufactured by}  (Acme generator company)
\end{quote}
The agent names are generic agents with scalar promises, and the bold
face associations are adjacencies.  To generate stories, one now only
needs to follow vector links from a particular initial position, as if
solving a difference equation.

{\em Inference rules} are relabellings of associations and pathways
that any agent can make, usually about the quasi-transitive vector
promises. Every agent assessing a promised association is free to
re-interpret a semantic assertion\cite{burgesskm}.

An agent that makes many scalar (material) promises defines a semantic
element type. An agent which makes many vector adjacency promises forms
part of a story structure.

\subsection{The semantics and dynamics of identity and context}\label{hierass}

The goal of promise theory, and indeed semantic spacetime, is to unify
the description of dynamics with semantics. The concept of {\em
  identity} lies at the heart of this unification. The promise of
identity comprises two aspects:
\begin{enumerate}
\item A distinguishing name or form\footnote{A name need not be a linguistic token, but a name in the promise theory sense, see \cite{promisebook}.} (semantics).
\item A local prominence, relative to its neighbourhood or context (dynamics).
\end{enumerate}
The semantics of {\em identity} associate symbolic labels, with
singular dynamical structures (singularities, fixed points, hubs,
delta distributions, etc); thus identities are where dynamics and
semantics meet. In a sense, semantics are associated with the opposite
of translation invariance and connectedness: they attach to things
that break maximal symmetry, and become identifiable. Consequently, we would
not expect complex semantics to be associated with long range effects.

In physics, there is a lot of attention given to invariances and
symmetries.  This leads to a meta-knowledge of uniform
expectation, without complex semantics. A lack of change means it
becomes easy to predict a trend.  However, all the `interesting'
phenomena, semantically, occur where symmetries break down.

\begin{figure}[ht]
\begin{center}
\includegraphics[width=10cm]{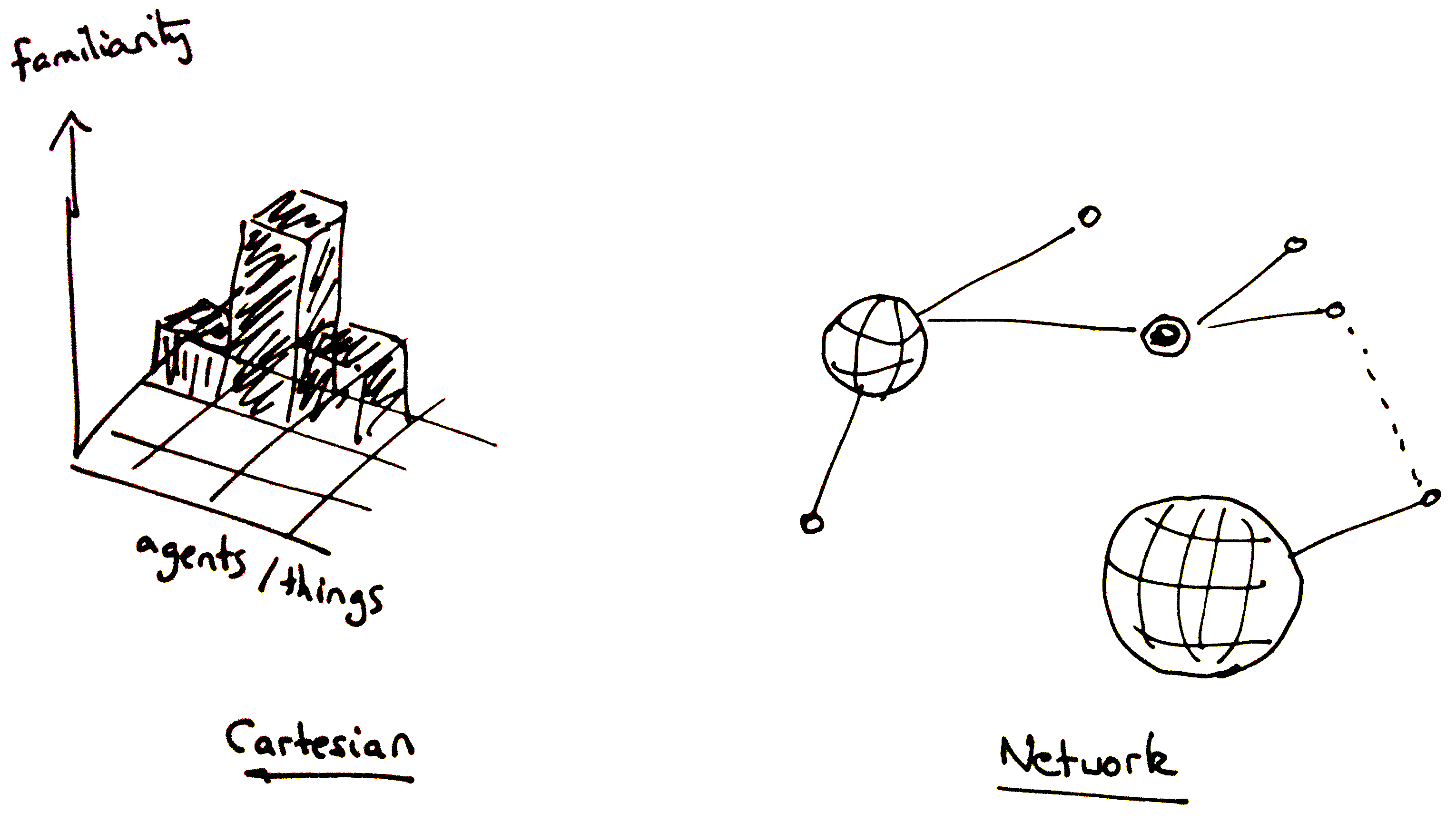}
\caption{The ranking of importance or familiarity is a promisable
  attribute which can be reinforced at each spacetime location by
  frequency of visitation. It plays a major role in identity, as
  familiar things `loom large' in the space.\label{familiar}}
\end{center}
\end{figure}

To claim to {\em know} something is a {\em
  familiarity} value judgement about information. Familiarity increases
the prominence of a location in a semantic space by repeatedly
visiting and reinforcing it (see figure \ref{familiar}). It mimics the
training of memory structures in brains and machine learning methods.
This is a stigmergic cooperation between agents. As a result, knowledge is associated
with signals and structures that stand out against a background noise
because they have grown in our subjective valuations.

The significance of localization explains why we build monuments, such
as obelisks and towers, that stand out against their backgrounds.  The
extreme low entropy is attributed with the cultural meaning of order, authority,
or control. Logos are singular symbols imbued with specific meaning,
like `brands', that stand out against pages of words and or images.

So how do we explain the importance of stories? Pathways are also
singular structures, like semantic world lines, that connect concepts.
Semantically meaningful paths are relatively rare within the
collection of all paths.  The fact that we are able to infer
reasonable stories from the many associations possible between
concepts suggests that the (dynamically based) transitivities
generalized in section \ref{3types} are automatically evaluated when
recognizing a concept. So it is not concepts that are hierarchically
evaluated, but associations. This is fascinating, though beyond the
scope of this work.

The concept of informational entropy is descriptive here. A symbol or
structure with high entropy represents a highly flat distribution of
much information, but little meaning.  Indeed, noise has the greatest
information content of all, just the lowest significance.  Conversely,
low entropy represents little information, but we attribute to this
great significance: it is something that stands up (like a Dirac delta
distribution) anomalously against the background of a statistical
average. Thus the association between information and meaning is an
inverse one.

The singularity of concepts makes them sound fragile. Indeed, singular
items are often considered fragile, because they are `single points of
failure' or `bottlenecks'.  Destroying one would inflict irreparable
change on the connectedness of a semantic space.  The way concepts become robust is by
associating with many example occurrences, and related.

If relationships were only between concepts and not the exemplary occurrences, then
the concept hubs would still bring fragility to a knowledge map. One of the advantages
of a spacetime perspective is in being able to redesign them so that associations can also
be made directly between the exemplars.

For a cluster of agents representing a concept, the complete
`identity' of the concept lies in the sum of associations as much as
any one of the agents in the cluster. This is reminiscent of
organizational membership and containment too, as in bi-graphs.  In other words, its
context defines its meaning as much as the promises it makes, through
its bindings.  The more interconnected it is, the more robust a hub of
semantics it becomes.  As long as the associations are fixed, the
meaning is constant. If none of the contextual bindings is preserved,
an agent essentially has a new identity.

\subsection{Low-entropy functional spaces, and componentization}

\begin{figure}[ht]
\begin{center}
\includegraphics[width=8cm]{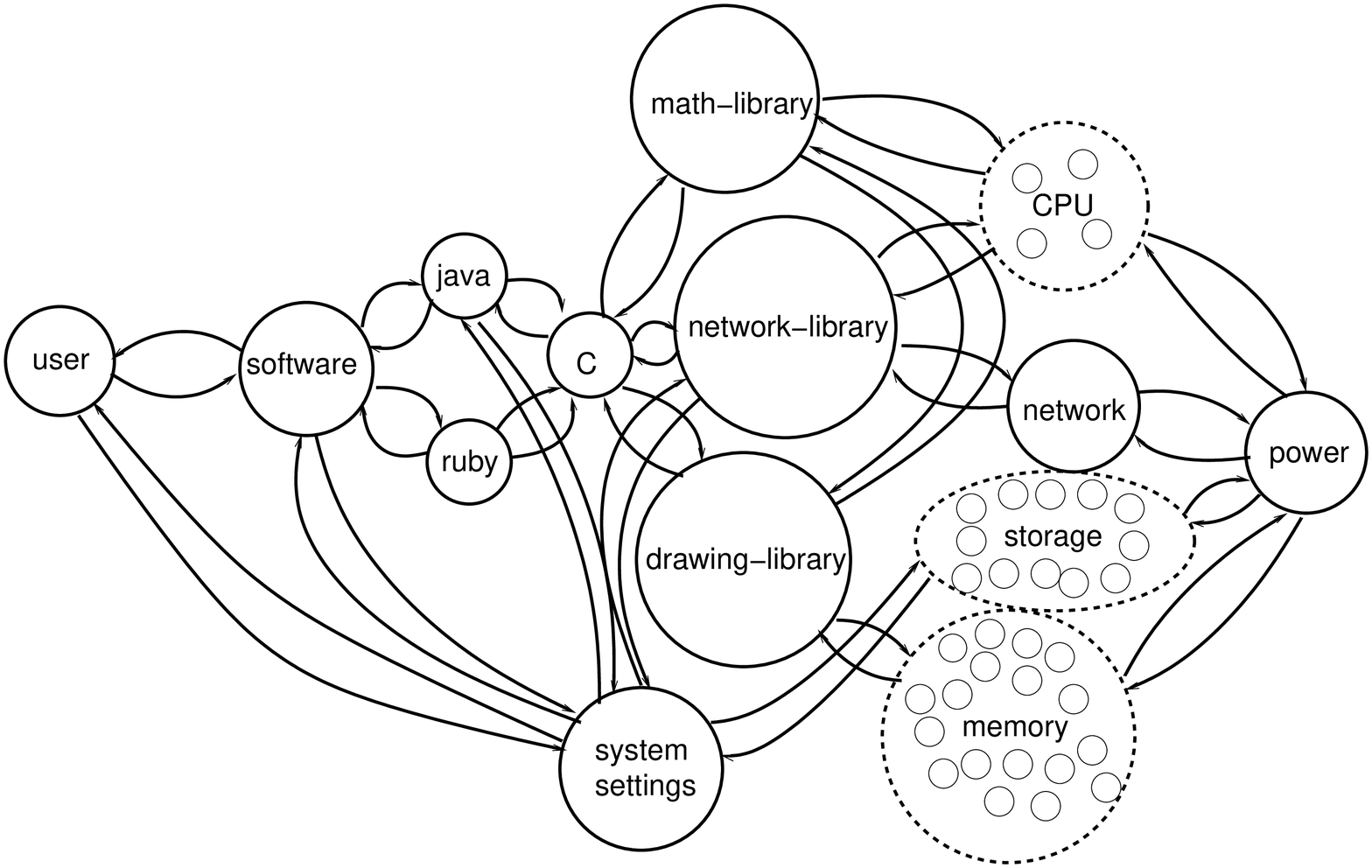}
\caption{A semantic spacetime of low entropy is an ecosystem of functional roles.\label{dependency}}
\end{center}
\end{figure}

Spacetimes, in which each agent is of unique and special significance,
represent specialization networks. One could characterize them as a
molecular phase, rather than gaseous or crystalline phase of atomic
agents, because there is local structure with global variation (see figure \ref{dependency}).

The entropy of agents' promise distribution is low in this kind
of structure, as most agents stand out uniquely in a special role.
A low spatial entropy network consists of mostly unique exemplars,
possibly indexed or organized by general concepts. 
They contrast with bulk materials of indistinguishable resources (see
section \ref{highS}). Thus a low entropy network has strong semantics,
but is fragile dynamically, as each element's uniqueness makes it a
possible point of fracture in the integrity of the whole.  Adjacency
combined with semantic uniqueness implies dependency.  For this
reason, low entropy implies `meaningful' as well as dynamically
fragile or potentially unstable.

Human organizations, shopping malls, organizational maps, electronic
circuitry, and computer flowchart algorithms are all examples of low
entropy structures, because a change at a single spatial element
would alter the perceived semantics of its observable purpose measurably.
See figure \ref{dependency}.

Databases also typically have low entropy, because each data record
has unique content. The coordinatization of databases is interesting
as it blends dynamical and semantic strategies. A simple numbering of
elements using an integer primary key is simple and reliable, but has
no interpretation. One therefore imposes a tabular matroid that has
coordinates based not only on an integer number, but also the
individual promises of content being represented: e.g.

\beq 
{\rm (key, name, address, occupation)} \label{dbtable}
\eeq 

Some guiding principles for modelling and coordinatizing semantic data
have been developed in the guise of the so-called {\em normal
  forms}\cite{date1,burgessbook2}, which attempt to avoid the
many-worlds problem of inconsistent branches.  Normal forms constrain
the similarity and redundancy of coded information within spatial
elements of the database (tables or records).  The normal forms are
thus entropy constraining patterns.

A database or space is in the {\em first normal form} if all of the
elements have the same size, shape and make the same type of promises.
Thus, one effectively ignores the uniqueness of the data (or the bodies
of the promises), paying attention to the similarity of form only, like the template
shown in (\ref{dbtable}).
This is similar to the idea of indexing, but without a separate map.

A first normal form might include repeated patterns, which are orthogonal degrees
of freedom forming a natural subspace, like say the addresses of
persons.  Semantically, one would like to extract any redundancy to
make a single point of change, because as a human tool, we don't want
to have to update the same information in more than one place. This
could lead to inconsistencies. Information inside one agent is in a
separate `world' to that of another, meaning that there is no causal
connection between the information. To assure the causality of a
change of address, one could factor it out as a dependency. 

\begin{figure}[ht]
\begin{center}
\includegraphics[width=8cm]{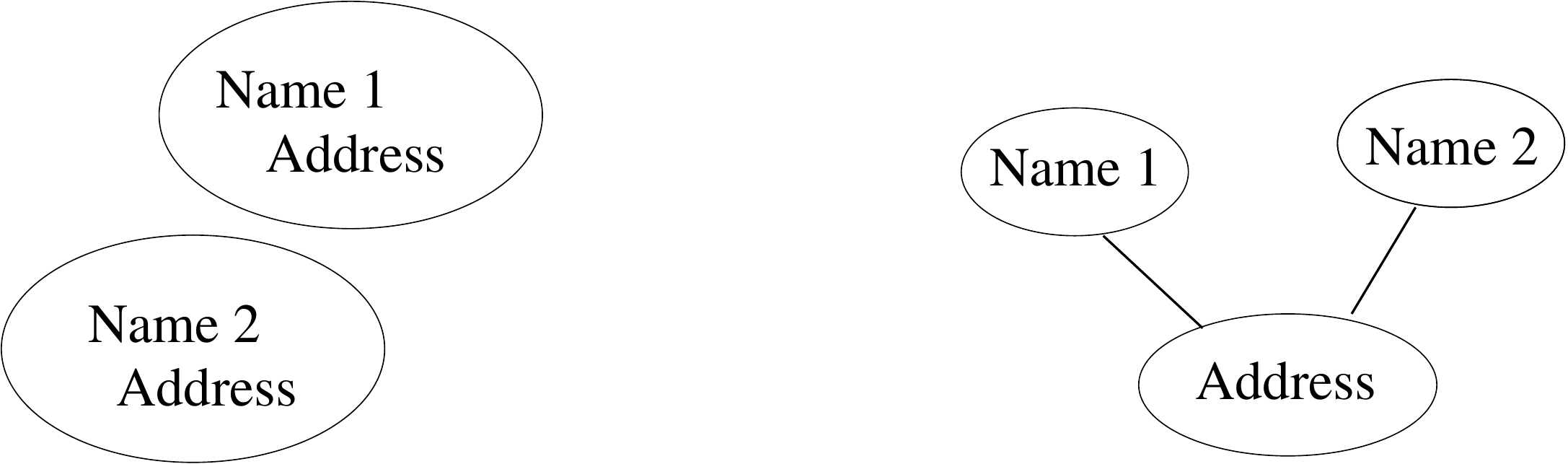}
\caption{Normalization is factoring out a common dependence. It
increases semantic focus at the expense of dynamical fragility.\label{nf}}
\end{center}
\end{figure}
For instance, several people can live at the same address, so one
could extract the address as an independent class of agents (role) so
that the promise that a person lives at an address is independent of
the address's details. This makes it more reliable and potentially
less work to make a change of address to one or more agents.

There are several more extended normal forms, which attempt to go even
further and separate out hidden dependences. From a spacetime
perspective, normal forms therefore turn agents into super-agent
clusters. From a promise theory perspective, they encourage the
separation of agency, and the avoidance of diverging worlds, at the
expense of increased complexity. As the number of agencies is
increased, the number of promises required to bind them in association
grows as the square of the number.

The many worlds problem can easily be overstated. It is fundamentally
about managing causality during change, and reliance on dependency
helps to make consistency inevitable. This is particularly important
because database change was originally expected to be entered by human
operators, prone to errors during repetitive tasks. In data
warehousing, conversely, duplication of information is not a problem,
as the database is populated by machinery from an independent source.
This duplication can be done without human error.  Similarly, in
database replication for backup and disaster recovery, duplication is
actually an intended strategy because of the fragility problem
discussed above. 

The many worlds problem haunts simple cloning of data, nonetheless, by
re-introducing the possibility that inconsistent promises might be
observed by agents working concurrently with the process of
duplication. Since a cloning is a change of promises in space, it is
also a change of time, so duplication cannot be an atomic operation,
without active concealment. At best copying can be gestated in
isolation, typically by mutex locking or temporary partitioning of
space, so that time is perceived as stopped for observers for the
duration of the process. See also the branching discussion in section
\ref{branching}.

\subsection{High-entropy load-sharing spaces and de-coordinatization}\label{highS}

Repetitive network structures are found in many redundant systems,
including biological tissues, physical materials, and artificial
communications networks. Unlike the low entropy ecosystems, the high
entropy regularity serves an effective equivalence
relation, based on low individual significance.  With low significance
comes higher resilience. No point in such a network has any particular
identity, and thus the items are interchangeable, and the spatial
structure is resilient to point disruptions.

The point, in practical terms, is that we don't need to identity
specific agents; finding one that is `good enough' is sufficient to
quench a need.  This return to translational invariance and high
entropy distributions suggests that coordinates are of less
significance. It might even be seen as a goal, from a technology
perspective, to do away with them altogether, even though humans
habitually name things we work with, as part of forming a working
relationship with them.

It is tempting to think that algorithms of information science, like
hashing functions, sorting trees, and so on, might remove the
need for coordinates by using the essence of their structure to locate
particular regions.
\begin{itemize}
\item Hashing functions assign a unique integer to every unique data pattern.
\item Sorting trees (like a coin counter) let data percolate left or right depending
on a criterion (e.g. size, alphabetic value, etc).
\end{itemize}
However, these algorithms merely transduce semantic names into spatial
processes faithfully. They only work if there is a coordinate system
to use as a pointer. They shift part of the coordinate problem into
the names themselves, like a code book (which is, of course, itself an
index). A token string gets converted into a numbered location by
mapping using a total function, somewhat like assigning hotel visitors
room numbers. This is not an elimination of coordinates, but rather a
spatial representation of semantics via a transformation function. It
is a form of possibly irregular auto-indexing (see work on
Batcher-Banyan networks \cite{banyan}).

One might completely eliminate identity from a spacetime, and still
interact with it. It would be like interacting with empty space; some
material promises are needed to grasp onto.  To eliminate all identity
would be to eliminate all information, and hence all structure from
spacetime, leaving only boundary conditions, which are a form of
identity.  There must be a source of information somewhere to
associate material promises with specific locations, data with
records, examples with concepts, else how could they come together
(e.g. the name of a body entering the featureless space)?  

The question becomes about how semantics are interpreted from
dynamical patterns. A few possibilities could be explored:

\begin{itemize}
\item Spacetime is fundamentally ordered as a lattice with long range order, or hierarchy.
Then a small tuple with coordinate numbering suffices to impose a map of locations. 
Within such a regime, we need to place `tenants' at locations:
\begin{itemize}
\item We use the data in the body of a promise to compute and impose the location (spatially ordered).
In this case the information lies in the coding (name) of the change events which
describe time. The source of this information is unknown, but it can be observed.

\item We stack items `first come first served' (FCFS) (time ordered imposition).

\end{itemize}

\item Spacetime is disordered, and something is mixing the gaseous
  mixture so that interaction might be perceived as a random
  collision. Now the source of the mixing information is unknown.
\end{itemize}

\begin{figure}[ht]
\begin{center}
\includegraphics[width=5cm]{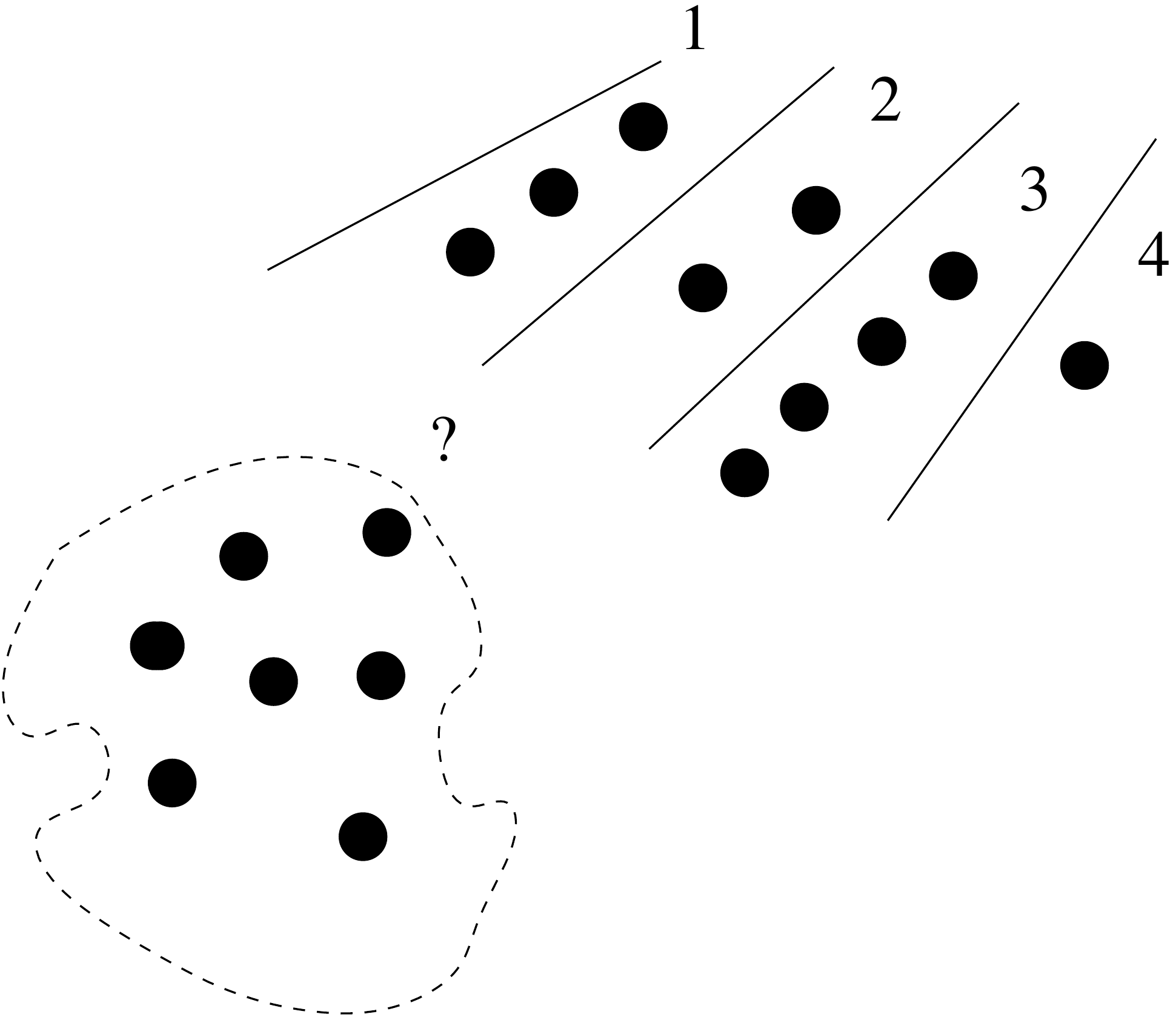}
\caption{A dispatcher is an indexing code book for mapping promises to locations.\label{dispatcher}}
\end{center}
\end{figure}

For example, consider a biological tissue as a homogeneous, high
entropy space of cellular agents. Blood, hormones and other
biochemical signals diffuse through the tissue with a chance of
absorption, to distribute the interaction load across the bulk.  In
order for this to work, batches of incoming promises (e.g. blood) have
to be finely grained so that they can be shared and bind to individual
receptors.

Another example of load assignment could be queueing lanes in a
supermarket or airport check-in counter. Servers are chosen by
customers on the basis of visual feedback of who is free, or by using
a single incoming funnel queue and dispatch process.  The agent
selecting locations for the incoming customers promises to show
queue lengths to customers, which in turn promise to use the
information to self-organize. The decision can be represented by a
dispatcher, which represents the algorithm for the decision (see
figure \ref{dispatcher}). In technology one often uses a `load balancer'
dispatch agent as a specialist semantic component. This breaks the
symmetry of the uniform load-bearing space with a singular point.
It has high semantic value, so it is easily understood, but it has low
dynamical integrity so it is a bottleneck and single point of failure.

A similar idea is a round-robin dispatcher, which counts the state of the next server to
allocate, modulo the total number of servers. A round-robin algorithm
is a trivial hashing function that always indexes new arrivals to the next
queue, like going around a clock. In either case, this is not a Markov
process. It requires memory to implement: memory which must be encoded
into space too.  Thus, the spatial world is used as the memory for
coordinating the process.

These examples are forms of collective {\em stigmergic cooperation},
in which the shared state of the process is written or encoded into
the environment itself as part of the total state.  Ants and other
insects use this approach to cooperate, with pheromone trails.  The
partitioning of regions of space for different purposes becomes part
of the general functional semantics of the space.  The processes
described require memory and a minimal identity to cooperate, whether
to maintain an order (as in FCFS) or a stigmergic trail.

An example of a technology, with a more sophisticated spacetime of
medium-to-high entropy, and which uses physical structure to navigate,
comes from the world of datacentres and resources. Network
architectures are designed to hard-code navigational properties
(analogous to spanning trees). However, they are designed with
functional semantics that ignore their symmetries as spaces.  When
drawing designs, the flatness of a printed page leads engineers to
draw networking structures in two dimensions, whereas a natural
coordinatization would be three dimensional. 

\begin{figure}[ht]
\begin{center}
\includegraphics[width=10cm]{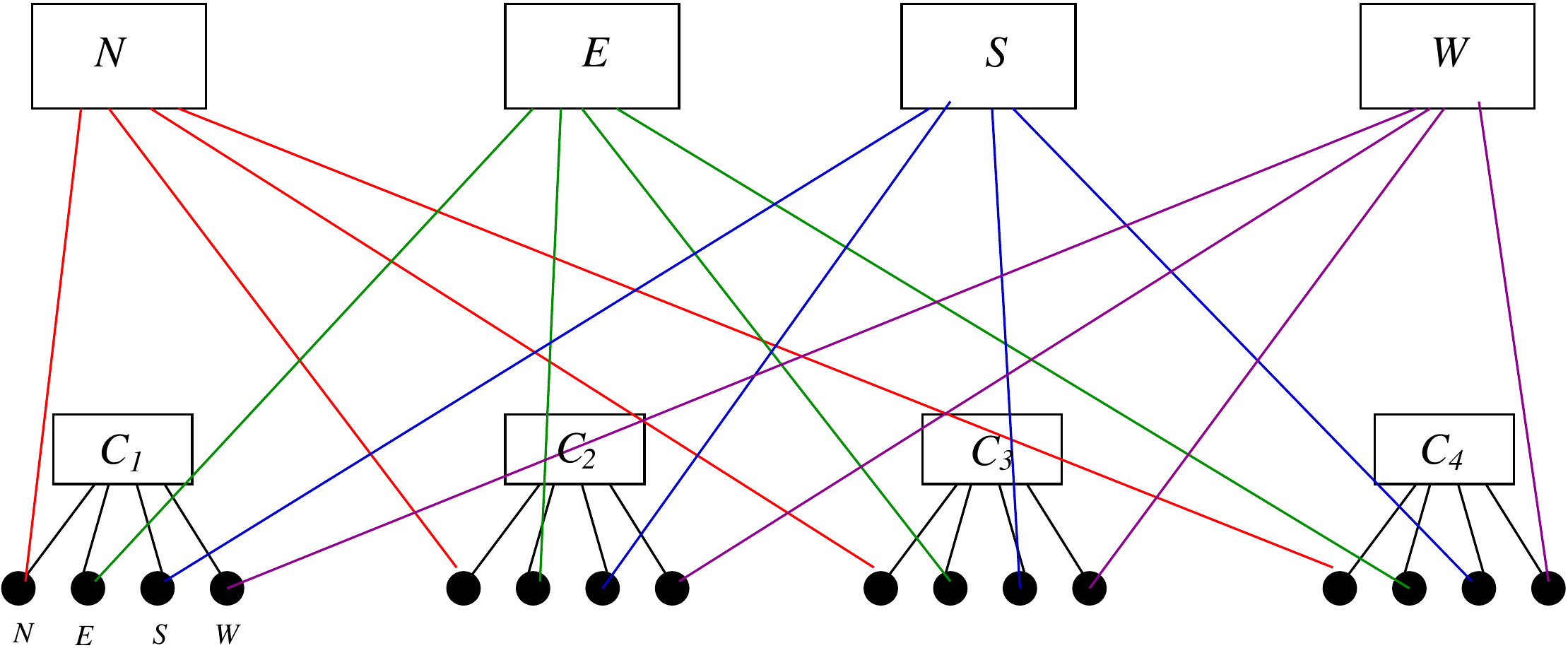}
\caption{A Body Centred Cubic (BCC) lattice. Regularity allows us to exploit symmetries.\label{bcube}}
\end{center}
\end{figure}

\begin{figure}[ht]
\begin{center}
\includegraphics[height=4cm, width=3cm]{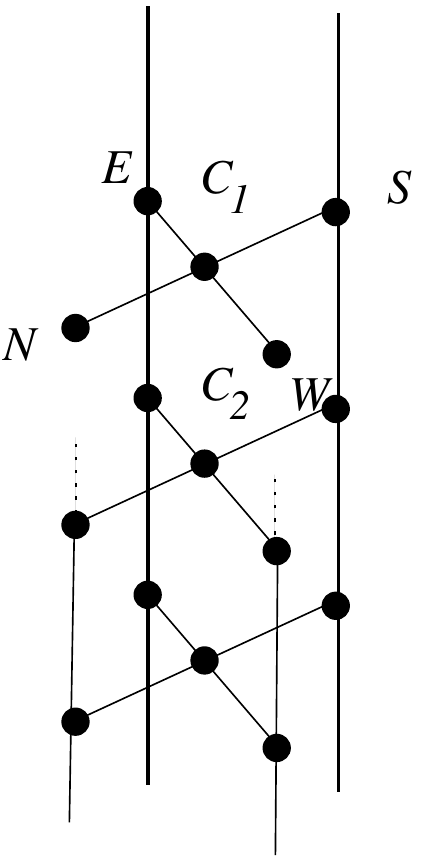}
\caption{The BCC lattice in three dimensions.\label{bcube2}}
\end{center}
\end{figure}

Consider the tree-like structure in figure \ref{bcube}, which has been
used in some cluster communications networks.
In two dimensions, it appears to be a slightly muddled tree.
However, if we re-draw it in three dimensions, it turns out to be a very well
known structure in material science: a body-centred cubic lattice,
similar (but not quite identical) to that of Iron (see Figure
\ref{bcube2}).

\begin{figure}[ht]
\begin{center}
\includegraphics[width=14cm]{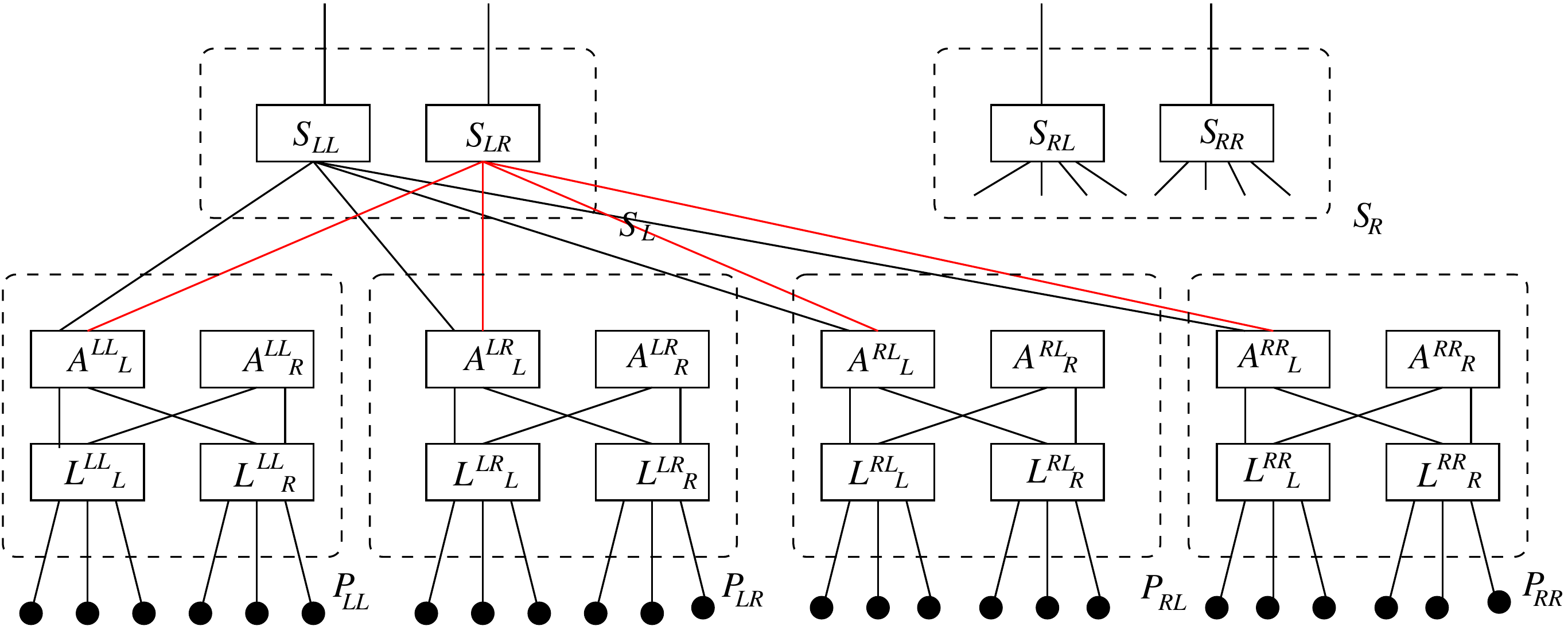}
\caption{A simple 2x2 redundant Clos non-blocking network or fat tree network.
Half the top links to the right hand nodes of the second level are missing for clarity.\label{clos}}
\end{center}
\end{figure}

Another example is the non-blocking networks, as invented for
telephony, which are now equally important in modern datacentres,
because they create regular load-sharing patterns that are extensible.
Consider the pattern in figure \ref{clos}.

The same argument applies here.  The inability to choose a natural
spacetime-motivated coordinate system, in three dimensions, leads to
the complexity of real and current datacentre designs. Racks and
servers are mounted in a three dimensional cubic lattice structure,
because this is how we design humans spaces, but the network devices
are connected in a three-dimensional tree-like form (Figure \ref{clos}),
with a radial symmetry. The amount of criss-crossed and folded cabling
required to perform this unholy union leads to mind-boggling cabling
complexity inside datacentres today.

\begin{figure}[ht]
\begin{center}
\includegraphics[width=14cm]{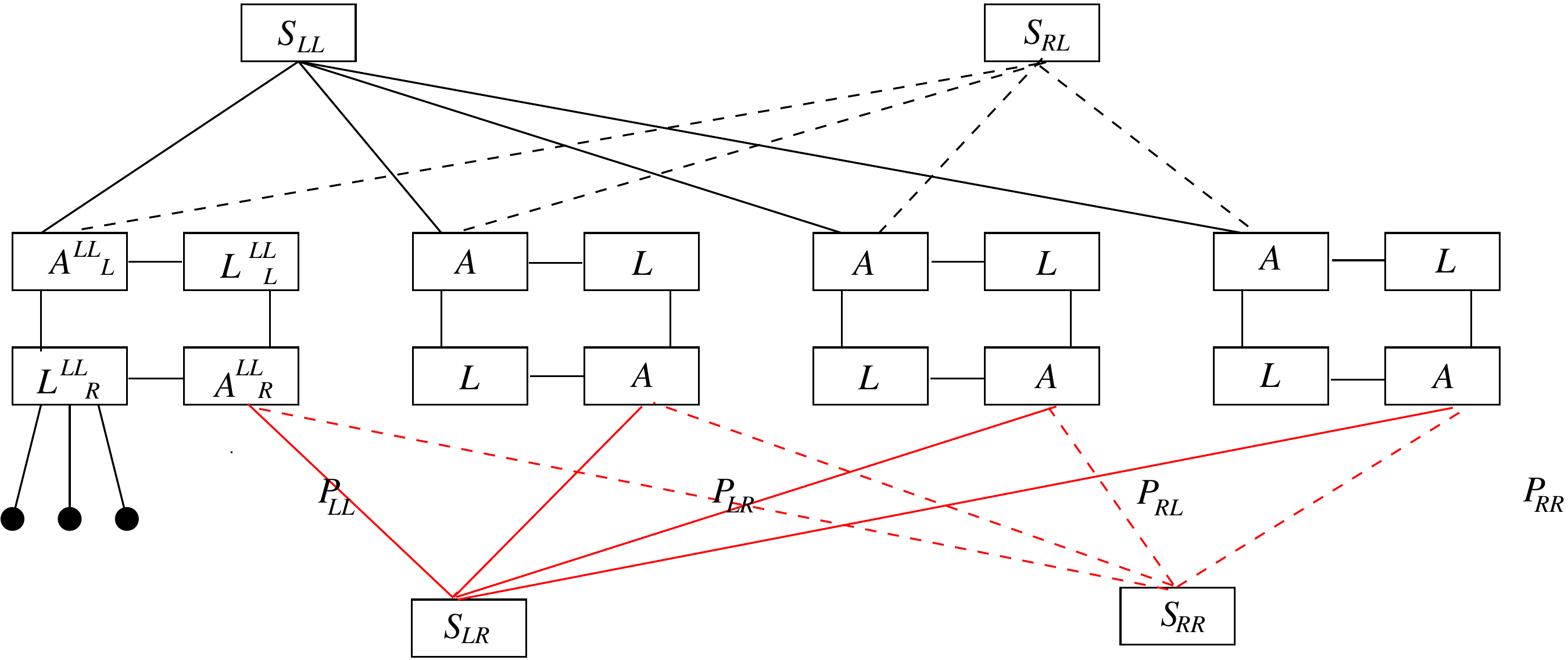}
\caption{Re-drawing the 2x2 Clos network, still in two dimensions.\label{clos2}}
\end{center}
\end{figure}

Could this be avoided? The simpler cubic lattice suggests that we
might have our cake and eat it. In fact, the more robust Clos network
can also be redrawn in a simpler geometry.  If one looks at Figure
\ref{clos}, any mathematician would immediately notice a regular
radial symmetry.  Indeed, if we begin to unravel the topology by
re-drawing and un-twisting the connections, in three dimensions, a
surprising thing happens.

\begin{figure}[ht]
\begin{center}
\includegraphics[width=9cm]{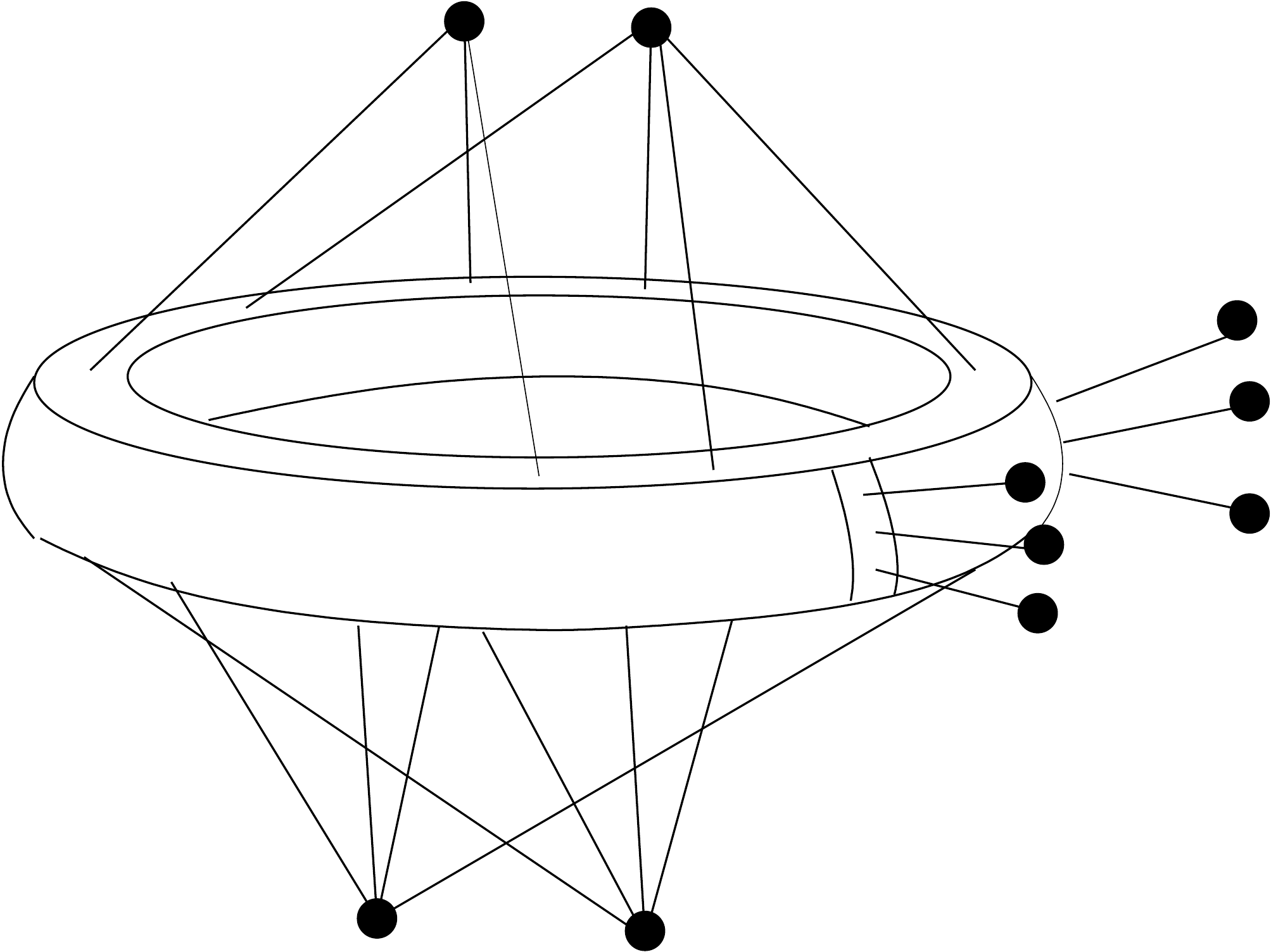}
\caption{The Clos network can now be unfolded into a radial geometry
  with line-of-sight connections that could be maintained by direct
  fibre-free laser optics.\label{clos4}}
\end{center}
\end{figure}

The first step is shown in Figure \ref{clos2}.
Instead of thinking about the network as a hierarchy from
top to bottom, and use the symmetry of the structure to avoid crossing
cables.  The remaining cables at top and bottom, which seem to cross,
do so because we are projecting a three dimensional structure into two
dimensions. If we bend the middle layers into a torus (doughnut) by rotating
them 90 degrees and arrange the outgoing connections,
then we can unfold the entire structure as a toroidal geometry as shown in Figure
\ref{clos4}.

The advantage of a radial design is that all nodes can reach on
another by line-of-sight connections, or perhaps with a simple mirror
to reflect back inside the inner annulus, thus avoiding the need for
expensive folding of fibre-optic waveguides (which need replacing as they grow
brittle).  Lasers could connect separate units directly. Perhaps such
datacentres will be built in the future.

If the leaf nodes are functionally distinguishable (low entropy), the only
interesting identity to assign names to is the port number, or leaf
address, but for practical fault finding it might be useful to encode
part of the path or story for a host.  A triplet such as (Spine, rack, node) 
would serve both needs. 

If leaves could be interchangeable (high entropy), then no coordinate
labels at all would be strictly necessary, just as one has no need to
label the atoms in a sheet of metal. The technological challenge then
is to distribute the load without imposing a technology based on
an unnatural and unnecessary set of identities. Self-routing fabrics,
are one example of this\cite{banyan}.

\subsection{Coordinatizing multi-phase space (a ubiquitous Internet of Things)}

If some agents are fixed and some are mobile or fluid in their
adjacencies, this presents a conundrum for the spanning sets. There
are two choices: either one fixes the coordinates of the gaseous
agents and de-couples the naming from the adjacency, or one redefines
the coordinate matroids continuously along with the flow of the mobile
agents.

There does not seem to be a precedent for this kind of labelling.
It would seem natural to put mobile agents in an entirely
orthogonal set of dimensions so that their wanderings in the other
dimensions can take place without affecting their numbering in
the gaseous phase. This wandering from fixed location to fixed location is
sometimes called `homing'. The mobile agents enter orbit around their home worlds.
If configurations are distinguishable in the gaseous
phase, then this generates a passing of time, in the sense already described.
This can only happen by changing their internal (scalar) properties.

A practical consequence of the spatial labelling, for knowledge spaces, lies in how we find
agents quickly. Without any symmetry to guide the finding of locations,
searching a space is a case of exploratory mapping.

A map, directory or index is area of space, which promises to
associate names with coordinates. To be useful, it should be
significantly smaller than the space it maps, so that searching the
map or index is less costly than searching the space itself.

\beq
{\rm index_i = (name_i, coordinate_i)}
\eeq

Indices are only effective it the list of significant names is less
than or equal to the list of spatial locations. In the worst case, any
space may be considered an index for itself. This is what we mean by
coordinates. The more we aggregate structures, and assign them non-unique names,
the more we can compress spatial structure into coarse-grained
locations.

In a two-phase space, the only thing that has to change is the addition of
extra dimensions that do not possess long range order.

\subsection{Proper time, branching, and many worlds}\label{branching}

The voluntary partitioning of space into semantically independent
regions, whether by boundary or voluntary abstention, is one of the
tools we use most often in technology, and nature itself has found
this to be a stable strategy in dynamical processes.  

Branching processes are processes in which a space partitions into two
or more disjoint sets, which then live out causally separate
histories. Sometimes, the separate entities might continue to interact
as with one another, or they might part company forever, recombine, or
remain weakly coupled.  Examples include:
\begin{itemize}
\item Cell mitosis (cell division) in biology,
\item Process forking or cloning in computer operating systems.
\item Worms and simple life-forms subjected to the guillotine.
\item Code version branching in software engineering
\item Involuntary loss of connectivity in a network due to breakage.
\end{itemize}
Once separated, the version history of each region is a separate
bundle of world lines which can evolve in independent ways, with their
own private space and time. 
They become separate worlds, in the Leibniz, Kripke, Everett sense.

Nothing guarantees that each world timeline will be unique.
Coincidentally, two branches might actually end up in the same state,
in which case they form indistinguishable worlds, but this is
unknowable to any branch\footnote{There is no reason to suppose the
  usual science fictional account in which many worlds branches all
  represent distinct realities.}. Two versions that are
indistinguishable in the same timeline represent the same time.

Once separated, each region is its own proper time clock. This is true
whether it occurs by voluntarily limiting its perception into
non-overlapping regions, or by physical separation. Thus each agent
measures time according to the information it receives.

As observers' worlds diverge, the space the see has to grow in order
to maintain time in their branched world.  In the ultimate case where
space completely partitions into a scenario of singular agents each in
their own world, time must stop in each branch, because the branching
reduces the number of states that can represent change. A perfect
static equilibrium between agents would arise.  Similarly, when a
promise has been kept, and no change of state can be registered due to
convergence, there are no changes of state by which to experience
time. Thus promised process narratives are the black holes of semantic
space: singularities from which timelines come to a static
equilibrium (an operator fixed point).

There is an simple relationship between many worlds and the breakage
of networks into partitions that cannot communicate\footnote{The
  divergence of these worlds would also be interpreted as
  inconsistencies, in the sense of Paxos or the CAP conjecture.},
though these breakages lead to merging of worlds with associated
collisions of intent. The same thing happens in software versioning
systems.  Software versions are not truly separate worlds, merely
embedded channels within a larger world which contains independent
clocks. As swimming lanes, embedded in a larger space, they are just
voluntarily longitudinal super agents, observable by external
agencies. This is why they can be indexed and why versions can be
separated from clock time.

Knowledge spaces, which are facsimiles of other processes, change
connections more dynamically than spaces with few semantics, as they
encode memories. They are often fed continuously with new information,
as representation of an independent, external process.  Each change or
addition leads to a new {\em version} of the knowledge, or a new {\em
  proper time} in the knowledge spacetime, but time is not usually a
measure of interest, except to know that it occurs. Knowledge spaces
{\em are} indices, if only of themselves.

The branching phenomenon can also happen at the meta-level of the
maintainer of the knowledge space. If one information base is cloned
and then each part develops in isolation, they will evolve away from
one another, as different species.

\section{Closing remarks}

This now quite lengthy review of ideas about spacetime attempts to compare and
contrast differing views, focusing on interpretational semantics.
Although, many pages were needed to sketch out these ideas, they only
scratch the surface of application to the worlds we inhabit and create.

The concepts of space and time used in these notes might seem
irritatingly contrary to convention from the standstead of physics,
yet they are in fact very ordinary, and seem entirely natural for
closed discrete systems. Quite possibly, one could construct a
continuum limit, for large number, to obtain results of the kind we
are more used to in natural science.  If that is the case,
intentionality must disappear in the continuum limit.

Some of the key ideas were:

\begin{itemize}
\item Agency plays a role in interpreting spacetime.
\item Semantics (identity and its dynamical prominence) play several roles in mapping out space and time.
\item Dynamics (magnitude and semantic interpretation) are heavily constrained by semantics of observers.
\item Coordinate systems (matroids) can be introduced for all topologies, to interpret space.
\item A discrete space with non-trivial semantics must be able to exist in different phases (gas, molecular, and solid).
\item the fidelity of transmitted information cannot be guaranteed in a space with non-trivial semantics.
\item The ability to define motion and speed is not to be taken for granted in a discrete transition system.
\item Branching into causally independent worlds has to be taken seriously both in natural science, and especially in artificial spaces.
\end{itemize}

With this review of concepts for reference, one may now proceed to
apply the ideas more rigorously to the study of intentional spaces,
such as information systems for the modern world.

\section*{Acknowledgement}

I am grateful to Paul Borrill for interesting discussions on the nature of time.

\bibliographystyle{unsrt}
\bibliography{spacetime} 

\appendix
\section{Graph bases and coordinatized dimensions}\label{appendix}

It is instructive to see examples of how to specify the dimensional
elements in a graph, using the notion of matroids or independent sets.
This puts the description of graphs on a par with that of lattices
and manifolds.

\begin{figure}[ht]
\begin{center}
\includegraphics[width=4.5cm]{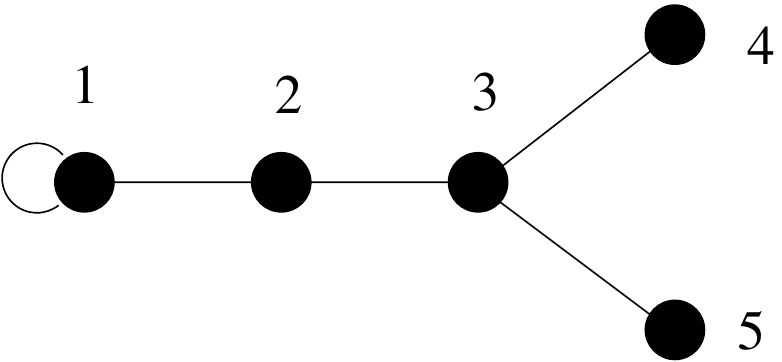}
\caption{A small graph - but how many dimensions does it have?\label{dimgraph0}}
\end{center}
\end{figure}

Consider the graph shown in fig. \ref{dimgraph0}. We may use this as a simple
example. The graph has a self loop and a tree structure.

The adjacency matrix for this graph is:
\beq 
A_{ij} = \left(
\begin{array}{ccccc}
1 & 1 & 0 & 0 & 0\\
1 & 0 & 1 & 0 & 0\\
0 & 1 & 0 & 1 & 1\\
0 & 0 & 1 & 0 & 0\\
0 & 0 & 1 & 0 & 0\\
\end{array} 
\right)
\eeq

This can be decomposed into a number of generators of spanning sets. If we choose rank $r$, then
\beq
A_{ij} = \sum_{a=1}^r I_a.
\eeq

\subsection{Example 1}\label{appex1}

\begin{figure}[ht]
\begin{center}
\includegraphics[width=6.5cm]{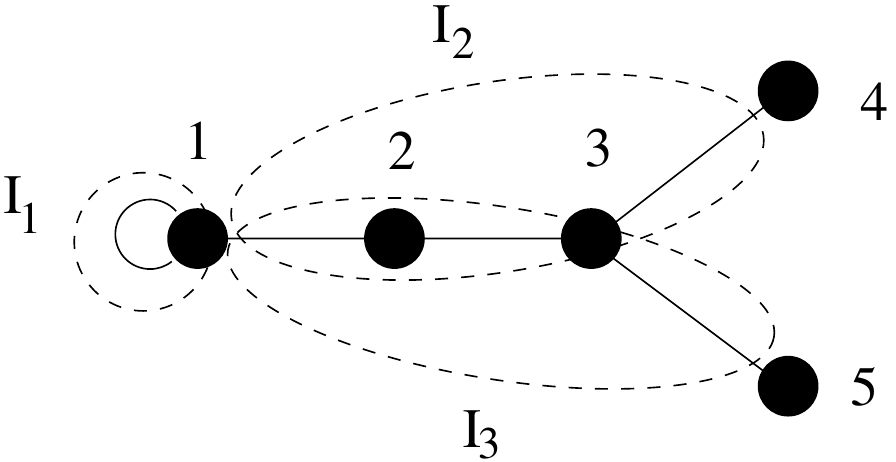}
\caption{A 3 dimensional spanning basis.\label{dimgraph1}}
\end{center}
\end{figure}

For example consider the matroid basis shown in fig. \ref{dimgraph1}.
This may be written as independent link sets as a linear decomposition of the adjacency matrix
with three independent sets to span the space.
\beq 
A = I_1 + I_2 + I_3
\eeq
\beq
A_{ij} = 
\left(
\begin{array}{ccccc}
1 & 0 & 0 & 0 & 0\\
0 & 0 & 0 & 0 & 0\\
0 & 0 & 0 & 0 & 0\\
0 & 0 & 0 & 0 & 0\\
0 & 0 & 0 & 0 & 0\\
\end{array} 
\right)
+
\left(
\begin{array}{ccccc}
0 & \2 & 0 & 0 & 0\\
\2 & 0 & \2 & 0 & 0\\
0 & \2 & 0 & 1 & 0\\
0 & 0 & 1 & 0 & 0\\
0 & 0 & 0 & 0 & 0\\
\end{array} 
\right)
+
\left(
\begin{array}{ccccc}
0 & \2 & 0 & 0 & 0\\
\2 & 0 & \2 & 0 & 0\\
0 & \2 & 0 & 0 & 0\\
0 & 0 & 0 & 0 & 1\\
0 & 0 & 0 & 1 & 0\\
\end{array} 
\right) \eeq The fractional factors show that the basis is not
orthogonal (i.e. disjoint). The spanning sets do overlap.
Interpreting the spanning sets as the imposed dimensionality, and
grouping related points together, we obtain tuples.  \beq
v_1 &=& (1,1,1)\\
v_2 &=& (0,1,1)\\
v_3 &=& (0,2,2)\\
v_4 &=& (0,3,0)\\
v_5 &=& (0,0,3) 
\eeq 
As the regions overlap and contain multiple
points, The three points lying along directions $I_2$ and $I_3$ are
labelled simply with coordinates 1,2,3 as consecutive elements. This
ordering is arbitrary, of course, but can be motivated by the action
of the adjacency sets as translation generators to generate the
ordering.  The names of the node in the graph need not match (they
belong to a global namespace).

Notice also that I wrote a value of 0 in the spirit of a linear
interpretation, meaning `does not depend on this direction', but this
is an arbitrary choice of ordinal that symbolizes that we are beyond
the edge of the space in this direction. One could easily write $b$
for `boundary' or `nw' for nowhere.  I like the way this enforced
dimensionalization shows how the appropriateness of the concept of
dimensionality is more closely related to regular symmetry than to
connectivity or the availability of a sense of {\em direction}.
Put another way, the semantics of direction may be freely chosen,
but not without the cost of some weirdness relative to a regular
lattice world.

\begin{figure}[ht]
\begin{center}
\includegraphics[width=6.5cm]{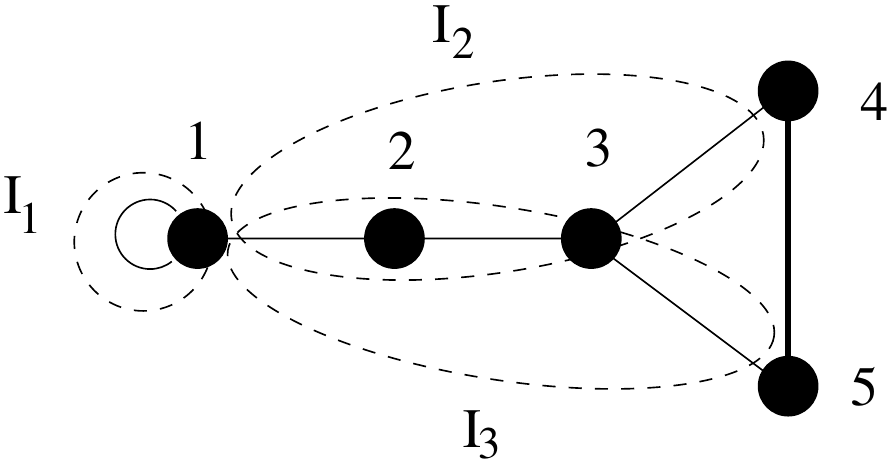}
\caption{Adding an edge to make a cycle does not change the basis, as loops cannot be included.\label{dimgraph4}}
\end{center}
\end{figure}

If we add an edge to the graph as in fig. \ref{dimgraph4}, so that we create a 
non-trivial cycle, then nothing changes in the coordinates: all the nodes as still
in the same place. All we've done is to change the underlying topology.
This shows that coordinates and topology are separate descriptions that do not
necessarily reflect one another, except in cases of assumed symmetry.

\subsection{Example2}

\begin{figure}[ht]
\begin{center}
\includegraphics[width=6.5cm]{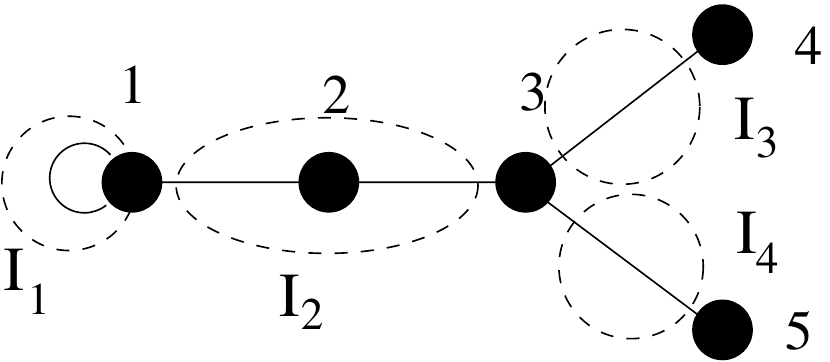}
\caption{A 4 dimensional spanning basis.\label{dimgraph2}}
\end{center}
\end{figure}

Next consider fig. \ref{dimgraph2}
This may be written as independent link sets as a linear decomposition of the adjacency matrix:
\beq 
A = I_1 + I_2 + I_3 + I_4
\eeq
\beq
A_{ij} = 
\left(
\begin{array}{ccccc}
1 & 0 & 0 & 0 & 0\\
0 & 0 & 0 & 0 & 0\\
0 & 0 & 0 & 0 & 0\\
0 & 0 & 0 & 0 & 0\\
0 & 0 & 0 & 0 & 0\\
\end{array} 
\right)
+
\left(
\begin{array}{ccccc}
0 & 1 & 0 & 0 & 0\\
1 & 0 & 1 & 0 & 0\\
0 & 1 & 0 & 0 & 0\\
0 & 0 & 0 & 0 & 0\\
0 & 0 & 0 & 0 & 0\\
\end{array} 
\right)
+
\left(
\begin{array}{ccccc}
0 & 0 & 0 & 0 & 0\\
0 & 0 & 0 & 0 & 0\\
0 & 0 & 0 & 1 & 0\\
0 & 0 & 1 & 0 & 0\\
0 & 0 & 0 & 0 & 0\\
\end{array} 
\right)
+
\left(
\begin{array}{ccccc}
1 & 0 & 0 & 0 & 0\\
0 & 0 & 0 & 0 & 0\\
0 & 0 & 0 & 0 & 0\\
0 & 0 & 0 & 0 & 1\\
0 & 0 & 0 & 1 & 0\\
\end{array} 
\right)
\eeq
In this basis of dimension 4 (which we observe subtly is labelled by links not nodes),
we write tuples for the vertices $v_1 \ldots v_5$ as:
\beq
v_1 &=& (1,1,0,0)\\
v_2 &=& (0,1,1,0)\\
v_3 &=& (0,1,1,1)\\
v_5 &=& (0,0,1,0)\\
v_5 &=& (0,0,0,1)
\eeq

\subsection{Example 3}

\begin{figure}[ht]
\begin{center}
\includegraphics[width=6.5cm]{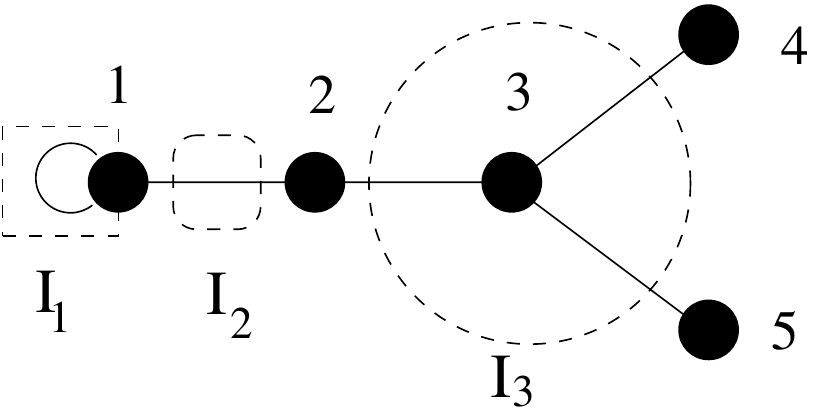}
\caption{Alternative 3 dimensional spanning basis.\label{dimgraph3}}
\end{center}
\end{figure}

In fig. \ref{dimgraph3} we see an alternative three dimensional 
matroid basis.
\beq
A_{ij} = 
\left(
\begin{array}{ccccc}
1 & 0 & 0 & 0 & 0\\
0 & 0 & 0 & 0 & 0\\
0 & 0 & 0 & 0 & 0\\
0 & 0 & 0 & 0 & 0\\
0 & 0 & 0 & 0 & 0\\
\end{array} 
\right)
+
\left(
\begin{array}{ccccc}
0 & 1 & 0 & 0 & 0\\
1 & 0 & 0 & 0 & 0\\
0 & 0 & 0 & 0 & 0\\
0 & 0 & 0 & 0 & 0\\
0 & 0 & 0 & 0 & 0\\
\end{array} 
\right)
+
\left(
\begin{array}{ccccc}
0 & 0 & 0 & 0 & 0\\
0 & 0 & 1 & 0 & 0\\
0 & 1 & 0 & 1 & 0\\
0 & 0 & 1 & 0 & 1\\
0 & 0 & 0 & 1 & 0\\
\end{array} 
\right) 
\eeq 
Interpreting the spanning sets again as the imposed dimensionality, and
grouping related points together, we obtain tuples.  
\beq
v_1 &=& (1,0,0)\\
v_2 &=& (0,1,1)\\
v_3 &=& (0,0,2)\\
v_4 &=& (0,0,3)\\
v_5 &=& (0,0,4) 
\eeq 
Four of the vertices now exist inside the direction $I_3$, whereas only one vertex
is in each of $I_1$ and $I_2$.

\subsection{Artificiality of dimensions}\label{embed}

\begin{figure}[ht]
\begin{center}
\includegraphics[width=8.5cm]{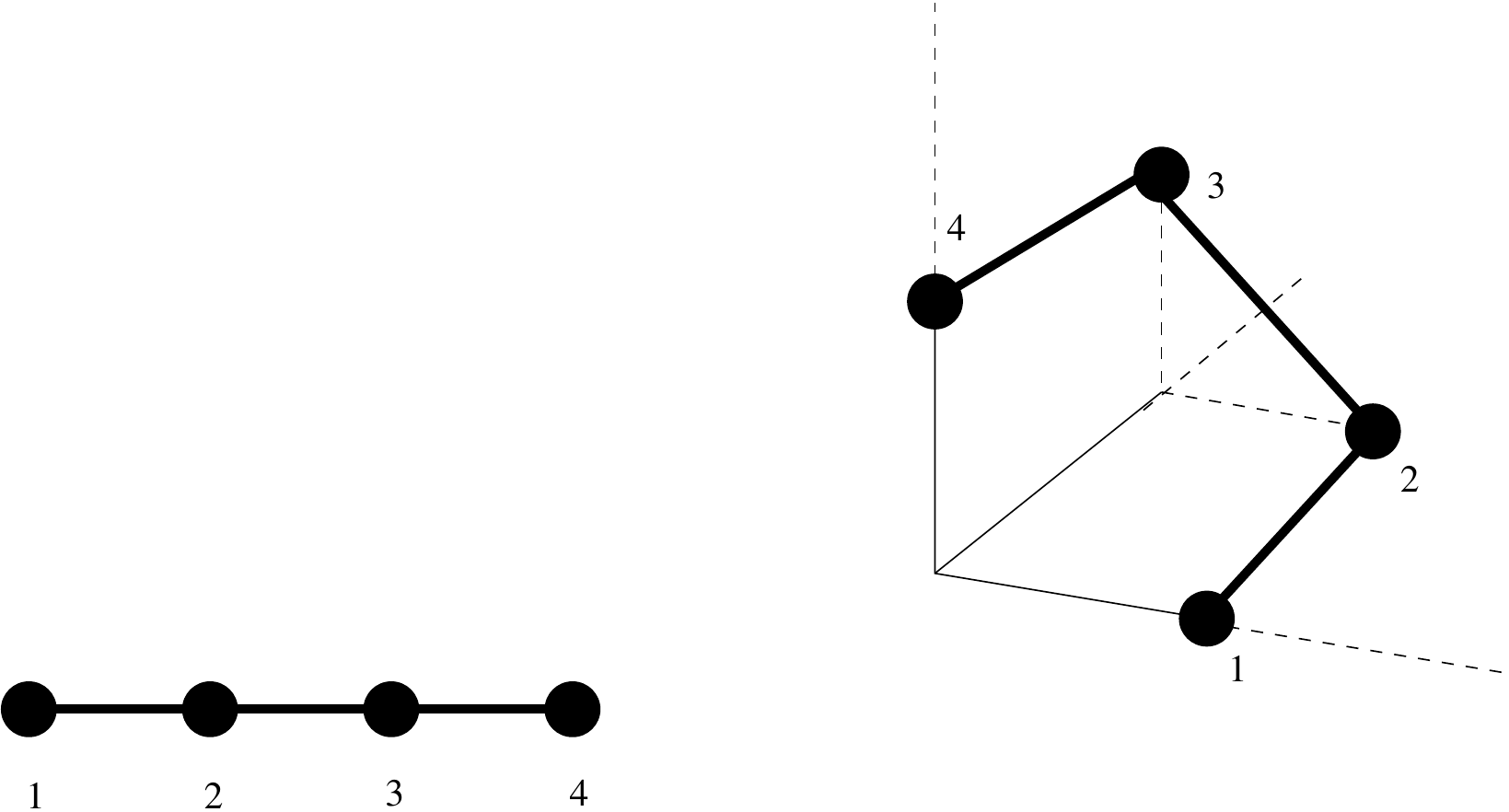}
\caption{A line bent into three dimensions\label{line}}
\end{center}
\end{figure}

It is instructive to take an extreme case of a linear graph and
reinterpret it in a three dimensional basis. To build sufficient rank,
we then need at least four vertices (see fig. \ref{line}).

Taking each link as a separate orthogonal direction, we obtain
the coordinates of the four points to be:
\beq
v_1 &=& (1,0,0)\\
v_2 &=& (1,1,0)\\
v_3 &=& (0,1,1)\\
v_4 &=& (0,0,1)\\
\eeq 
Viewing these in a pseudo 3d lattice shows that the constraints simply turn this
into a bent string. It is in the nature of graphs to be bounded, though this one
is rather extreme. What's important to see is that the degrees of freedom
are no longer related to the dimensionality, but rather to the connectivity
constraints.

With such coordinatizations, the computation of distance no more
convenient.  There is no simple Pythagoras formula, or line of sight
distance, because we cannot interpolate the existence of smooth continuous
expanse.

\end{document}